 \documentclass[preprint2,longabstract]{aastex}
\usepackage{natbib}
\usepackage{color}
\usepackage{longtable}

\shorttitle{12C-and 13C-methyl formate}
\shortauthors{Favre et al.}
\begin{document}

\title{$^{13}$C--methyl formate: observations of  a sample of high mass star-forming regions including Orion--KL and spectroscopic
characterization\thanks{This publication is based on data acquired with the Atacama Pathfinder Experiment (APEX). APEX is a collaboration between the Max-Planck-Institut fur Radioastronomie, the European Southern Observatory, and the Onsala Space Observatory (under programme ID 089.F-9319).}}

\author{C\'ecile Favre}
\affil{Department of Astronomy, University of Michigan, 500 Church St., 
    Ann Arbor, MI 48109, USA;}
\email{cfavre@umich.edu}

\and

\author{Miguel Carvajal}
\affil{Dpto. F\'{\i}sica Aplicada, Unidad Asociada CSIC, Facultad
  de Ciencias Experimentales, Universidad de Huelva, 21071, Spain}
\email{miguel.carvajal@dfa.uhu.es}

\and

\author{David Field}
\affil{Department of Physics and Astronomy, University of Aarhus, Ny Munkegade 120, 8000 Aarhus C, Denmark}

\and

\author{Jes~K. J\o rgensen, Suzanne~E. Bisschop}
\affil{Centre for Star and Planet Formation, Niels Bohr Institute, University of Copenhagen, Juliane Maries Vej 30, 2100 Copenhagen \O, Denmark}
\affil{Natural History Museum of Denmark, University of Copenhagen, {\O}ster Voldgade 5-7, 1350 Copenhagen K., Denmark}
  
\and

\author{Nathalie Brouillet, Didier Despois, Alain Baudry}
\affil{Univ. Bordeaux, LAB, UMR 5804, F-33270, Floirac, France}
\affil{CNRS, LAB, UMR 5804, F-33270, Floirac, France}

\and

\author{Isabelle Kleiner}
\affil{Laboratoire Interuniversitaire des Syst\`emes Atmosph\'eriques
  (LISA), CNRS, UMR 7583, Universit\'e de Paris-Est et Paris Diderot,
  61, Av. du G\'en\'eral de Gaulle, 94010 Cr\'eteil Cedex, France}

\and

\author{Edwin A. Bergin, Nathan R. Crockett, Justin L. Neill}
\affil{Department of Astronomy, University of Michigan, 500 Church St., 
    Ann Arbor, MI 48109, USA;}
    
\and

\author{Laurent Margul\`es, Th\'er\`ese R. Huet, Jean Demaison}
\affil{Laboratoire de Physique des Lasers, Atomes et Mol\'ecules, UMR
  CNRS 8523, Universit\'e Lille I, 59655 Villeneuve d'Ascq Cedex, France}

%
%
\begin{abstract}
We have surveyed a sample of massive star-forming regions located over a range of distances from the Galactic centre for methyl formate, HCOOCH$_{3}$, and its isotopologues H$^{13}$COOCH$_{3}$ and HCOO$^{13}$CH$_{3}$. The observations were carried out with the APEX telescope in the frequency range 283.4--287.4~GHz. Based on the APEX observations, we report tentative detections of the $^{13}$C-methyl formate isotopologue HCOO$^{13}$CH$_{3}$ towards the following four massive star-forming regions: Sgr~B2(N-LMH), NGC~6334~IRS~1, W51 e2 and G19.61-0.23.
In addition, we have used the 1~mm ALMA science verification observations of Orion-KL and  confirm the detection of the $^{13}$C-methyl formate species in Orion-KL and image its spatial distribution. Our analysis shows that the  $^{12}$C/$^{13}$C isotope ratio in methyl formate toward Orion-KL Compact Ridge and Hot Core-SW components (68.4$\pm$10.1 and 71.4$\pm$7.8, respectively) are,  for both the $^{13}$C-methyl formate isotopologues, commensurate with the average $^{12}$C/$^{13}$C ratio of CO derived toward Orion--KL. Likewise, regarding the other sources, our results are consistent with the $^{12}$C/$^{13}$C in CO. We also report the spectroscopic characterization, which includes a complete partition function, of the complex H$^{13}$COOCH$_{3}$ and HCOO$^{13}$CH$_{3}$ species. New spectroscopic data for both isotopomers H$^{13}$COOCH$_{3}$ and HCOO$^{13}$CH$_{3}$, presented in this study, has made it possible to measure this fundamentally important isotope ratio in a large organic molecule for the first time.
\end{abstract}


%
\section{Introduction}
\label{sec:introduction}

Determination of elemental isotopic ratios is valuable for understanding the chemical evolution of interstellar material.
In this light, carbon monoxide $^{12}$C/$^{13}$C, can be an important tracer of process of isotopic fractionation.
Numerous measurements of the $^{12}$C/$^{13}$C ratios towards Galactic sources have been carried out using simple molecules such as CO, CN and H$_{2}$CO \citep{Langer:1990,Langer:1993,Wilson:1994,Wouterloot:1996,Milam:2005}. These studies have shown that the $^{12}$C/$^{13}$C ratio becomes larger with increasing distance from the Galactic Center. More specifically, 
\citet{Wilson:1999} gives a mean $^{12}$C/$^{13}$C ratio of 69$\pm$6 in the Local ISM, 53$\pm$4 at 4~kpc (the molecular ring) and of about 20 toward the Galactic center, showing a strong gradient that can be given for CO by \citep{Milam:2005}:
\begin{equation}
\rm
^{12}C/^{13}C=5.41(1.07)D_{GC}+19.03(7.90)\label{eq:ratio1}\end{equation} with $D\rm_{GC}$ the distance from the Galactic Center in kpc.
Furthermore, \citet{Milam:2005} have shown that the $^{12}$C/$^{13}$C gradient for the CO, CN and H$_{2}$CO molecular species can be defined by:
\begin{equation}
\rm
^{12}C/^{13}C=6.21(1.00)D_{GC}+18.71(7.37),\label{eq:ratio}\end{equation}
with $D\rm_{GC}$ the distance from the Galactic Center in kpc. This makes these carbon isotopologue species valuable indicators of Galactic chemical evolution:  although they are formed through different chemical pathways and present different chemical histories, they do not show significantly different $^{12}$C/$^{13}$C ratios.

Until now the $^{12}$C/$^{13}$C ratio has predominantly been measured in simple species that form mostly via reactions in the gas phase.
In contrast, complex molecules are believed to form, for the most part, on grain surfaces, although gas phase formation cannot be ruled out \citep[e.g.][] {Herbst:2009,Charnley:2005}.  In this case, for complex species, the isotopic ratios might betray evidence of the grain surface formation as the ratio would differ from pure gas phase formation since gas phase processes, such as selective photodissociation and fractionation in low-temperature ion-molecule reactions, would impact the $^{12}$C/$^{13}$C ratio which is then implanted in larger species \citep{Charnley:2004,Wirstrom:2011}. Indeed, \citet{Wirstrom:2011} have shown that the isotopic $^{12}$C/$^{13}$C ratio in methanol (CH$_3$OH) can be used to distinguish a gas-phase origin from an ice grain mantle one. Methanol is believed to be formed on dust grains from hydrogenation of CO \citep[e.g.][]{Cuppen:2009}. If this is the case, the measured $^{12}$C/$^{13}$C ratios in CO and  CH$_3$OH should be similar. Otherwise,  the isotopic $^{12}$C/$^{13}$C ratio in methanol should be higher than the one in CO due to fractionation of species that rely on the atomic `carbon isotope pool' for formation \citep[see][]{Wirstrom:2011,Langer:1984}.

In that light, we extend the $^{12}$C/$^{13}$C investigation to interstellar methyl formate (HCOOCH$_{3}$, hereafter MF or $^{12}$C--MF), which is among the most abundant complex molecules detected in massive star-forming regions \citep[e.g.][]{Liu:2001,Remijan:2004,Bisschop:2007,Demyk:2008,Shiao:2010,Favre:2011,Friedel:2008,Friedel:2012}. Also, the detection of  both the $^{13}$C--MF isotopologues, H$^{13}$COOCH$_{3}$ (hereafter, $^{13}$C$_1$--MF) and HCOO$^{13}$CH$_{3}$ (hereafter,$^{13}$C$_2$--MF) have been reported toward Orion--KL by \citet{Carvajal:2009} based on IRAM 30m-antenna observations. More specifically, we suggest that the $^{12}$C/$^{13}$C ratio in methyl formate could also be used as an indicator of its formation origin. This since methyl formate may be efficiently formed close to the surface of icy grain mantles during the hot core warm up phase via reactions involving mobile radical species, such as CH$_3$O and HCO, that are produced by cosmic--ray induced photodissociation of methanol ices and ultimately owe their origin to hydrogenation of CO \citep[e.g.][]{Bennett:2007,Horn:2004,Neill:2011,Garrod:2006,Garrod:2008,Herbst:2009}.  In this instance and in agreement with \citet{Wirstrom:2011}, if the $^{12}$C/$^{13}$C ratios in methyl formate, methanol and CO are similar, that would likely suggest a formation on grain surfaces.

In this paper we investigate the carbon isotopic ratio for methyl formate isotopologues and therefore address the issue whether  the $^{12}$C/$^{13}$C ratio is the same for both simple and large molecules.  Our analysis is based on recent spectroscopic and laboratory measurements of both the common isotopologue and the $^{13}$C isotopologues \citep[see,][and this study]{Carvajal:2007,Carvajal:2009,Carvajal:2010,Ilyushin:2009,Kleiner:2010,Margules:2010,Haykal:2014}. We would particularly like to stress that in order to derive a $^{12}$C/$^{13}$C ratio with accuracy and  to significantly reduce uncertainties, homogeneous data are a necessity. In Section \ref{sec:observations}, we present the ALMA Science Verification observations of Orion-KL along with the APEX observations of our massive star-forming regions sample.  Spectroscopic characterization of the  $^{13}$C-methyl formate molecules is presented in Section \ref{sec:spectro}. Data modeling, results and analysis are presented and discussed in Sects.~\ref{sec:resultsa}, \ref{sec:results} and~\ref{sec:discussion}, with conclusions set out in Sect.~\ref{sec:Conclusions}.

%
\section{Observations and data reduction}
\label{sec:observations}

%
\begin{deluxetable}{llllllll}
\tablewidth{0.pt}
\tablecolumns{8}
\tablecaption{List of sources observed with the APEX telescope.\label{tab1}}
\tablehead{	
Source &Observed & $\rm\alpha_{J2000}$&$\rm\delta_{J2000}$& V$\rm_{LSR}$ & Distance from &Distance from& \\
& Date& & & & the Sun&the GC\\
& &($\rm^{h}:^{m}:\fs$) &($\degr:\arcmin:\farcs$)& (km~s$^{-1}$)& (kpc) &(kpc) }
\startdata
Sgr~B2(N-LMH) & 2012 April 02, 03	&17:47:19.9 & -28:22:19.5 & 64.0&7.1&0.1\tablenotemark{a}\\
G24.78+0.08	& 2012 April 02 & 18:36:12.6     & -07:12:11.0&111.0& 7.7&3.7\tablenotemark{b}\\
G29.96-0.02	& 2012 April 01& 18:46:04.0     & -02:39:21.5& 98.8& 6.0 & 4.6\tablenotemark{c} \\
G19.61-0.23	& 2012 March 28 &18:27:38.1    &  -11:56:39.0& 40.0&3.5&4.8\tablenotemark{d}  \\
NGC~6334~IRS~1 & 2012 March 28	& 17:20:53.0& -35:47:02.0&-8.0& 1.7 & 6.8\tablenotemark{e} \\
				&	2012 August 17, 18			&			&		&	&	&	\\
W51 e2		& 2012 April 02 & 19:23:43.9  &    $+$14:30:34.8 &55.3 &5.41&8.3\tablenotemark{f}\\
Orion-KL		& 2012 April 01 &05:35:14.2	& -05:22:36.0 & 8.0& 0.4 & 8.9\tablenotemark{g} \\
				&	2012 April 04, 05		&			&		&	&	&	\\
 \enddata
\tablenotetext{a}{\citet{Milam:2005}.}
\tablenotetext{b}{\citet{Beltran:2011}.}
\tablenotetext{c}{\citet{Pratap:1999}.}
\tablenotetext{d}{\citet{Remijan:2004}.}
\tablenotetext{e}{\citet{Kraemer:1998}.}
\tablenotetext{f}{\citet{Sato:2010}.}
\tablenotetext{g}{\citet{Remijan:2003}.}
     \end{deluxetable} 
     
\subsection{ALMA Science Verification observations}
Orion-KL was observed with 16 antennas (each of 12~m in diameter) on January 20, 2012, as part of the ALMA Science Verification (hereafter, ALMA-SV) program. The observations cover the frequency range 213.7~GHz to 246.6~GHz in band 6. The phase-tracking centre was $\alpha_{J2000}$ = 05$^{h}$35$^{m}$14$\fs$35, $\delta_{J2000}$ = -05$\degr$22$\arcmin$35$\farcs$00. The observational data consist of 20 spectral windows, each with 488~kHz channel spacing resulting in 3840 channels across 1.875~GHz effective bandwidth.

We used the public release calibrated data that are available through the ALMA Science Verification Portal\footnote{http://almascience.eso.org/almadata/sciver/OrionKLBand6/}. Data reduction and continuum subtraction were performed using the Common Astronomy Software Applications (CASA) software\footnote{http://casa.nrao.edu}.  
More specifically, the continuum emission was estimated by a zeroth order fit to the line-free channels within each spectral window (hereafter spw) and subtracted.
Finally, the spectral line data cleaning was performed using the \citet{Clark:1980} method and a pixel size of 0.4$\arcsec$. Also, a Briggs weighting with a robustness parameter of 0.0 was applied giving a good trade-off between natural and uniform weighting \citep{Briggs:1995}.
The resulting  synthesized beam sizes are: 
\begin{description}
\item[- ] 1.6$\arcsec$ $\times$ 1.1$\arcsec$ (P.A. of about -176-- 4$\degr$) for the spw  0, 1, 4, 5, 8, 9, 12, 13, 16 and 17, 
\item[- ] 1.7$\arcsec$ $\times$ 1.2$\arcsec$ (P.A. of about -1-- -11$\degr$) for the spw 2, 3, 6, 7, 10, 11, 14, 15, 18 and 19.
\end{description}

\subsection{APEX observations}
%

\subsubsection{Source sample}

Our survey is composed of a sample of seven high-mass star-forming regions that are listed in Table~\ref{tab1}  together with their respective coordinates, LSR velocities and distances from the Sun as well as from the Galactic Center. The 7 sources were primarily selected upon the following criteria: 
\textit{i)} the previous detection of the main HCOOCH$_{3}$ isotopologue, based on single-dish and/or interferometric observations \citep[e.g.][]{Liu:2001,Remijan:2004,Bisschop:2007,Demyk:2008,Friedel:2008,Shiao:2010,Favre:2011,Belloche:2009,Widicus-Weaver:2012,Fontani:2007,Olmi:2003,Beuther:2007,Beuther:2009,Kalenskii:2010,Requena-Torres:2006,Hollis:2000,Mehringer:1997}, with a derived column density in the range 10$^{16}$-10$^{17}$~cm$^{-2}$ depending on the source and the assumed source size, and  \textit{ii)} covering a wide range in distance from the Galactic Center, here from 0.1~kpc to 8.9~kpc (see Table~\ref{tab1}).

\subsubsection{Observations}
\label{sec:observationspart2}

The observations were performed with the APEX telescope on Llano de Chajnantor, Northern Chile, between March and August 2012 (see Table~\ref{tab1}). The Swedish Heterodyne Facility Instrument (SHeFi) APEX-2 receiver, which operates with an IF range of 4--8~GHz, was used in single sideband mode in connection to the eXtended bandwidth Fast Fourier Transform Spectrometer (XFFTS) backend in the frequency range 283.4~GHz -- 287.4~GHz. This frequency range was chosen from line intensity predictions based on the Orion--KL study by  \citet{Carvajal:2009}. The half-power beam size is 22$\arcsec$ for observations at 285.4~GHz.
The image rejection ratio is 10~dB over the entire band\footnote{http://www.apex-telescope.org/heterodyne/shfi/}. Also, the XFFTS  backend covers 2.5~GHz bandwidth instantaneously with a spectral resolution of about 0.08~MHz (corresponding to 0.08~km/s). 
However, noting that line-widths of the target lines are estimated to be between 4 and 8~km~s$^{-1}$ based on earlier methyl formate observations referred to above, the spectra were smoothed to a spectral resolution of 1.5~km~s$^{-1}$.
Further, in this paper, the spectra are reported in units of the main beam temperature (T$\rm_{MB}$), that is given by 
\begin{equation}
\rm
T_{MB} = \frac{\eta_{f}}{\eta_{MB}}T_{A}^{*},
\end{equation}
where $\eta_{f}$T$_{A}^{*}$ is the antenna temperature outside the atmosphere, $\eta_{f}$ the forward efficiency ($\eta_{f}$=0.97 for the APEX-2 instrument\footnote{see http://www.apex-telescope.org/telescope/efficiency/}) and $\eta_{MB}$ the main beam efficiency ($\eta_{MB}$=0.73 for the APEX-2 instrument$^{5}$).

G29.96-0.02,  G19.61-0.23, NGC 6334~IRS~1 (2012 March 28) and W51~e2 observation data were taken in wobbler switching with a throw of 150$\arcsec$ in azimuth and a wobbling rate of 0.5~Hz in symmetric mode.
Regarding Orion-KL, NGC~6334~IRS~1 (2012 August 17 and 18), Sgr~B2(N-LMH) and G24.78+0.08, the data were performed in position switching mode using the reference OFF positions that are listed in Table~\ref{tab2}.  

%
\begin{table}
\begin{center}
\caption{Reference position for Position Switching mode.\label{tab2}}
\begin{tabular}{lll}
\tableline\tableline
Source &OFF position\tablenotemark{a}\\
\tableline
Orion-KL &  EQ[-500$\arcsec$,0.0$\arcsec$] \\
NGC~6334~IRS~1 & EQ[-500$\arcsec$,0.0$\arcsec$] \\
Sgr~B2(N-LMH) & EQ[-752$\arcsec$,342$\arcsec$] \\
G24.78+0.08 & EQ[7071$\arcsec$,-947$\arcsec$] \\
\tableline
\tablenotetext{a}{The coordinates are given in the equatorial (EQ) system.}    
\end{tabular}
\end{center}
 \end{table}

The tuning frequency was set to 285.370~GHz for all the observed sources. Also, additional scans at a tuning frequency of 285.400~GHz were performed toward Orion-KL and Sgr~B2(N-LMH) as complementary observations in order to likely identify lines that are coming from the image side-band.
The three following strong contaminants have been identified:
\begin{description}
\item [- ] the sulfur monoxide line, SO $^{3}\Sigma$ (v = 0, 6$_7$--5$_6$), at 296.550~GHz,
\item [- ]  the sulfur monoxide line, $^{33}$SO (7$_7$--6$_6$), at 298.246~GHz,
\item [- ]  and the sulfur dioxide line, SO$_{2}$ (v = 0, 9$_{2, 8}$--8$_{1, 7}$) at 298.576~GHz.
\end{description}

\noindent These contaminants have been identified from the detailed model of molecular emission towards Orion--KL that matched emission from $\sim$100~GHz to 1.9~THz \citep[hereafter HIFI spectral fit, see][]{Crockett:2010,Crockett:2014}.
Accounting for a 10~dB rejection, the lines mentioned above could contaminate the observed Orion-KL spectrum with a signal greater than 2~K in T$\rm_{MB}$ scale (i.e $\ge$1.5~K in T$\rm_{A}^{*}$). Such contamination would be significant in our observations. 
In addition, two unidentified lines from image side-band were present in the observed spectrum.
Therefore, in each data set and for each source, the channels corresponding to the emission from all these lines have been removed.
Our data should thus be free from contamination by lines which are coming from the rejected side-band. Nonetheless, we stress that other lines from the image side-band may still, unfortunately, pollute the observed spectra. 

%
\begin{deluxetable}{rrrr}
\tablewidth{0.pt}
\tablecolumns{4}
\tablecaption{Rotational-torsional-vibrational partition function\tablenotemark{a} for $^{13}$C$_1$--MF, $^{13}$C$_2$--MF and $^{12}$C--MF.\label{tab3}}
\tablehead{
T(K) & $^{13}$C$_1$--MF & $^{13}$C$_2$--MF& $^{12}$C--MF}
\startdata
300.0  &252230.47   &255988.58  &249172.44 \\
225.0  &105303.23   &106847.86 &  104015.96 \\
150.0  & 36879.42   & 37442.27    &  36433.43 \\
 75.0  &  9003.31   &  9162.12  &  8894.06\\
 37.50 &  2920.56   &  2971.20 & 2885.3  \\
 18.75 &  1027.71   &  1045.22 &  1015.31  \\
  9.375&   364.76   &   370.95&  360.33 \\
 \enddata
\tablenotetext{a}{The nuclear spin degeneracy was not considered in these calculations (see Appendix \ref{AP:PF}).}
     \end{deluxetable}

%
\section{Spectral characterization for the $^{13}$C-methyl formate isotopologues} 
\label{sec:spectro}

The interstellar identifications of $^{13}$C$_1$--MF, $^{13}$C$_2$--MF were carried out from their spectral predictions in the frequency range of the facilities.  These predictions were computed through the Hamiltonian parameters of the $^{13}$C$_2$--MF isotopologue provided by \citet{Carvajal:2009} and of the $^{13}$C$_1$--MF isotopologue from \citet{Carvajal:2010}. The dipole moments used in the intensity calculation were given by \citet{Margules:2010}.

The spectroscopic characterization of $^{13}$C--MF isotopologues were carried out starting with millimeter-- and submillimeter--wave recordings in the laboratory and followed by their spectral analysis and the assignments of the transition lines through an established fitting procedure. 
The effective Hamiltonian used for the global spectroscopic analysis of both isotopologues is based in the so-called Rho-Axis Method (RAM) \citep{Herbst:1984,Hougen:1994,Kleiner:2010} applicable for molecules with a CH$_3$ rotor.
 The BELGI version of the RAM code used  in this study is available online\footnote{The source code BELGI along with an example of input data file and a readme file are available at the Web site: http://www.ifpan.edu.pl/\~\,kisiel/introt/introt.htm\#belgi, managed by Dr. Zbigniew Kisiel. For extended versions of the code, please contact Isabelle Kleiner or Miguel Carvajal.}. Further details regarding its application to the methyl formate isotopologues are described by \citet{Carvajal:2007}.

The Hamiltonian parameters were fitted to the experimental data of $^{13}$C$_2$--MF ($\sim 940$ lines) which were provided only for the ground torsional state $v_t=0$ \citep{Carvajal:2009}. New experimental data for the $v_t=0$ ground and $v_t=1$ first excited torsional state are presently being processed \citep{Haykal:2014}. A more extensive set of experimental data ($\sim 7500$ transition lines) of the ground and first excited states of $^{13}$C$_1$--MF has been used in the fit of the RAM Hamiltonian. The complete set of available experimental data \citep[see][]{Willaert:2006,Carvajal:2009,Maeda:2008,Maeda:2008a} was compiled in \citet{Carvajal:2010}.

\subsection{Partition functions}

To calculate the observed intensities of the spectral lines, the populations of each level must be estimated using an accurate partition function in order to provide reliable estimates of the temperatures and column densities of the different regions in the ISM. With this goal in mind, a convergence study for the partition functions of $^{13}$C-isotopologues, which ensures that high enough energy levels have been included for a particular temperature, has been carried out in this work. The partition function calculations are described in the \verb"Appendix" \ref{AP:PF}.
Table ~\ref{tab3} summarizes  the rotational-torsional-vibrational partition function values that are used here for $^{13}$C$_1$--MF and $^{13}$C$_2$--MF.

%
\section{Data analysis}
\label{sec:resultsa}

\subsection{Database and MF, $^{13}$C--MF frequencies}

We used the measured and predicted transitions coming from both the table of \citet{Ilyushin:2009} and the JPL database\footnote{http://spec.jpl.nasa.gov/home.html} \citep{Pickett:1992,Pickett:1998} for the MF line assignments, as in \citet{Favre:2011}. 
Regarding the methyl formate isotopologue $^{13}$C$_1$--MF and $^{13}$C$_2$--MF line assignments, our present analysis is based on this study (see Section \ref{sec:spectro}) and on the  spectroscopic characterization performed by \citet{Carvajal:2007,Carvajal:2009,Carvajal:2010}. Likewise, the measured and predicted transitions of the species $^{13}$C$_1$--MF species \citep{Carvajal:2010} are now available on the CDMS database\footnote{http://www.astro.uni-koeln.de/cdms} \citep{Muller:2001,Muller:2005}  and at Splatalogue\footnote{www.splatalogue.net} \citep{Remijan:2007}.
Current spectroscopic data for MF and $^{13}$C--MF treat both the two torsional substates  -- with A and E symmetries --  simultaneously. 

As we aim to derive accurate isotopic ratios, we should be confident with the intensity calculation of the molecular species at different temperatures. Therefore, the isotope ratio accuracy will depend, on one hand, on the spectroscopic determination of transition frequencies,
assignments and line strengths and, on the other hand, on the partition function approximation considered. 
Accurate spectroscopic characterizations of the main isotopologue was carried out previously \citep{Ilyushin:2009} using the RAM method, while for the $^{13}$C--MF isotopologues we used the same values for the electric dipole moments as for the $^{12}$C--MF species (see Section 3.1).
This assumption would not affect the line strengths more than $\sim 1$\%. Hence, the accurate derivation of the abundance ratio between different isotopologues will rely, as far as the spectroscopic data are concerned, on the partition function. This was computed under the same level of approximation for all the molecular species under study.   
Table \ref{tab3} shows the values of the partition function used for H$^{12}$COOCH$_3$. These values were computed on the basis of the new calculations in this manuscript to fuller account for the effect of vibrationally excited levels. In the JPL catalog entry, only contributions of the $v_t=0$ and $v_t=1$ level are incorporated into the partition function, while here we account for all torsional--vibrational energy states. This results in a higher inferred $^{12}$C methyl formate abundance than would be derived using the value in the JPL catalog since the partition function  is now larger than in the JPL tables.  
The partition function used here is higher than that in the JPL catalog by a factor of 1.2 at a temperature of 150~K, and by a factor of 2.5 at 300~K. 
Our more accurate partition functions yield a more accurate abundance of methyl formate than is reported in earlier publications.
 
It is also worthwhile to remark that the methyl formate partition function provided in the JPL catalog file has an extra factor of $2$ in its formula with respect to ours. This factor arises from the product of the reduced nuclear spin and K-level degeneracy statistical weights $g_{I} $, $g_{k}$ \citep{Turner:1991,Favre:2011}.
As for methyl formate the statistical weights are cancelled in the intensity mathematical expression \citep[see e.g. Eq.(1) of][]{Turner:1991}, they have not been considered in the partition function calculation of this paper. This means that when the comparison between the partition function of this work and the one provided in the JPL catalog was established, this latter was divided out by the factor of $2$.

Also, our spectral line analysis of the ALMA-SV observations of Orion-KL takes into account the $^{12}$C methyl formate transitions that are both in the ground and first torsionally excited states since they seem to probe a  similar temperature toward this region \citep[see][]{Favre:2011,Kobayashi:2007}. However, regarding the sources observed with the APEX telescope, we have only considered methyl formate transitions in their ground torsional states $v_t = 0$. More specifically, the number of detected transitions in the  $v_t = 1$ state (4 lines with a similar upper energy level) is insufficient to determine any trend with respect to transitions emitting in the ground state ($v_t = 0$).

\subsection{XCLASS modeling and Herschel/HIFI spectral fit}
Assuming local thermodynamic equilibrium (LTE),  we have modeled all the methyl formate isotopologue emission by using the XCLASS\footnote{http://www.astro.uni-koeln.de/projects/schilke/XCLASS} program along with the HIFI spectral fit that are based on the observations of Orion-KL acquired with Herschel/HIFI as part of the Herschel Observations of Extra-Ordinary Sources key program \citep{Bergin:2010, Crockett:2014}. This allows us to make reliable line identifications and determine where potential line blends may exist.
Further details regarding the XCLASS modeling of the Herschel/HIFI Orion-KL spectral scan, along with fit parameters, can be found in \citet{Crockett:2014}.

\begin{deluxetable}{cccc|ccc}
\tablewidth{0.pt}
\tablecolumns{7}
\tablecaption{Number of detected transitions of $^{12}$C--MF and $^{13}$C-MF in the ALMA-SV data of Orion--KL\tablenotemark{a}.\label{tab4}}
\tablehead{Spw\tablenotemark{b} & \multicolumn{3}{c}{Compact Ridge} & \multicolumn{3}{c}{Hot Core--SW} \\ 
&HCOOCH$_3$ & H$^{13}$COOCH$_3$ & HCOO$^{13}$CH$_3$& HCOOCH$_3$ & H$^{13}$COOCH$_3$ & HCOO$^{13}$CH$_3$}
\startdata
0 &1 & 1& 2 &1 &-- &2\\
1&3 & --& 6&3& -- &1\\
2& 19& 10& 3&19&7&3\\
3& 5&6 & 3&5&3&3\\
4& 11& 7& 4&10&3&4\\
5&20 &13 &6 &19&5&5\\
6& 5&12 & --&5&4&--\\
7& 6&2 &13 &4&--&6\\
8& 11& 11 &1 &8&5&--\\
9&10 & 13& 2&9&10&2\\
10& 4& 1& 5&4&1&3\\
11& 20&10 &4 &20&5&2\\
12& 1& --& 15&--&--&6\\
13&2 &4 & 17&2&3&11\\
14& 3& 9&2&3&6&--\\
15& 6&4 & --&6&2&--\\
16& 15&12 & 8 &15&6&6\\
17& 23 &16 & 3&21&12&2\\
18& 2& 3& 17&2&1&5\\
19& -- & 1&11&-- &--&6\\
 \enddata
 \tablenotetext{a}{The corresponding line frequencies are given in Figures \ref{fga1}, \ref{fga2}, \ref{fga3}, \ref{fgb1}, \ref{fgb2} and \ref{fgb3}.}
\tablenotetext{b}{The ALMA-SV data consist of 20 spectral windows (see Section~\ref{sec:observations}).}
     \end{deluxetable}

In the present analysis, we assumed that the $^{13}$C$_1$--MF and $^{13}$C$_2$--MF species emits within the same source size, at the same rotational temperature and velocity, and with the same line--width as the methyl formate molecule. The only adjustable parameter is the molecular column density.
To initialize the model of the ALMA-SV observations of Orion--KL, we used as input parameters (source size, rotational temperature, column density, $v$$_{LSR}$ and $\Delta$$v$$_{LSR}$), the values derived by our previous Plateau de Bure Interferometer (PdBI) observations, which were performed with a similar angular resolution  \citep[1.8$\arcsec$ $\times$ 0.8$\arcsec$, see][]{Favre:2011}.
Regarding the APEX observations, we used previously related and reported values derived from single-dish (JCMT, IRAM--30m, Herschel) and/or interferometric observations (BIMA, CARMA) as starting values to initialize the fitting. More specifically, we used the values derived by \citet{Bisschop:2007} for G24.78+0.08, by \citet{Zernickel:2012} and \citet{Bisschop:2007} for NGC6334I, by \citet{Demyk:2008} for W51~e2, by \citet{Shiao:2010} for G29.96--0.02, by \citet{Belloche:2009} for SgrB2(N), by \citet{Remijan:2004} and \citet{Shiao:2010} for G19.61--0.23 and by \citet{Tercero:2012}, \citet{Carvajal:2009} and \citet{Crockett:2014} for Orion-KL.

%
\section{Results}
\label{sec:results}
In the following section we report the main results for each observing facility.

\subsection{ALMA-SV observations of Orion-KL}
\label{sec:resultsori}
%
\subsubsection{Emission maps}
\begin{figure*}
\epsscale{1.}
\plotone{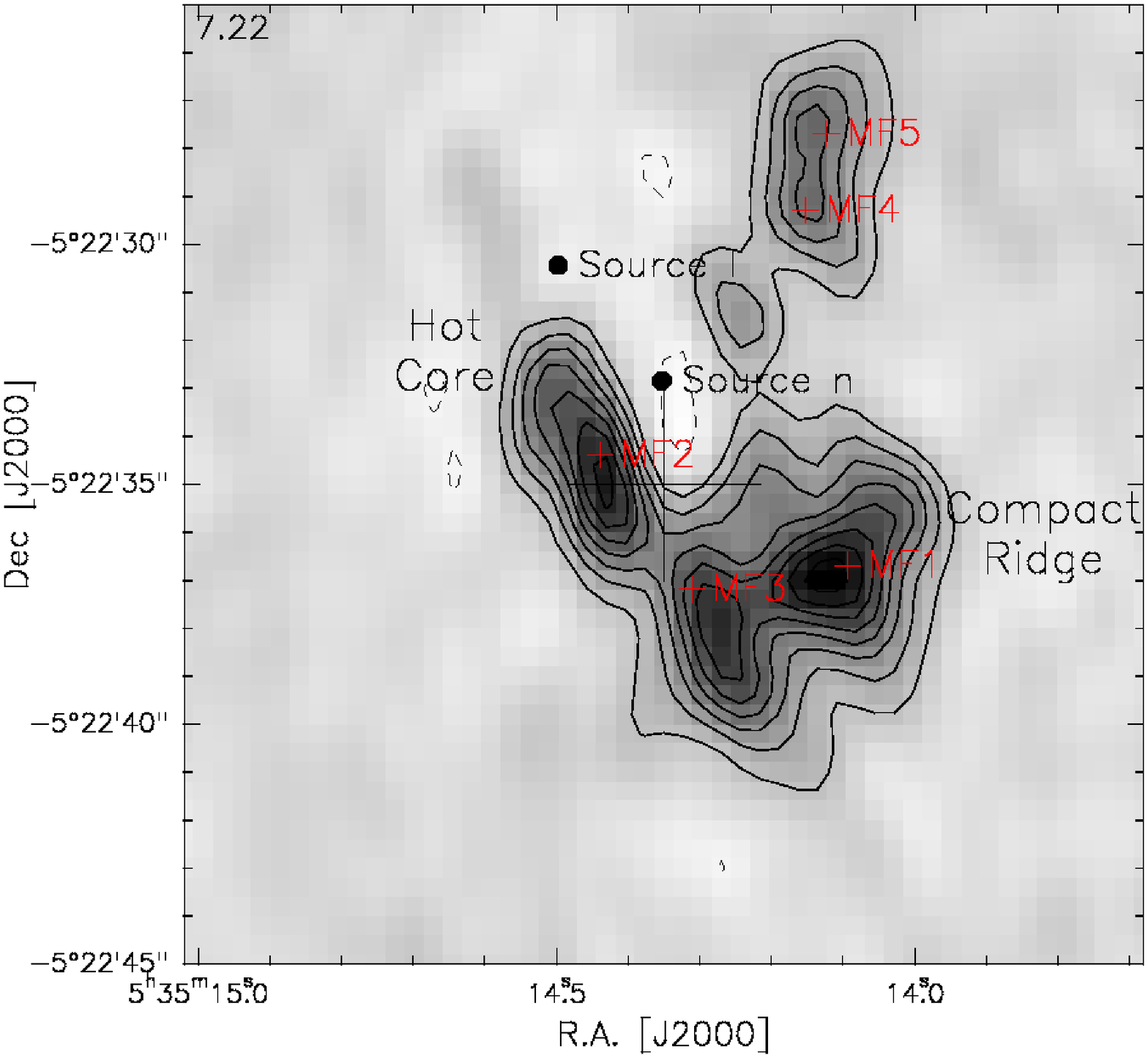}
\plotone{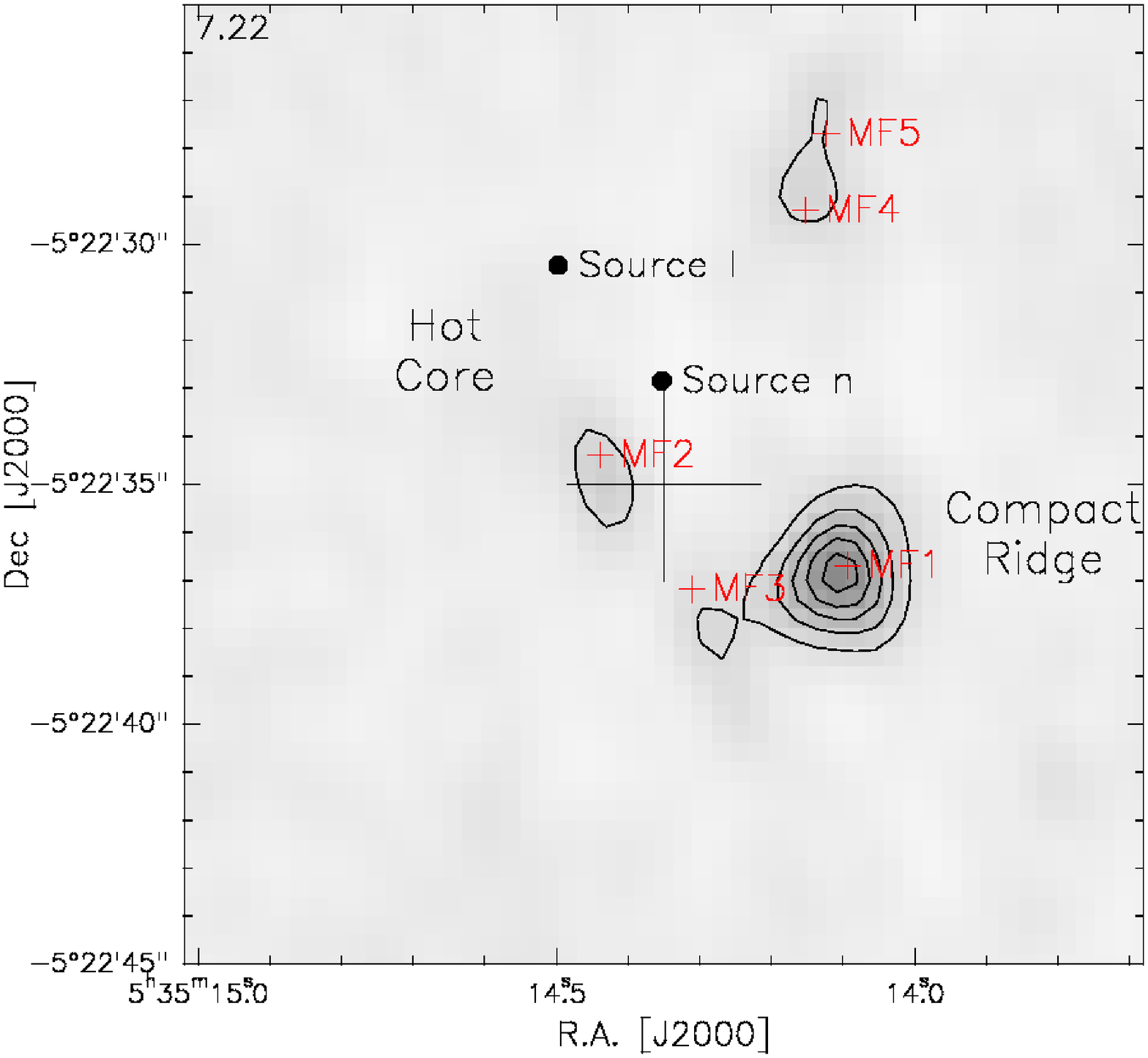}
\plotone{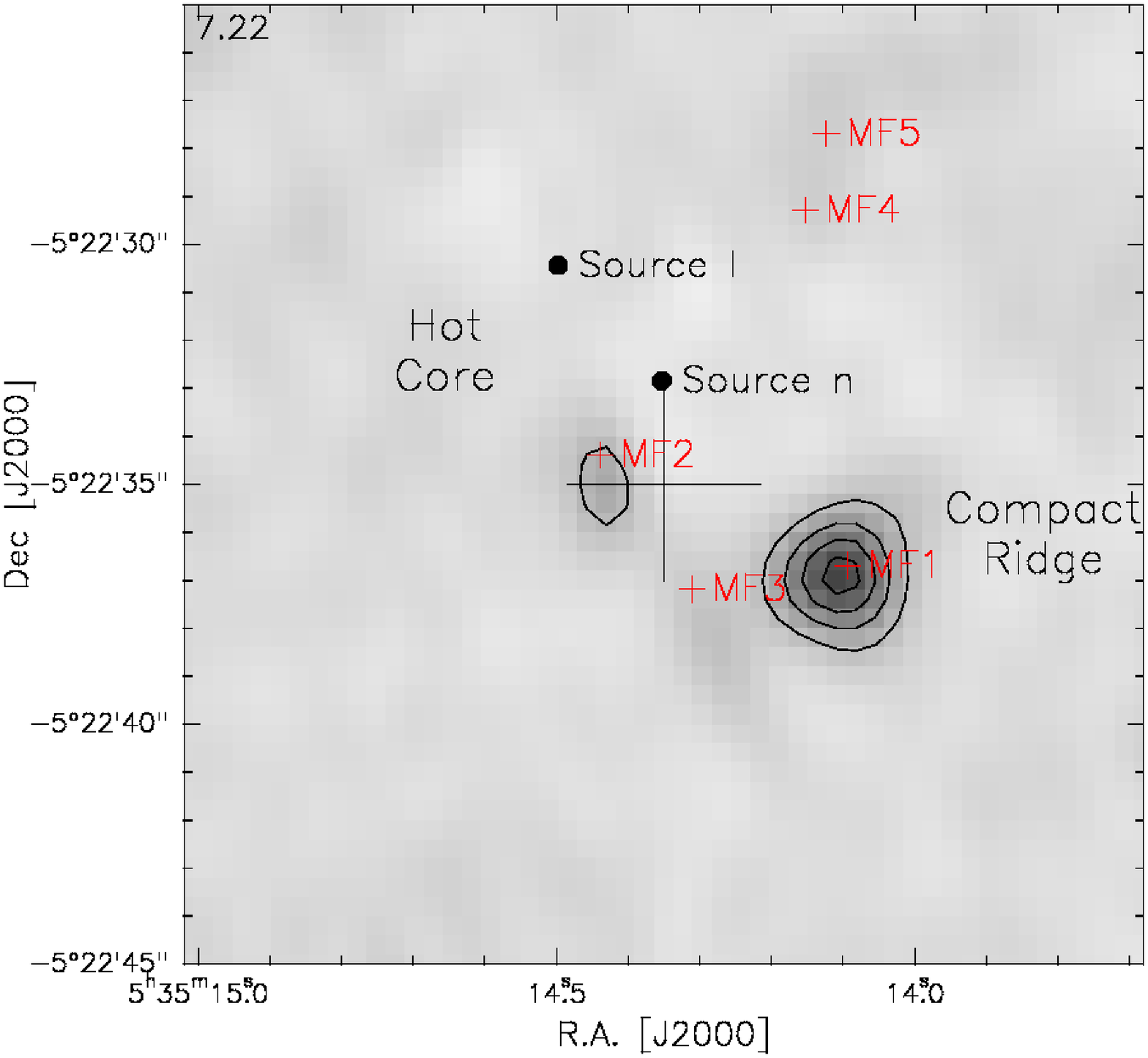}
\caption{HCOOCH$_{3}$ ($^{12}$C--MF, 234124~MHz, E$_{up}$=179~K, S$\mu$$^{2}$=35~D$^{2}$, top left panel), H$^{13}$COOCH$_{3}$ ($^{13}$C$_1$--MF, 220341~MHz, E$_{up}$=153~K, S$\mu$$^{2}$=37~D$^{2}$, top right panel) and HCOO$^{13}$CH$_{3}$ ($^{13}$C$_2$--MF, 216671~MHz, E$_{up}$=152~K, S$\mu$$^{2}$=36~D$^{2}$, bottom panel) emission channel maps at 7.22~km~s$^{-1}$ as observed with ALMA. The first contour and level step are 500~mJy~beam$^{-1}$ ($\sim$14$\sigma$) and 100~mJy~beam$^{-1}$ ($\sim$5$\sigma$) for $^{12}$C--MF and $^{13}$C--MF, respectively. The synthesized beam size is 1.7$\arcsec$ $\times$ 1.1$\arcsec$. The black cross indicates the centered position of the observations. The main HCOOCH$_3$ emission peaks (MF1 to MF5) identified by \citet{Favre:2011} are indicated.\label{fg1}}
\end{figure*}

The mean velocity for emission observed towards the Compact Ridge and Hot Core-SW regions is around 7.3~km~s$^{-1}$ for all the methyl formate isotopologues. We also observed a second velocity component around 9~km~s$^{-1}$ toward the Compact Ridge in HCOOCH$_3$  \citep[as reported by][]{Favre:2011}. This velocity component is not observed in $^{13}$C--MF. Figure~\ref{fg1} shows maps of the MF, $^{13}$C$_1$--MF and $^{13}$C$_2$--MF emission in the 7.2~km~s$^{-1}$ channel measured  at 234124~MHz, 220341~MHz and 216671~MHz, respectively.
The HCOOCH$_3$ distribution shows an extended V--shaped molecular emission that links the radio source I to the BN object as previously observed in methyl formate by \citet{Favre:2011} and \citet{Friedel:2008}. Likewise, as reported by \citet{Favre:2011}, the main molecular peaks are located toward the Compact Ridge and the Hot Core-SW \citep[respectively labeled MF1 and MF2 in Fig.\ref{fg1}; for more details see][]{Favre:2011}. Also we note that from the optically thick HCOOCH$_3$ lines, another cold component (T$\sim$40--50~K) arising from the vicinity of the source IRC7 is observed.  We did not analyze this component in the present study, however. 
Finally, the $^{13}$C--MF isotopologues are mainly detected toward the Compact Ridge and the Hot Core-SW (5$\sigma$ detection level, see Fig~\ref{fg1}). 

%
%

\subsubsection{Spectra}
Numerous transitions of  $^{12}$C--MF and $^{13}$C--MF, with S$\mu$$^{2}$ $\ge$ 10 D$^{2}$ and from upper energy levels of 166~K up to 504~K for the main molecule and E$_{up}$ of 99~K up to 330~K for the $^{13}$C--MF species, are present in the ALMA data. We have modeled each spectral window individually (see Section 6.1.3).
Table \ref{tab4} provides the number of clearly detected MF and $^{13}$C--MF transitions per spectral window toward both the Compact Ridge and the Hot Core-SW. Table \ref{tabB1} in \verb Appendix  \verb B \space summarizes the line parameters for all detected, blended, or not detected transitions of $^{12}$C--MF, $^{13}$C$_1$--MF  and $^{13}$C$_2$--MF in all ALMA spectral windows. Furthermore, figures~\ref{fga1}, \ref{fga2} and \ref{fga3} in \verb Appendix  \verb C \space show the $^{12}$C--MF, $^{13}$C$_1$--MF and $^{13}$C$_2$--MF transitions that are detected and/or partially blended in the ALMA-SV data  along with our best XCLASS models toward the Compact Ridge. In addition, Figures \ref{fgb1}, \ref{fgb2} and \ref{fgb3} in the \verb Appendix  \verb D \space show the emission of the same transitions, along with our models, toward the Hot Core-SW.
The quality of our models is based on the reduced $\chi^{2}$, which lies in the range 0.3--2.4, depending on the fit\footnote{The optically thick lines, although shown in appendix, are excluded from our model due to optical depth problem in the model.}. More specifically, the bulk of the emission is best reproduced for: 
\begin{description}
\item[-] a source size of 3$\arcsec$ (in agreement with the ALMA-SV observations) towards both the Compact Ridge and the Hot-Core--SW,
\item[-] a rotation temperature of 80~K toward the Compact Ridge and of 128~K toward the Hot-Core--SW,
\item[-] a $v$$_{LSR}$ of 7.3~km~s$^{-1}$ for both components,
\item[-] and a line-width of 1.2~km~s$^{-1}$ toward the Compact Ridge and of 2.4~km~s$^{-1}$ toward the Hot-Core--SW.
\end{description}
Only the column density differs within the different spectral windows, between the spatial components associated with Orion--KL and between the isotopologues. The $^{12}$C--MF models include the observed second velocity component well reproduced for a $v$$_{LSR}$ of 9.1~km~s$^{-1}$, a source size of 3$\arcsec$, a rotation temperature of 120~K and a column density of 7 $\times$ 10$^{16}$ cm$^{-2}$.

%
\begin{figure*}
\center
\includegraphics[angle=90,width=14cm]{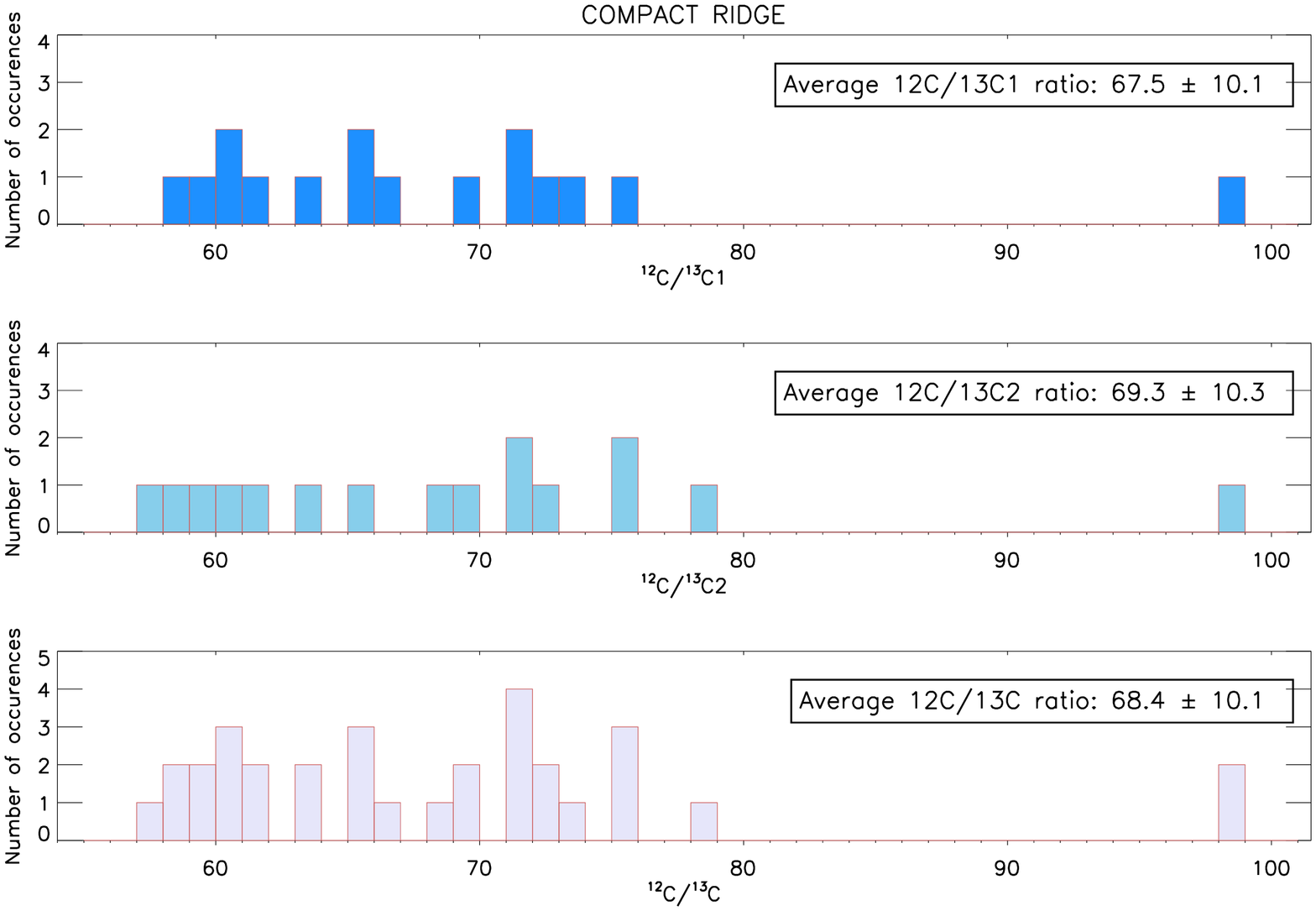}
\includegraphics[angle=90,width=14cm]{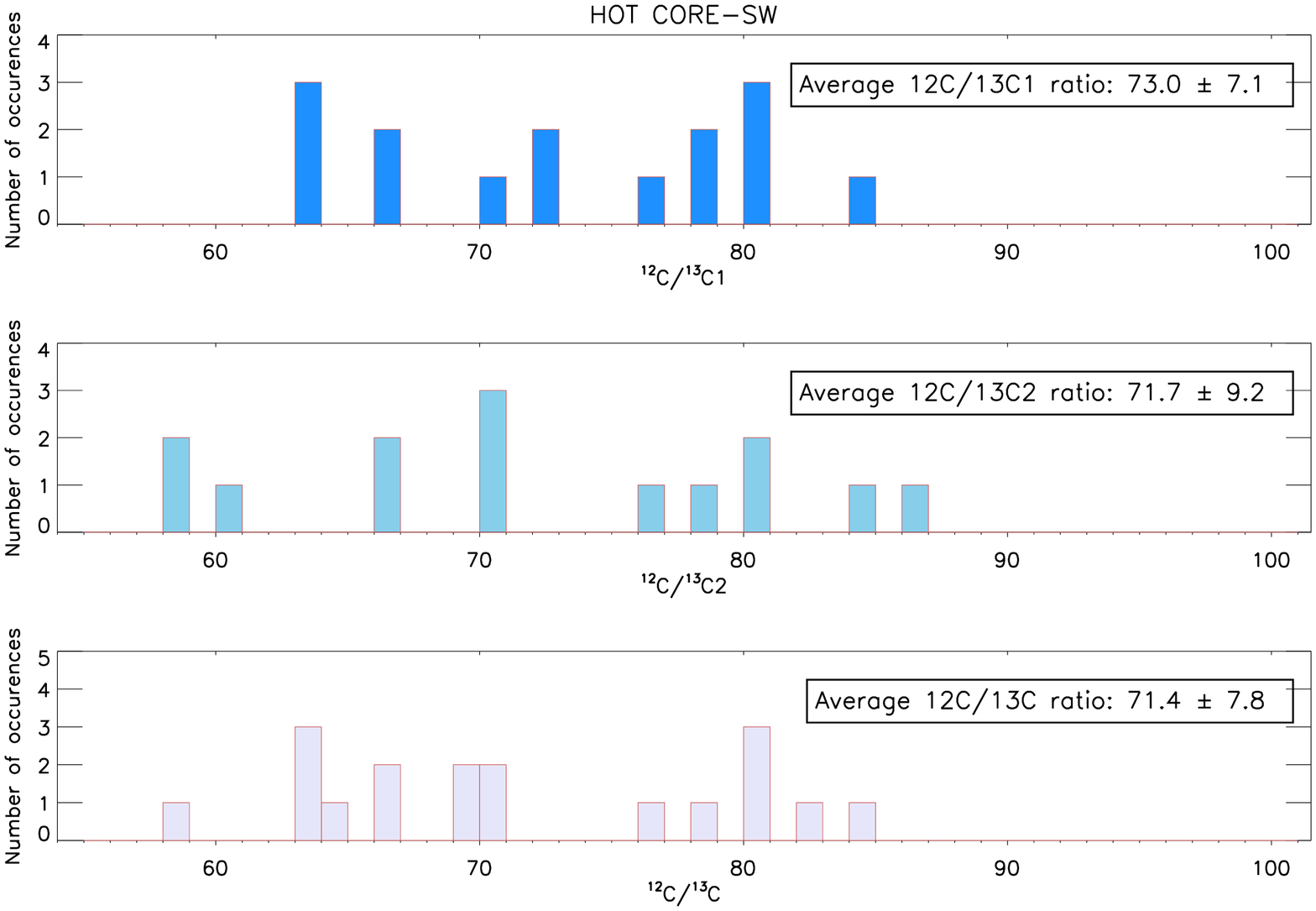}
\caption{Isotopic ratio distribution of the methyl formate isotopologues within each ALMA spectral window as derived toward the Orion-KL Compact Ridge (Top panel) and Hot Core-SW (bottom panel). {\em Top sub-panels:} Isotopic ratio distribution for the $^{12}$C/$^{13}$C$_1$ ratio. {\em Middle sub-panels:} Isotopic ratio distribution for the $^{12}$C/$^{13}$C$_2$ ratio. {\em Bottom sub-panels:} Isotopic ratio distribution for the $^{12}$C/$^{13}$C ratio, assuming the two $^{13}$C--MF isotopologues have similar abundances. The derived average isotopic ratio is indicated in each sub-panel. \label{fg2}}
\end{figure*}

\subsubsection{Isotopic $\rm^{12}C/^{13}C$ ratio}
Figure~\ref{fg2} shows the distribution of the $^{12}$C/$^{13}$C$_1$--MF~, $^{12}$C/$^{13}$C$_2$--MF and $^{12}$C/$^{13}$C--MF isotopic ratios we derived within each ALMA spectral window towards both the Compact Ridge\footnote{An outlier at $^{12}$C/$^{13}$C $\sim$ 100 is seen towards the Compact Ridge in each distribution. It result from MF measurement performed in the spw $\#$16. The outlier likely doesn't impact the derived isotopic ratio since either we include or exclude the value because the derived ratio remains the same within the uncertainties.} and the Hot Core-SW. 
The average $^{12}$C/$^{13}$C$_1$ and $^{12}$C/$^{13}$C$_2$ ratios are 67.5$\pm$10.1 and 69.3$\pm$10.3 in direction of the Compact Ridge, and 73.0$\pm$7.1 and 71.7$\pm$9.2 in direction of the Hot Core-SW.
If we assume the ratio to be the same for both isotopologues, meaning there is no significant difference, we derive an average $^{12}$C/$^{13}$C isotopic ratio in methyl formate of 68.4$\pm$10.1 and of 71.4$\pm$7.8 towards the Compact Ridge and the Hot Core-SW, respectively.

\subsection{APEX observations of all the sources}
\label{sec:resultsall}
\begin{figure*}
\begin{center}
\includegraphics[angle=0,width=8cm]{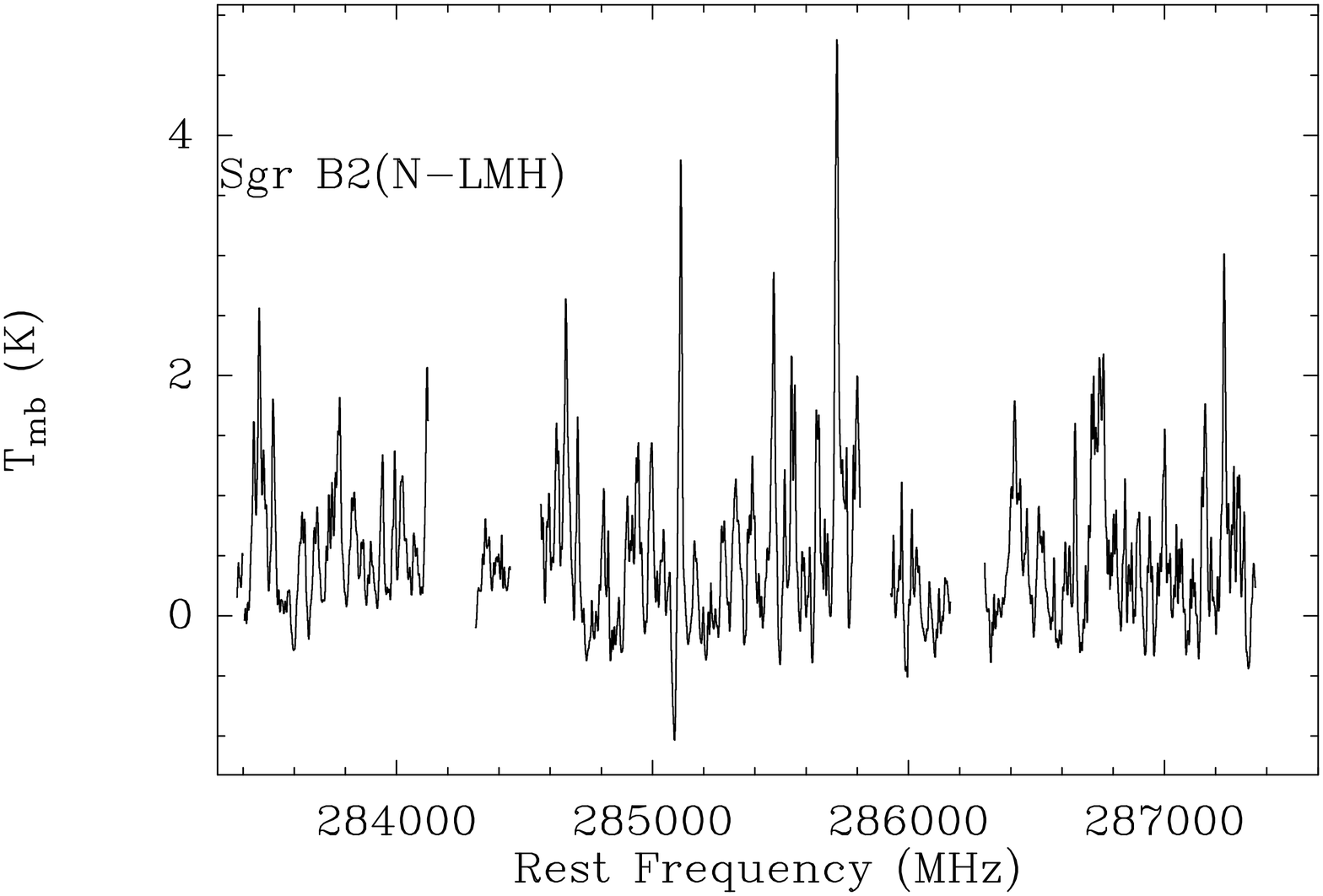}
\includegraphics[angle=0,width=8cm]{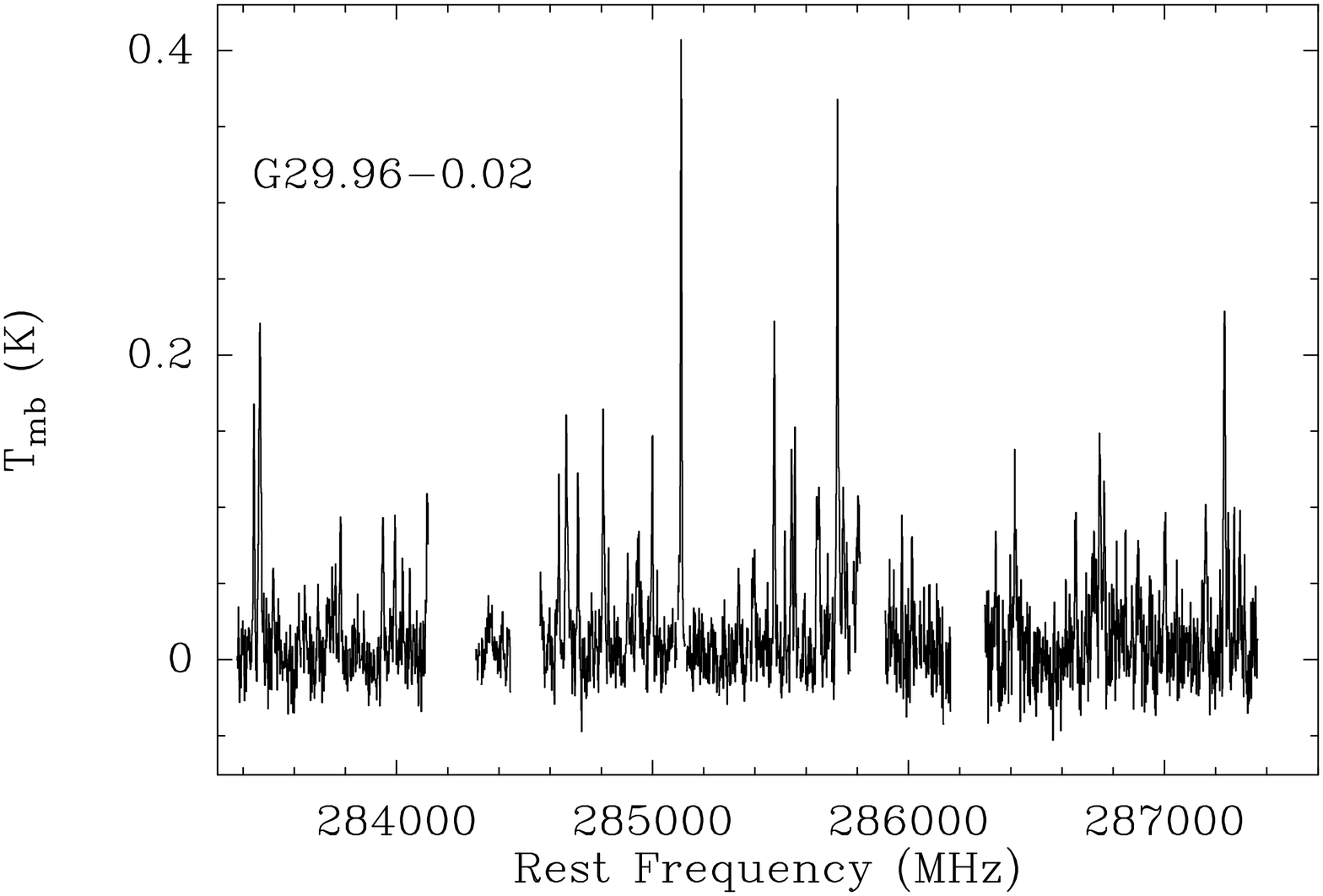}
\includegraphics[angle=0,width=8cm]{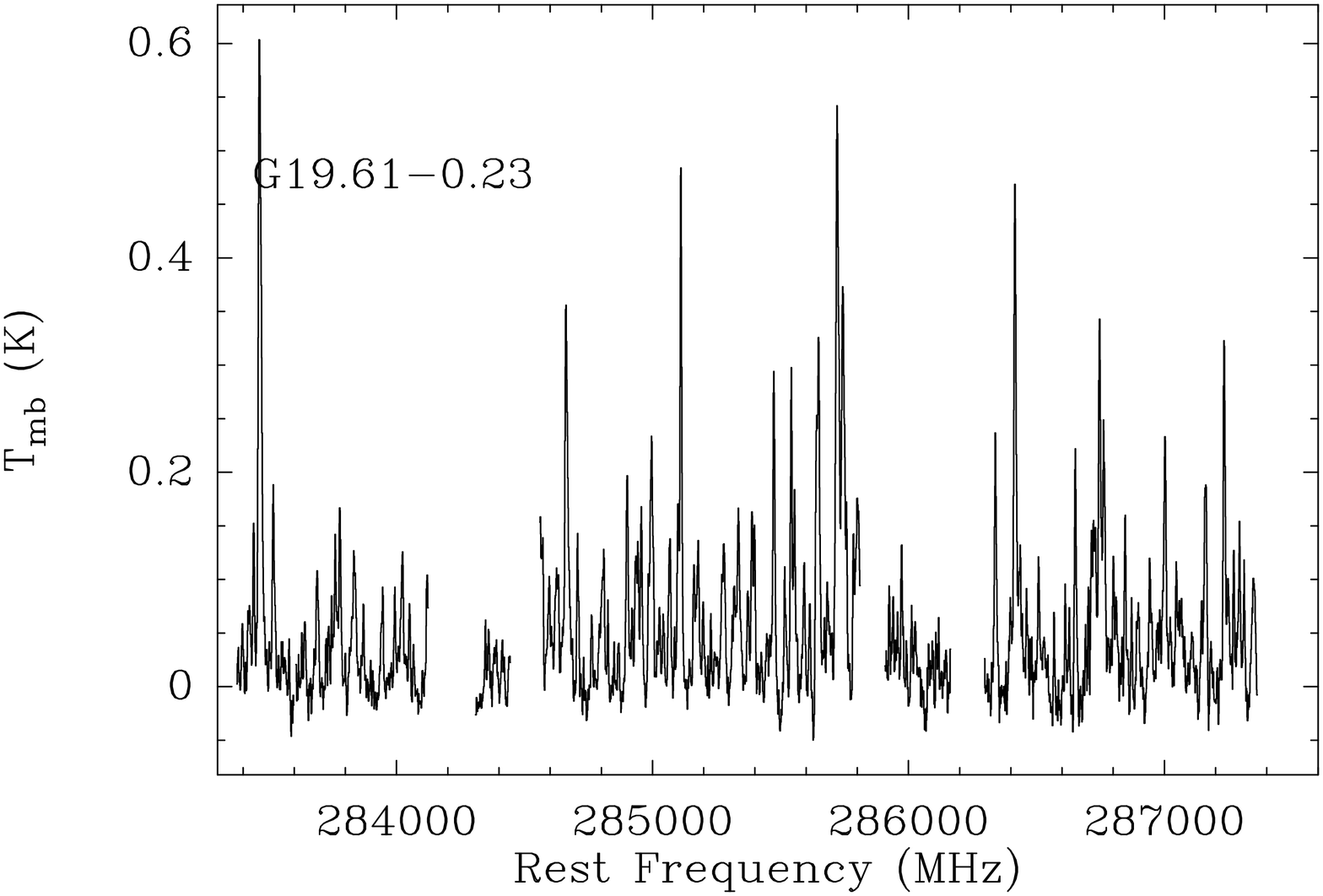}
\includegraphics[angle=0,width=8cm]{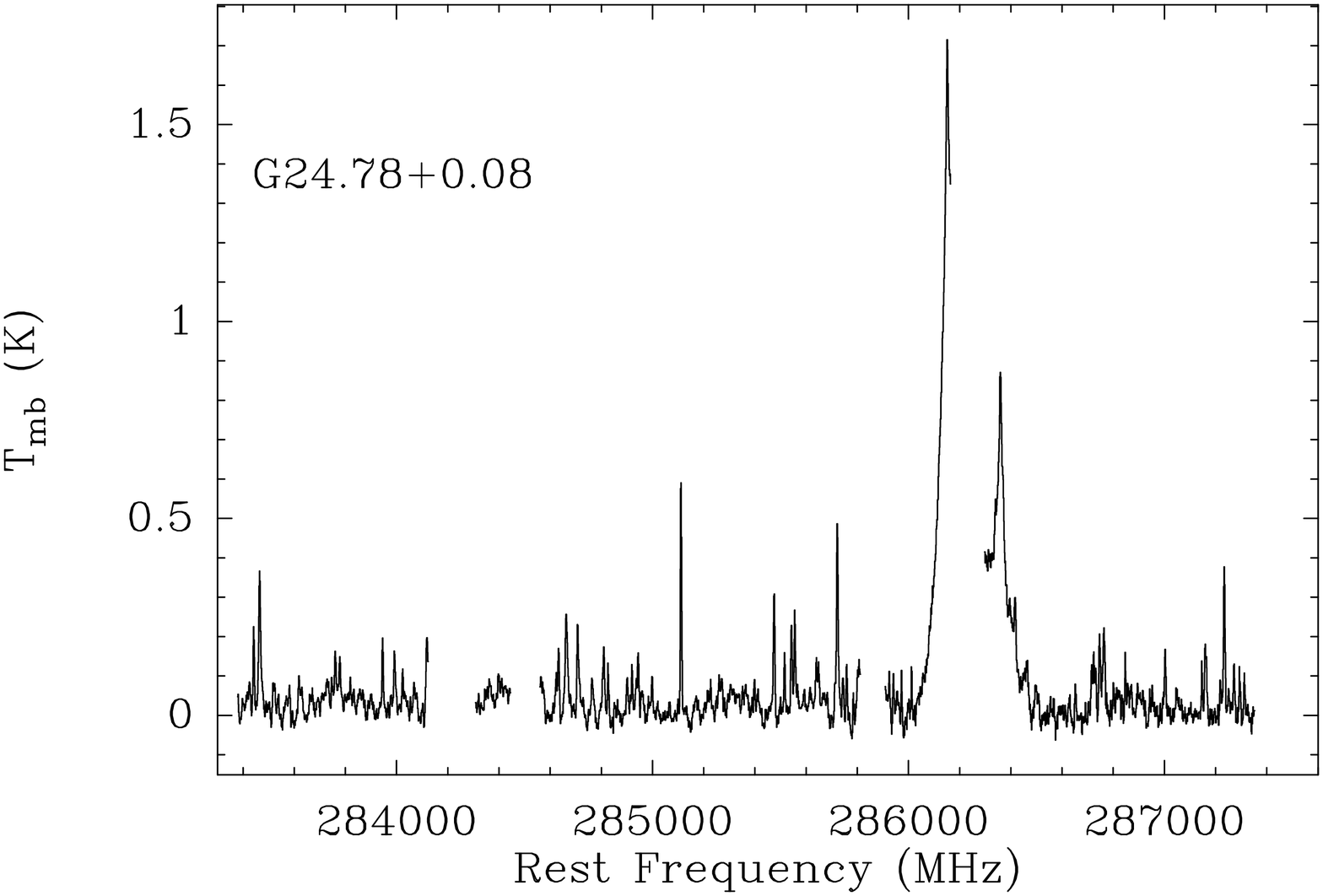}
\includegraphics[angle=0,width=8cm]{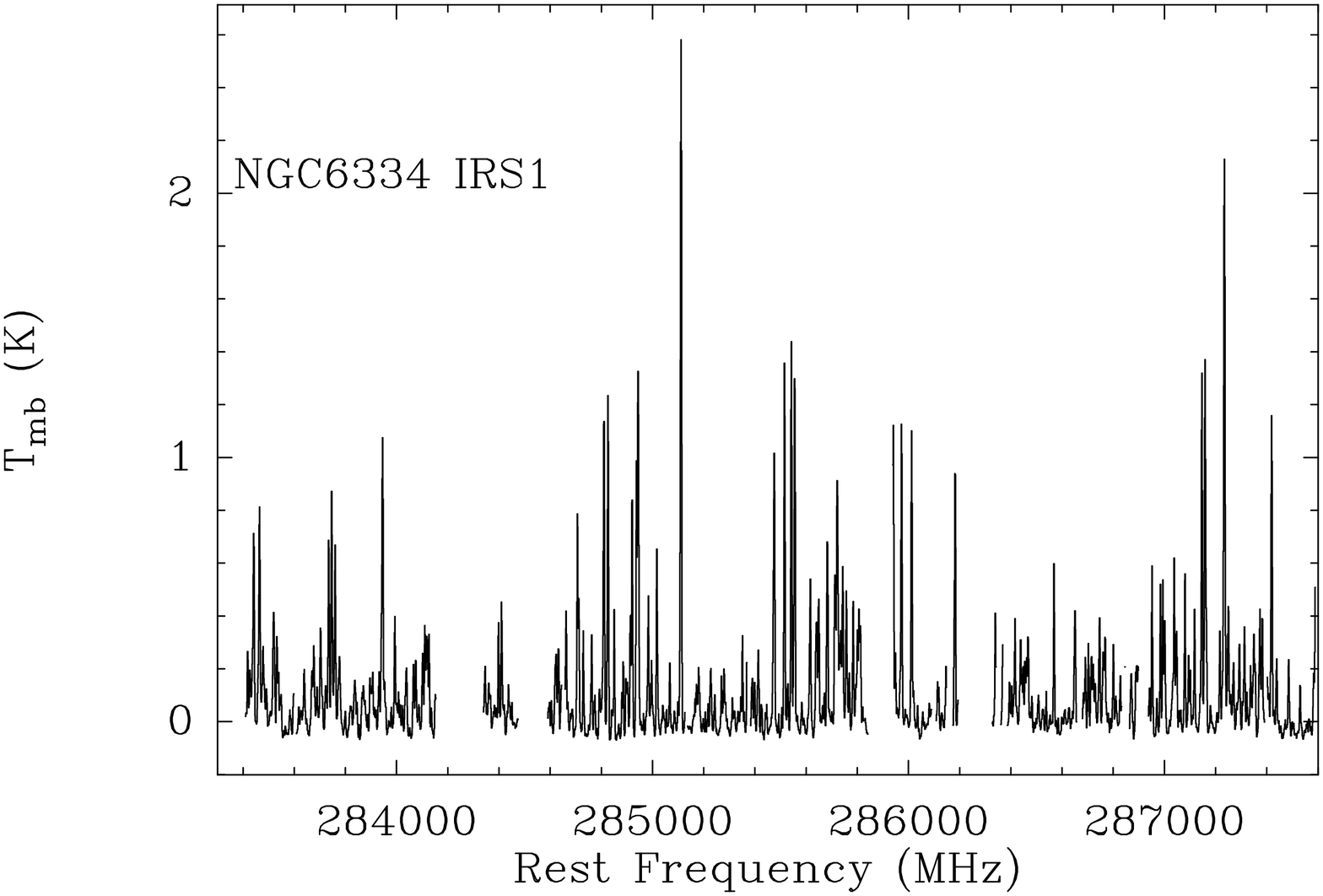}
\includegraphics[angle=0,width=8cm]{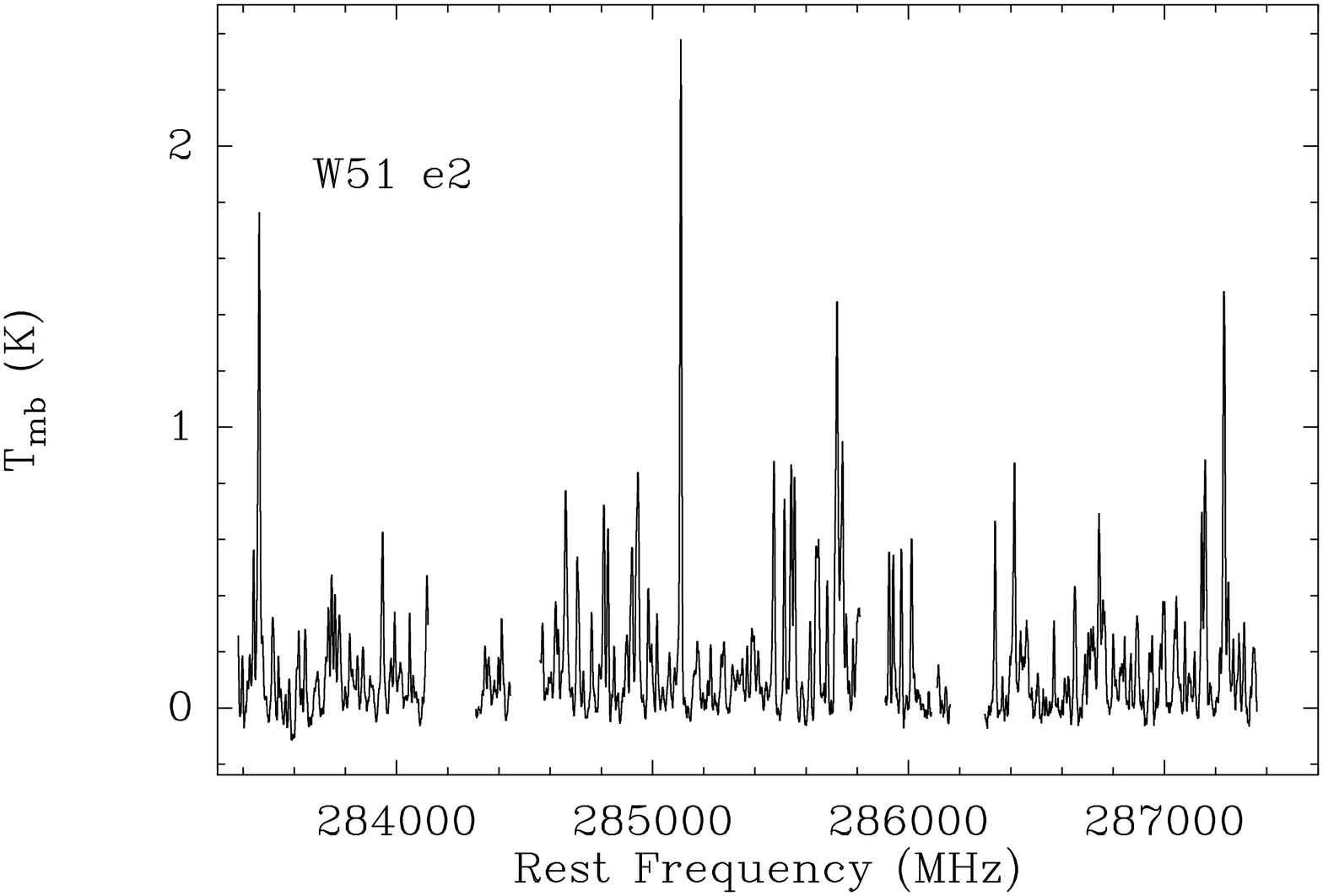}
\caption{Spectra centered at 285.450~GHz as observed with the APEX telescope for all the sources. The name of each observed sources is indicated on each plot. The spectral resolution is smoothed to 1.5~km~s$^{-1}$. Line assignment is shown in the Orion--KL spectrum (bottom panel) in red for the detected methyl formate $^{12}$C--MF and $^{13}$C-MF transitions and in grey for the other molecules \citep[based on the Herschel/HIFI spectral fit to Orion--KL for the latter, see][]{Crockett:2014}. \label{fg3}}
\end{center}
\end{figure*}

Figure \ref{fg3} shows the spectra observed with the APEX telescope toward our sample of 7 sources (see Tab.~\ref{tab1}). Lines which have been identified through the Herschel/HIFI spectral fit are indicated in the Orion-KL spectrum (see bottom panel on Fig.~\ref{fg3}). The molecular richness of the observed sources is clearly seen. Also, the different spectra illustrate the problem of the spectral confusion for the weaker emissive lines.

\subsubsection{Main isotope: HCOOCH$_{3}$}
\begin{figure*}[h!]
\includegraphics[angle=0,width=10cm]{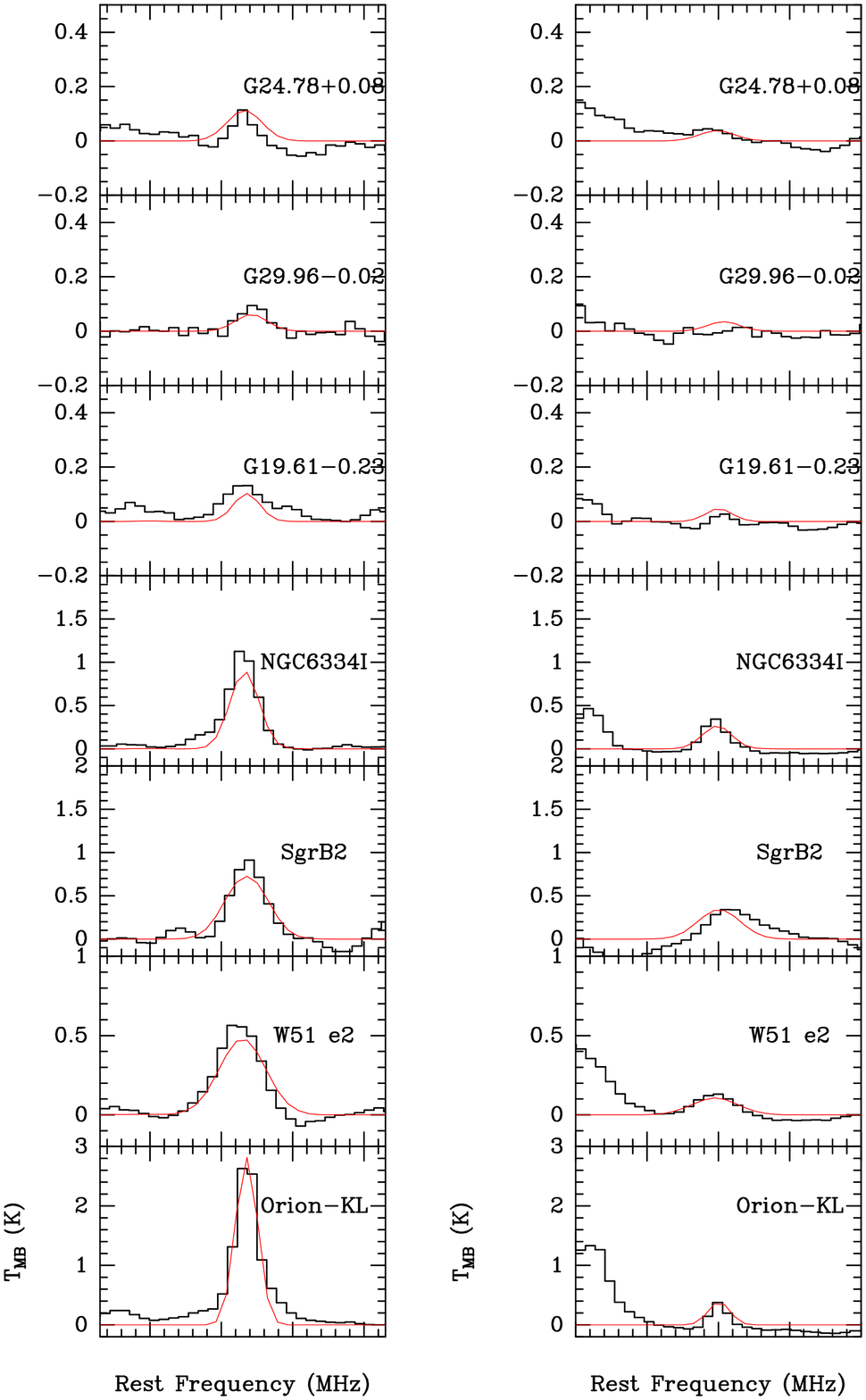}
\caption{HCOOCH$_3$ ($^{12}$C--MF, transition at 285973.267~MHz, left panel) and HCOO$^{13}$CH$_3$ ($^{13}$C$_2$--MF, transition at 284730.102~MHz, right panel)  synthetic spectra (in red) overlaid on the observed spectrum (in black) as observed with APEX. The name of the sources is indicated on each plot. \label{fg4}}
\end{figure*}

Table \ref{tab5} lists the detected or partially blended methyl formate transitions, with S$\mu$$^{2}$ $\ge$ 2.5 D$^{2}$ and E$_{up}$ up to 304~K as observed with APEX towards  the different sources.  Note that for some partially blended lines, the emission arising from the contaminant has been identified through the Herschel template spectra, in which emission from 35 molecules has been modeled assuming LTE \citep{Crockett:2014}. The following procedure was used: \textit{1)} superposing the Herschel resulting model to the APEX observations and identifying the potential contaminant(s) and  \textit{2)}, adjusting the observational parameters (e.g. velocity, typical line-width) to the model, checking the coherence over the full spectrum. The adopted parameters (source size, rotational temperature, column density, velocity and line-width) which were used to model the APEX observations are given in Table \ref{tab6} for each source. The quality of our models is based on the reduced $\chi^{2}$, which lies in the range 0.23--4.75.
In addition, Figure~\ref{fg4} shows the observed methyl formate spectrum of the transition at 285973.267~MHz ($\rm 23_{8,	 15}-22_{	8, 14	}$,E) along with our models for each source. 

The main observational results for the methyl formate molecule are briefly summarized below for the individual sources.  

\textit{Orion-KL:} We detected sixteen HCOOCH$_{3}$ lines and observed fifteen transitions that are partially blended, with S$\mu$$^{2}$ $\ge$ 4 D$^{2}$ (see Table~\ref{tab5}). The LSR velocity is 7.7~km~s$^{-1}$ and the derived column density is 9.7 $\times$ 10$^{16}$~cm$^{-2}$.

\textit{W51~e2:} We detected eleven HCOOCH$_{3}$ lines and observed six transitions that are partially blended. Fourteen transitions (with S$\mu$$^{2}$ $\le$ 12 D$^{2}$) are too faint to be detected (which is commensurate with our model of the source). The spectra display a $v$$_{LSR}$ of 55.6~km~s$^{-1}$ and we derived a column density of 9.0 $\times$ 10$^{16}$~cm$^{-2}$.

\textit{G19.61-0.23:}  We detected ten HCOOCH$_{3}$ lines and observed seven transitions that are partially blended while fourteen transitions were too faint to be detected. The $v$$_{LSR}$ is 39.7~km~s$^{-1}$ and the derived column density is 7.0 $\times$ 10$^{16}$~cm$^{-2}$.

\textit{G29.96-0.02:}  We detected ten HCOOCH$_{3}$ lines and observed seven transitions that are partially blended. Fourteen transitions were too faint to be detected. Spectra display a  $v$$_{LSR}$ of 97.8~km~s$^{-1}$ and we derived a column density of 3.5 $\times$ 10$^{15}$~cm$^{-2}$.

\textit{G24.78+0.08:}  We detected ten HCOOCH$_{3}$ lines and observed seven transitions that are partially blended while fourteen transitions were too faint to be detected. The $v$$_{LSR}$ is 111~km~s$^{-1}$ and we derived column density of 6.0 $\times$ 10$^{15}$~cm$^{-2}$.

\textit{NGC~6334~IRS~1:}  We detected eleven HCOOCH$_{3}$ lines and observed six transitions that are partially blended with fourteen transitions were too faint to be detected. Spectra exhibit a $v$$_{LSR}$ of $-$8~km~s$^{-1}$ and we derived a column density of 4.5 $\times$ 10$^{17}$~cm$^{-2}$.

\textit{Sgr~B2(N):}  We detected seven HCOOCH$_{3}$ lines and observed nine transitions that are partially blended while fourteen transitions were too faint to be detected. The $v$$_{LSR}$ is around 63.7~km~s$^{-1}$. We derived a column density of 3.0~$\times$ 10$^{17}$~cm$^{-2}$.

\subsubsection{H$^{13}$COOCH$_{3}$}
The $^{13}$C$_1$--MF lines all appear to be blended or just at or below the confusion limit level. We note that some transitions overlap with lines from strongly emissive molecules such as ethyl cyanide \citep[whose presence is known through the Herschel template spectra to Orion--KL,][]{Crockett:2014}, which might hide faint emission. We therefore do not detect the $^{13}$C$_1$--methyl formate toward any of the observed sources, excluding Orion--KL and that only in the supplementary ALMA data.

%
\clearpage 

\begin{figure*}
\figurenum{3}
\includegraphics[angle=0,width=\textwidth]{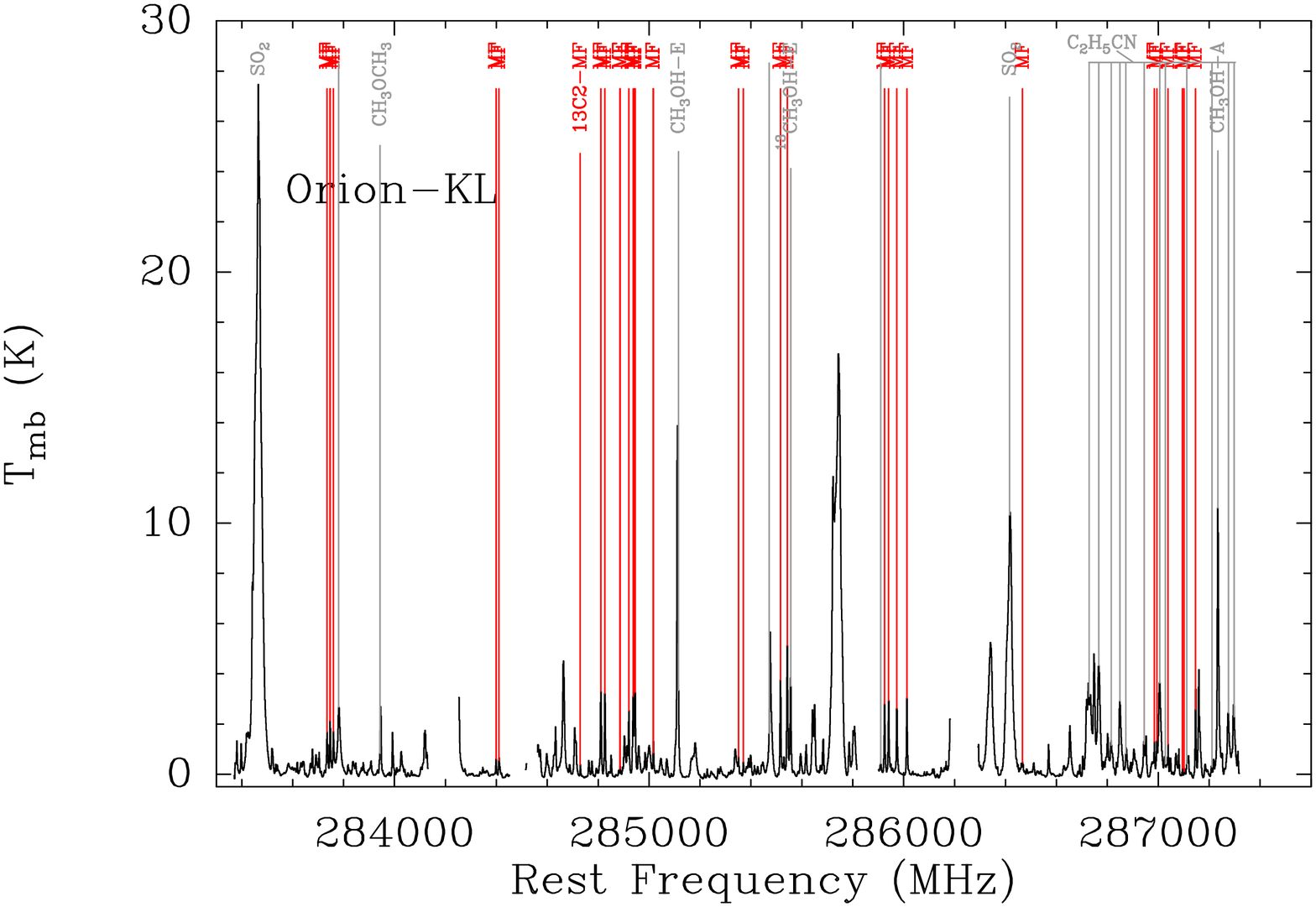}
\caption{Continue.}
\end{figure*} 

\clearpage
%

\clearpage

%
\begin{deluxetable}{rrrr|r|r|r|r|r|r|r}
\tablewidth{0pt}
\tabletypesize{\footnotesize}
\rotate
\tablecolumns{11}
\tablecaption{Transitions of $^{12}$C and $^{13}$C-methyl formate observed with the APEX telescope.\label{tab5}}
\tablehead{
Frequency & Transition & E$_{up}$ &  S$\mu$$^{2}$  & \multicolumn{7}{c}{Note\tablenotemark{a}}\\
(MHz) & & (K) & (D$^{2}$)&&&&&&&\\
&&&&Orion--KL& W51 e2 & G19.61-0.23& G29.96-0.02 & G24.78+0.08&NGC~6334~IRS~1& Sgr~B2(N-LMH)\\}
\startdata
\multicolumn{11}{c}{HCOOCH$_3$}\\
\hline
283734.887	&	$	23	_{	11	,	12	}$	--	$	22	_{	11	,	11	}$	E	&	243.2	&	45.4	&D&D&D&D&PB&D&D\\
283746.738	&	$	23	_{	11	,	13	}$	--	$	22	_{	11	,	12	}$	A	&	243.2	&	45.4	&PB&PB&PB&PB&PB&PB&PB\\
283746.738	&	$	23	_{	11	,	12	}$	--	$	22	_{	11	,	11	}$	A	&	243.2	&	45.4	&PB&PB&PB&PB&PB&PB&PB\\
283760.086	&	$	23	_{	11	,	13	}$	--	$	22	_{	11	,	12	}$	E	&	243.2	&	45.4	&D&D&D&D&D&D&PB\\
284398.680	&	$	13	_{	5	,	8	}$	--	$	12	_{	4		9	}$	E	&	70.4	&	2.5	&D&ND&ND&ND&ND&ND&ND\\
284410.529	&	$	13	_{	5	,	8	}$	--	$	12	_{	4	,	9	}$	A	&	70.4	&	2.7	&PB&ND&ND&ND&ND&ND&ND\\
284810.313	&	$	23	_{	5	,	19	}$	--	$	22	_{	5	,	18	}$	E	&	180.8	&	55.8	&PB&PB&PB&PB&PB&PB&PB\\
284826.396	&	$	23	_{	5	,	19	}$	--	$	22	_{	5	,	18	}$	A	&	180.8	&	55.8	&D&D&D&D&D&D&D\\
284885.531	&	$	27	_{	11	,	16	}$	--	$	27	_{	10	,	17	}$	A	&	303.7	&	7.6	&PB&ND&ND&ND&ND&ND&ND\\
284886.320\tablenotemark{b}	&	$	27	_{	11	,	16	}$	--	$	27	_{	10	,	17	}$	E	&	303.7	&	6.1	&PB&ND&ND&ND&ND&ND&ND\\
284920.240	&	$	23	_{	9	,	14	}$	--	$	22	_{	9	,	13	}$	E	&	217.0	&	49.8	&D&D&D&D&D&D&D\\
284937.218	&	$	23	_{	9	,	15	}$	--	$	22	_{	9	,	14	}$	A	&	217.0	&	49.9	&PB&PB&PB&PB&PB&PB&PB\\
284942.751	&	$	23	_{	9	,	14	}$	--	$	22	_{	9	,	13	}$	A	&	217.0	&	49.9	&PB&PB&PB&PB&PB&PB&PB\\
284945.147	&	$	23	_{	9	,	15	}$	--	$	22	_{	9	,	14	}$	E	&	216.9	&	49.8	&PB&PB&PB&PB&PB&PB&PB\\
285016.270	&	$	23	_{	4	,	20	}$	--	$	22	_{	3	,	19	}$	E	&	173.3	&	6.3	&PB&ND&ND&ND&ND&ND&ND\\
285016.977	&	$	23	_{	4	,	20	}$	--	$	22	_{	3	,	19	}$	A	&	173.3	&	6.3	&PB&ND&ND&ND&ND&ND&ND\\
285351.819	&	$	24	_{	3	,	21	}$	--	$	23	_{	4	,	20	}$	E	&	187.0	&	6.8	&PB&ND&ND&ND&ND&ND&ND\\
285370.140	&	$	24	_{	3	,	21	}$	--	$	23	_{	4	,	20	}$	A	&	187.0	&	6.8	&D&ND&ND&ND&ND&ND&ND\\
285515.739	&	$	22	_{	5	,	17	}$	--	$	21	_{	5	,	16	}$	E	&	169.4	&	53.7	&D&D&D&D&D&D&D\\
285542.584	&	$	22	_{	5	,	17	}$	--	$	21	_{	5	,	16	}$	A	&	169.4	&	53.7	&D&D&PB&PB&D&D&PB\\
285924.822	&	$	23	_{	8	,	16	}$	--	$	22	_{	8	,	15	}$	A	&	205.9	&	51.8	&D&D&D&D&D&D&--\\
285940.794	&	$	23	_{	8	,	16	}$	--	$	22	_{	8	,	15	}$	E	&	205.9	&	49.9	&D&D&D&D&D&D&D\\
285973.267	&	$	23	_{	8	,	15	}$	--	$	22	_{	8	,	14	}$	E	&	206.0	&	49.9	&D&D&D&D&D&D&D\\
286012.485	&	$	23	_{	8	,	15	}$	--	$	22	_{	8	,	14	}$	A	&	205.9	&	51.8	&D&D&D&D&D&D&D\\
286467.129\tablenotemark{b}	&	$	25	_{	11	,	14	}$	--	$	25	_{	10	,	15	}$	A	&	272.2	&	5.5	&PB&ND&ND&ND&ND&ND&ND\\
286984.997	&	$	11	_{	6	,	5	}$	--	$	10	_{	5	,	5	}$	E	&	62.8	&	3.2	&PB&ND&ND&ND&ND&ND&ND\\
286994.417	&	$	11	_{	6	,	6	}$	--	$	10	_{	5	,	5	}$	A	&	62.8	&	3.2	&PB&ND&ND&ND&ND&ND&ND\\
287038.552	&	$	11	_{	6	,	5	}$	--	$	10	_{	5	,	6	}$	A	&	62.8	&	3.2	&D&ND&ND&ND&ND&ND&ND\\
287094.976	&	$	24	_{	11	,	14	}$	--	$	24	_{	10	,	15	}$	E	&	257.4	&	6.3	&D&ND&ND&ND&ND&ND&ND\\
287101.120	&	$	24	_{	11	,	13	}$	--	$	24	_{	10	,	14	}$	E	&	257.4	&	6.3	&D&ND&ND&ND&ND&ND&ND\\
287146.630	&	$	23	_{	7	,	17	}$	--	$	22	_{	7	,	16	}$	E	&	196.4	&	53.4	&D&D&D&D&D&D&PB\\
\hline
\multicolumn{11}{c}{H$^{13}$COOCH$_3$}\\
\hline
 283853.856&	$	23	_{	8	,	16	}$	--	$	22	_{	8	,	15	}$ E & 204.3  & 52.9&ND$^{*}$&ND$^{*}$&ND$^{*}$&ND$^{*}$&ND$^{*}$&ND$^{*}$&ND$^{*}$\\
 286035.726&	$	23	_{	7	,	16	}$	--	$	22	_{	7	,	15	}$ A & 195.0  & 56.7&ND$^{*}$&ND$^{*}$&ND$^{*}$&ND$^{*}$&ND$^{*}$&ND$^{*}$&ND$^{*}$\\
\hline
\multicolumn{11}{c}{HCOO$^{13}$CH$_3$}\\
\hline
284729.511\tablenotemark{c} &	$	27	_{	1	,	27	}$	--	$	26	_{	1	,	26	}$ E & 194.2 & 71.1&D&D&TD&ND$^{*}$&ND$^{*}$&D&TD\\
284729.537\tablenotemark{c} &	$	27	_{	0	,	27	}$	--	$	26	_{	0	,	26	}$ E & 194.2 & 71.1&D&D&TD&ND$^{*}$&ND$^{*}$&D&TD\\
284730.102\tablenotemark{c} &	$	27	_{	1	,	27	}$	--	$	26	_{	1	,	26	}$ A & 194.2 & 71.1&D&D&TD&ND$^{*}$&ND$^{*}$&D&TD\\
284730.127\tablenotemark{c} &	$	27	_{	0	,	27	}$	--	$	26	_{	0	,	26	}$ A & 194.2 & 71.1&D&D&TD&ND$^{*}$&ND$^{*}$&D&TD\\
 \enddata
\tablenotetext{a}{D: detected, TD: tentative detection, PB: partial blend, ND and ND$^{*}$: not detected (too faint emission). Also, ND$^{*}$ indicates the $^{13}$C$_{1}$-- and $^{13}$C$_{2}$--HCOOCH$_3$ transitions that are emitting with an intensity less or equal to 3 times the noise level and that we used to constrain our model. The symbol "--" indicates that  part of the spectrum has been removed (see Section~\ref{sec:observationspart2}).}
\tablenotetext{b}{Those two transitions are predicted, i.e. not measured.} 
\tablenotetext{c}{Pile-up of these 4 lines. Also the frequencies presented in this table are only computed (i.e. not measured) because there are not experimental data for these transitions available yet.} 
     \end{deluxetable}

\clearpage

%
%
\begin{deluxetable}{lcc|ccc|ccc}
\tablewidth{0pt}
\tablecolumns{9}
\tablecaption{$^{12}$C--MF and $^{13}$C--MF XCLASS model parameters (source size, rotational temperature, column density, velocity and line-width) which reproduce best the APEX spectra.\label{tab6}}
\tablehead{
Source & $\theta_s$ & T$_{rot}$ & \multicolumn{3}{c}{N$_{tot}$} & $v$$_{LSR}$ & $\Delta$$v$  \\
& ($\arcsec$) & (K) & \multicolumn{3}{c}{(cm$^{-2}$)} & (km~s$^{-1}$) &(km~s$^{-1}$) \\
&  & &HCOOCH$_3$ & H$^{13}$COOCH$_3$&HCOO$^{13}$CH$_3$& &}
\startdata
Sgr~B2(N-LMH)\tablenotemark{b} & 4  & 80 & 3.0$\times$10$^{17}$&$\le$1.8$\times$10$^{16}$&$\le$1.8$\times$10$^{16}$& 63.7 & 7.0 \\
G24.78+0.08 \tablenotemark{c}& 10&121 &6.0$\times$10$^{15}$ &$\le$3.0$\times$10$^{14}$&$\le$3.0$\times$10$^{14}$&111.0 & 6.0\\
G29.96-0.02 \tablenotemark{c} & 10& 150 & 3.5$\times$10$^{15}$ & $\le$3.0$\times$10$^{14}$& $\le$3.0$\times$10$^{14}$& 97.8&5.5 \\
G19.61-0.23 \tablenotemark{b} & 3.3 & 230 & 7.0$\times$10$^{16}$& $\le$5.0$\times$10$^{15}$&$\le$5.0$\times$10$^{15}$ & 39.7&4.5\\
NGC~6334~IRS~1\tablenotemark{a}&  3&115 &4.5$\times$10$^{17}$ &$\le$2.0$\times$10$^{16}$&$\le$2.0$\times$10$^{16}$&-8.0 & 5.0\\
W51 e2\tablenotemark{a}   &  7&176 &9.0$\times$10$^{16}$ &$\le$3.0$\times$10$^{15}$&$\le$3.0$\times$10$^{15}$&55.6 & 8.0\\
Orion--KL\tablenotemark{a} & 10& 100 &9.7$\times$10$^{16}$&$\le$1.82$\times$10$^{15}$ &$\le$1.82$\times$10$^{15}$& 7.7 & 3.7\\ 
 \enddata
\tablenotetext{a}{Observed sources where one transition of HCOO$^{13}$CH$_3$ is detected.}
\tablenotetext{b}{Observed sources where HCOO$^{13}$CH$_3$ is tentatively detected.}
\tablenotetext{c}{Observed sources where HCOO$^{13}$CH$_3$ is not detected.}
     \end{deluxetable}

\subsubsection{HCOO$^{13}$CH$_{3}$}
We report the detection of one transition of $^{13}$C$_2$--MF toward Orion--KL, W51~e2, NGC~6334~IRS~1 and (tentatively) G19.61-0.23, and Sgr~B2(N) (see Fig. \ref{fg4} and Table~\ref{tab5}).  More specifically, the spectral feature that we assigned to $^{13}$C$_2$--HCOOCH$_3$ is a pile-up of four lines  with frequencies lying in the range 284729--284730~MHz, with a line-strength of 71~D$^{2}$ and an upper energy level of 194~K (Table~\ref{tab5}). 
Figure~\ref{fg4} exhibits the observed spectrum of the HCOO$^{13}$CH$_3$ transition emitting at 284730~MHz along with our XCLASS for each sources (reduced $\chi^{2}$ of about 0.23--0.69).
 We infer that it is difficult to attribute this spectral feature to another molecule on the basis  that: 
 \begin{enumerate}
\item The line rest frequencies of these four lines are predicted with an uncertainty of  0.012~MHz (which corresponds to 0.012~km~s$^{-1}$).
\item Several $^{13}$C$_2$--methyl formate transitions with similar S$\mu$$^{2}$ and E$_{up}$ are detected in the ALMA-SV data of Orion-KL (see Section 5.~1.). Furthermore their excitation level in Orion--KL is consistent with the emission level of this line (and non detection of other lines) in the APEX data.
\item This is the strongest line in the APEX band and the two next highest S$\mu^{2}$ lines (at 67 and 61~D$^{2}$) are respectively blended and at the confusion limit level.
\end{enumerate}

We note  that given the sensitivity limit in all the sources except Orion--KL (because of the ALMA-data) we cannot claim a definitive detection of this molecule in the APEX observations.
 
\subsubsection{Isotopic $\rm^{12}C/^{13}C$ ratio}
Since we do not have definitive detections of the $^{13}$C--MF, Table~\ref{tab7} lists  the lower limits of the isotopic $^{12}$C/$^{13}$C ratio that are estimated assuming that the two $^{13}$C--MF isotopologues have similar abundances. Please note that the upper limits of the $^{13}$C--MF column density have been set by adjusting the observational parameters to the model with a resulting fit constrained by a 3$\sigma$ upper limit. The quality of our models being still based on the reduced $\chi^{2}$ calculations\footnote{The reduced $\chi^{2}$ roughly gives a measure of how the model fit the data over the bandpass.}.

%
\begin{deluxetable}{lccl}
\tablewidth{0.pt}
\tablecolumns{4}
\tablecaption{$^{12}$C/$^{13}$C--HCOOCH$_3$ ratio as measured with ALMA and APEX, respectively.\label{tab7}}
\tablehead{
Source & $^{12}$C/$^{13}$C for CO & $^{12}$C/$^{13}$C for CO & $^{12}$C/$^{13}$C--HCOOCH$_3$\tablenotemark{c}  \\
& CN and H$_{2}$CO\tablenotemark{a} & only\tablenotemark{b} &}
\startdata
\multicolumn{4}{c}{ALMA observations}\\
\hline
Orion--KL  -- Compact Ridge & 74$\pm$16& 67$\pm$17 &68.4$\pm$10.1\\
Orion--KL  -- Hot Core--SW & 74$\pm$16& 67$\pm$17 &71.4$\pm$7.8\\
\hline
\multicolumn{4}{c}{APEX observations}\\
\hline
Sgr~B2(N-LMH)\tablenotemark{e} &19$\pm$7 &20$\pm$8& $\ge$17 \\
G24.78+0.08\tablenotemark{f} & 42$\pm$11&39$\pm$12& $\ge$20 \\
G29.96-0.02\tablenotemark{f} &47$\pm$12 &44$\pm$13&$\ge$11 \\
G19.61-0.23\tablenotemark{e} & 49$\pm$12 &45$\pm$13&  $\ge$14\\
NGC~6334~IRS~1\tablenotemark{d} & 61$\pm$14& 56$\pm$15& $\ge$23 \\
W51 e2\tablenotemark{d}  & 70$\pm$16 &64$\pm$17& $\ge$30\\
Orion--KL\tablenotemark{d} & 74$\pm$16& 67$\pm$17 &$\ge$53 \\
 \enddata
\tablenotetext{a}{Based on the following equation for CO, CN and H$_{2}$CO $\rm {^{12}C/^{13}C=6.21(1.00)D_{GC}+18.71(7.37)}$ from \citet{Milam:2005} (see Eq.\ref{eq:ratio}).}
\tablenotetext{b}{Based on the following equation for CO  $\rm {^{12}C/^{13}C=5.41(1.07)D_{GC}+19.03(7.90)}$ from \citet{Milam:2005} (see Eq.\ref{eq:ratio1}).}
\tablenotetext{c}{This study, assuming that both $^{13}$C$_{1}$--MF and $^{13}$C$_{2}$-MF isotopologues have similar abundances (i.e.  $\frac{^{12}C-MF}{^{13}C_{1}-MF} = \frac{^{12}C-MF}{^{13}C_{2}-MF}$, see Sec.~\ref{sec:resultsori} and Fig.~\ref{fg2}).}
\tablenotetext{d}{Observed sources where one transition of HCOO$^{13}$CH$_3$ is detected.}
\tablenotetext{e}{Observed sources where HCOO$^{13}$CH$_3$ is tentatively detected.}
\tablenotetext{f}{Observed sources where HCOO$^{13}$CH$_3$ is not detected.}
       \end{deluxetable}
       
%
\section{Discussion}
\label{sec:discussion}

{\subsection{Measurement caveats}
The analysis above relies on some assumptions. In the following section we discuss whether they could modify the interpretation of our derived $^{12}$C/$^{13}$C ratio.

{\subsubsection{LTE and radiative pumping effects}
It is important to note that the above analysis hinges upon the assumption that methyl formate is in LTE, which applies at the high densities in hot cores. 
We assumed that LTE is a reasonable approximation given that the model fit to the Herschel observations of methyl formate in Orion--KL contained over a thousand emissive transitions which are closely fit using an LTE model \citep{Crockett:2014}. A strong IR radiation field could affect the LTE analysis. Among the observed sources, Orion--KL and SgrB2(N) have the strongest IR radiation field. Therefore, if present radiative pumping effects would be the strongest toward those sources. The Herschel observations and analysis of  Orion--KL and SgrB2(N) \citep{Crockett:2014,Neill:2014} have shown that \textbf{i)} pumping is not needed to fit the lines as LTE closely matches the observed emission; \textbf{ii)} there was evidence for radiative pumping in emission lines of other molecules, in particular methanol, but not for methyl formate.

{\subsubsection{Contamination from strong absorption lines in SgrB2(N)}
Contamination from strong absorption lines in SgrB2(N) may also affect methyl formate emission. There are two potential levels of contamination that could be an issue. First is absorption of methyl formate that lies in the foreground envelope. However, all the transitions that we have detected in this study cover fairly high energy levels that are not populated in the envelope \citep[e.g.][]{Neill:2014}. Another issue would be contamination from other species with ground state transitions that have similar frequencies; we see no evidence for this in our data.

{\subsubsection{Scattering on the isotopic ratio of the methyl formate isotopologues within each ALMA spectral window}
Another possible caveat of our $^{12}$C/$^{13}$C ratio estimate is the individual modeling of each sub-band of the ALMA-SV observations of Orion--KL. In our exploration of the ALMA--SV data we found that a large source of uncertainty is an about 10$\%$ difference in calibration between sub-bands; that is some sub-bands have a slightly different calibration than other sub-bands. We infer that this is due to some structure (that could be a slope, a curvature or a frequency dependence) in the calibration that affects the band pass and results in this slight measurement uncertainty that exists within a given sub-band.

Based upon this and due to the fact that different sub-bands have different number of lines (see Table~\ref{tab4}) we have chosen to fit the $^{12}$C/$^{13}$C ratio in each sub-band individually and to use the relative errors in the fit from those bands to set the absolute uncertainty to our measurement. Nevertheless, it is important to note that a single set of parameters (not shown here) also fit all the data and give rise, within the uncertainties, to a similar isotopic abundance ratio.

\subsection{Comparison of the derived  column densities with previous studies}
In this section we relate our results to previous studies performed towards our source sample. Our models are not unique and some differences with previously reported result can appear. This is in part due to the different (and more accurate) $^{12}$C--MF partition functions used here (see discussion in Section~4.~1). Also, we note that for all the observed sources, the observed $v$$_{LSR}$ and $\Delta$$v$$_{LSR}$ are also consistent with those in the literature.

\textit{Orion-KL:} From the ALMA-SV observations of Orion--KL we derived a methyl formate column density over all the spectral windows of 5--8.5 $\times$10$^{17}$ cm$^{-2}$ toward the Compact Ridge and of 3.3--4.3 $\times$10$^{17}$ cm$^{-2}$ toward the Hot Core--SW. 
These results are higher by a factor 2--5 with our reported values obtained from observations using the Plateau de Bure Interferometer and performed with a similar synthesized beam \citep[1.8$\arcsec$ $\times$ 0.8$\arcsec$, see][]{Favre:2011}. This discrepancy can be explained by the fact that in this study we used a different partition function that, as discussed in Section~4.~1, results in a higher inferred $^{12}$C--MF abundance compared to the  $^{12}$C--MF abundance derived using the JPL catalog partition function used by \citet{Favre:2011}.  
The spatial and the velocity distribution are in agreement with previous observations \citep{Favre:2011,Friedel:2008}. We refer to \citet{Favre:2011} for a detailed comparison with previous related interferometric and single-dish studies performed by \citet{Friedel:2008,Beuther:2005,Liu:2002,Remijan:2003,Hollis:2003,Blake:1996,Blake:1987,Schilke:1997,Ziurys:1993}.

For the Orion--KL observations carried out with the APEX telescope, the bulk of the HCOOCH$_3$ emission is well reproduced by a single component model with a rotational temperature of 100~K, a source size of 10$\arcsec$ and a column density of 9.6 $\times$10$^{16}$ cm$^{-2}$. Our derived APEX rotational temperature and column density agree with the Herschel/HIFI observations \citep{Crockett:2014} as well as with \citet{Favre:2011} in which the authors do not separate the two HCOOCH$_{3}$ velocity components (T$\rm_{rot}$ of 101~K, N$\rm_{HCOOCH3}$ = 1.5$\times$10$^{17}$ cm$^{-2}$). 

Regarding the H$^{13}$COOCH$_{3}$ and HCOO$^{13}$CH$_{3}$ species, their detection towards Orion-KL has previously been reported by \citet{Carvajal:2009} based on IRAM-30m observations. The authors used a source size of 15$\arcsec$, a column density of 7 $\times$10$^{14}$ cm$^{-2}$ and rotational temperature of 110~K to reproduce the emission arising from the Compact Ridge component associated with Orion-KL. The $^{13}$C$_2$--MF column density, derived from the APEX observations, lies in the range 7--9.8 $\times$10$^{15}$ cm$^{-2}$ and differs from the one derived by \citet{Carvajal:2009}. This is likely due to the fact we use a different partition function (see above) along with different assumptions with regard to the beam filling factor  (source size of 3$\arcsec$  and a T$_{rot}$ of 80~K for our best models). 

\textit{W51~e2:} From the APEX observations, we derived a slightly lower (factor 1.8) methyl formate column density in comparison to the measured column density reported by \citet{Demyk:2008}.

\textit{G19.61-0.23:}  Using CARMA observations (2$\arcsec$ resolution), \citet{Shiao:2010} have reported a derived column density of (9$\pm$2) $\times$10$^{16}$ cm$^{-2}$ given a rotation temperature of 161~K. Likewise, BIMA observations, \citet{Remijan:2004} derived from a source average of 2.8$\arcsec$ a column density in methyl formate of 3.4 $\times$10$^{17}$~cm$^{-2}$ given a temperature of 230~K. In our analysis, we have adopted the HCOOCH$_3$ rotation temperature derived by \citet{Remijan:2004} rather than the one reported by \citet{Shiao:2010}. This choice is based upon the fact that \citet{Shiao:2010} used a temperature derived from ethyl cyanide observations whereas \citet{Remijan:2004} used a rotation temperature based on methyl formate observations themselves. Our best fit results in a methyl formate column density of 7$\times$10$^{16}$ cm$^{-2}$ with is commensurate with the value derived from the CARMA observations. Regarding the BIMA observations, the difference between the derived column densities is likely due to beam dilution.

\textit{G29.96-0.02:}  Using 2$\arcsec$ resolution CARMA observations, \citet{Shiao:2010} have reported a derived column density of (4$\pm$1) $\times$10$^{16}$ cm$^{-2}$ which is in agreement with our results (see Table~\ref{tab6}) taking into account the different assumptions on the source size with respect to beam.

\textit{G24.78+0.08:} The HCOOCH$_3$ column density derived from the APEX observations (see Table~\ref{tab6}) differs from the one derived by \citet{Bisschop:2007} likely due to different assumptions with regard to the beam filling factor.

\textit{NGC~6334~IRS~1:} The deviation observed with values reported by \citet{Bisschop:2007} are also likely due to different assumptions with regard to the beam filling factor. Nonetheless, our value of 4.5$\times$10$^{17}$ cm$^{-2}$ is consitent with the value reported by \citet{Zernickel:2012} (N=7$\times$10$^{17}$ cm$^{-2}$ for a 3$\arcsec$ source size) from Herschel/HIFI observations of this region.

\textit{Sgr~B2(N):} Using IRAM--30m observations of SgrB2(N), \citet{Belloche:2009} modeled methyl formate emission using two velocity components associated with two sources separated by only 5.3~$\arcsec$ \citep[based on PdBI and ATCA observations, see][]{Belloche:2008a}. These components differ by about 9~km~s$^{-1}$. Our best model includes only the component emitting at the systemic velocity of the source (i.e. 63.7~km~s$^{-1}$) and our derived parameters are in agreement with the study performed by \citet{Belloche:2009}.

\subsection{Isotopologue detection and sensitivity}
Our analysis points out the need for high sensitivity  to detect isotopologues of complex molecules. Indeed, due to lack of sensitivity in our APEX observations only one $^{13}$C$_2$--MF line  (pile-up of four $^{13}$C$_2$ transitions with S$\mu$$^{2}$ of 71~D$^{2}$) is detected and, most of the $^{13}$C$_1$--MF transitions emit below and/or at the confusion limit level. In contrast, both  $^{13}$C-MF isotopologues are detected in observations performed with higher sensitivity \citep[e.g.][and this study for the supplementary ALMA data.]{Carvajal:2009}. 

 \subsection{The $^{13}$C budget in the galaxy}
From Equation~\ref{eq:ratio1} for CO and Equation~\ref{eq:ratio} for CO, CN and H$_2$CO \citep[][]{Milam:2005} we have calculated the $^{12}$C/$^{13}$C ratio in CO for each sources observed with the APEX telescope and for ALMA-SV observations of Orion--KL. These values are given in Table~\ref{tab7}. Our study shows that the derived lower limits for the APEX $^{12}$C/$^{13}$C--methyl formate ratios are consistent within the uncertainties with the $^{12}$C/$^{13}$C ratio in CO for each source. The same conclusion applies for the isotopic ratios derived towards the Orion--KL hot core--SW and compact ridge positions (ALMA-SV data).

\subsection{Implications}
Numerous measurements of the $^{12}$C/$^{13}$C isotopic ratio have been performed through several molecular tracers, such as CO and OCS, toward Orion-KL. For example, from OCS and H$_2$CS isotopologue observations \citet{Tercero:2005} have reported an average ratio of 45$\pm$20. Using methanol observations, \citet{Persson:2007} have found a $^{12}$C/$^{13}$C isotopic ratio 57$\pm$14. \citet{Savage:2002} derived a ratio of 43$\pm$7 in CN. From $^{13}$CO observations, \citet{Snell:1984} have reported an average $^{12}$CO/$^{13}$CO isotopic ratio of 74$\pm$9 in the high-velocity outflow of Orion-KL. Likewise, from infrared measurement performed with the Kitt peak Mayall 4m telescope, \citet{Scoville:1983} obtained a $^{12}$CO/$^{13}$CO isotopic ratio of 96$\pm$5.
Using C$^{18}$O observations, \citet{Langer:1990} and  \citet{Langer:1993} derived ratios of 63$\pm$6 and 74$\pm$9 according to the observed position. These finding suggest that the gas in Orion-KL does not seem to be heavily fractionated since the $^{12}$C/$^{13}$C ratio in most simple species is almost the same. Our results are consistent with this finding since: 
 \begin{enumerate}
\item for each $^{13}$C--MF isotopologues, the derived isotopic ratios (68.4$\pm$10.1 toward the Compact Ridge and of 71.4$\pm$7.8 toward the Hot Core-SW, see Fig.~\ref{fg2}) are consistent with each other.
\item These results are consistent within the error bars with the values derived for CH$_3$OH and for CO by \citet{Persson:2007}, \citet{Snell:1984} and \citet{Scoville:1983} toward Orion--KL. 
\end{enumerate}
Therefore, the present observations do not support methyl formate formation in gas--phase formation from  $^{12}$C/$^{13}$C fractionated gas.
In addition, regarding methyl formate gas-phase formation mechanisms, \citet{Horn:2004} have shown that there are no very efficient gas--phase pathways to form methyl formate, meaning that there are no efficient primary pathways to form the $^{13}$C--MF isotopologues either.  One possibility  that could lead to the gas phase formation of the $^{13}$C-MF isotopologues  would be ÒsecondaryÓ fractionation process involving the $^{12}$C--methyl formate itself and  $^{13}$C$^{+}$ (E. Herbst, private communication). Such reaction, however are unlikely to occur since  high barriers are expected. This would also argue against the possibility of methyl formate gas--phase formation from  $^{12}$C/$^{13}$C fractionated gas.
This finding combines with the hypothesis of \citet{Wirstrom:2011} strongly suggests that grain surface reactions are likely the main pathways to form methyl formate ($^{12}$C and $^{13}$C).

%
\section{Conclusions}
\label{sec:Conclusions}
We have investigated the $^{12}$C/$^{13}$C isotopic ratio in methyl formate toward a  sample of massive star-forming regions located over a range of distances from the Galactic center, through observations performed with the APEX telescope. In addition, we have measured the  $^{12}$C/$^{13}$C-methyl formate ratio towards Orion-KL using the ALMA-SV observations. 
Also, we reported new spectroscopic measurements of the H$^{13}$COOCH$_3$ and HCOO$^{13}$CH$_3$ species. Our study is based on this laboratory spectral characterization and points out the importance of these data in deriving accurate partition functions and therefore abundances of methyl formate. Our analysis also points out that to accurately derive a reliable abundance ratio between different species, it is necessary use a homogeneous observational database.

We have performed LTE modeling of the observational data. A multitude of $^{13}$C$_{1}$--MF and $^{13}$C$_{2}$--MF transitions have been detected in the ALMA-SV observations carried out toward Orion-KL, \textit{i)} confirming the previous detection of the $^{13}$C--MF isotopologues reported by \citet{Carvajal:2009} and, \textit{ii)} imaging their spatial distribution for the first time. Assuming that the two $^{13}$C--MF isotopologues have similar abundances, we reported a $^{12}$C/$^{13}$C isotopic ratio in methyl formate of 68.4$\pm$10.1 and  71.4$\pm$7.8 toward the Compact Ridge and Hot Core-SW components, respectively. A salient result is that those measurements are consistent with the  $^{12}$C/$^{13}$C ratio measured in CO and in CH$_{3}$OH.
Our findings suggest that grain surface chemistry very likely prevails in the formation of methyl formate main and $^{13}$C isotopologues.

Regarding the APEX observations, we have reported a tentative detection ($\ge$3$\sigma$ level) of the $^{13}$C$_2$-MF isotopologue towards the following four massive star-forming regions: Sgr~B2(N-LMH), NGC~6334~IRS~1, W51 e2 and G19.61-0.23.  The derived lower limits for the $^{12}$C/$^{13}$C--methyl formate ratio are consistent with the  $^{12}$C/$^{13}$C ratio measured in CO  showing a increasing ratio with distance from the Galactic centre. 
A larger source sample and further observations with high sensitivity are essential to confirm this trend.

In addition, we used the Herschel/HIFI spectral tools, that are available to the community \citep{Crockett:2014}, to make reliable line identifications and to appreciate where potential line blends may exist. The current work illustrates how to we can merge the legacy of Herschel with other telescopes such as ALMA.

%
\acknowledgments

This work was supported by the National Science Foundation under grant 1008800. We are grateful to the {\it Ministerio de Econom\'{\i}a y Competitividad} of Spain by the financial support through the Grant No. FIS2011-28738-C02-02 and to the French Government through the Grant No. ANR-08-BLAN-0054 and the French PCMI (Programme National de Physique Chimie du Milieu Interstellaire). This paper makes use of the following ALMA data: ADS/JAO.ALMA$\#$2011.0.00009.SV. ALMA is a partnership of ESO (representing its member states), NSF (USA) and NINS (Japan), together with NRC (Canada) and NSC and ASIAA (Taiwan), in cooperation with the Republic of Chile. The Joint ALMA Observatory is operated by ESO, AUI/NRAO and NAOJ. C. F. thanks Dahbia Talbi, Eric Herbst and Anthony Remijan for enlightening discussions. Finally, we thank the anonymous referee for the helpful comments.  \\


{\it Facilities:} \facility{APEX}, \facility{Herschel/HIFI}, \facility{ALMA}

\clearpage
 
%
 \appendix

\section{Partition functions calculations}
\label{AP:PF}

Several approximations for the partition function have been used, following e.g., \citet{Blake:1987}, \citet{Turner:1991}, \citet{Oesterling:1999}, \citet{Groner:2007}, \citet{Demyk:2008}, \citet{Maeda:2008}, \citet{Favre:2011} and \citet{Tudorie:2012}. In the present work, the partition function was approximated as the product of the rotational ($Q_{rot}$), torsional ($Q_{tor}$)  and vibrational (without the torsional contribution, $Q_{vib}$) contributions \citep{Herzberg:1991} and, is given by:
\begin{equation}
\label{Q-general}
Q= g_{ns} \sum_i (2 \, J_i + 1) e^{- {E_i \over k_B T}} \approx g_{ns} \,
Q_{rot} \, Q_{tor} \, Q_{vib},
\end{equation}
\noindent where $J_i$ stands for the rotational angular momentum of level $i$,
$k_B$ is the Boltzmann constant, $T$ is the temperature, and $E_i$ is the vibrational-torsional-rotational energy which is referred to the ground vibrational-torsional state as the zero point energy. The nuclear spin degeneracy $g_{ns}$ does not need to be taken into account in the calculations of the partition function because $g_{ns}$ is the same for all MF symmetry states ($A_1$, $A_2$ and $E$). Based on the ${\cal C}_{3v}(M)$ symmetry \citep{Bunker:1998} $g_{ns}=16$ for both $^{13}$C--MF isotopologues of methyl formate.

When observed intensities are estimated, $g_{ns}$ in the numerator is canceled with that of the denominator inside the partition function. Thus $g_{ns}$ will henceforth be ignored and excluded in the comparisons of partition functions of $^{13}$C--MF isotopologues.

\subsection{Rotational partition function}
The rotational partition function $Q_{rot}$ was obtained using equation 9 of \citet{Groner:2007}, that is:

\begin{equation}
\label{Q-rot}
Q_{rot} = \sum_{J,K_a,Kc} (2 \, J +1) \, e^{-{E^{(rot)}_i \over k_B
    T}},
\end{equation}
where $E^{(rot)}_i$ are the rotational energies, that is, those for the rotational states only in the $A$-symmetry ground torsional state. The RAM model~\citep{Herbst:1984,Hougen:1994,Kleiner:2010} was used to predict the torsional-rotational states as explained before. Also, the torsional-rotational states ($v_t=0$ and A-symmetry) up to $J=79$ were included in Eq.~(\ref{Q-rot}), which is enough for the convergence study mentioned above.

A comparison was done with the asymmetric top approximation for the rotational partition function of \citet{Herzberg:1991}. For sufficiently high temperatures (or small rotational constants):

\begin{equation}
\label{Qrot-approx}
Q_{rot}^{appr} \approx \sqrt{{\pi \over A^{PAM} \, B^{PAM} \, C^{PAM}}
  \left({ k_B T \over h}\right)^3}
\end{equation}
where the rotational constants are referred to the principal axis system, not to the Rho-Axis System.
Therefore, an appropriate transformation was performed from the rotational parameters given in \citet{Carvajal:2009,Carvajal:2010}.

The rotational partition function computed as a direct sum (Eq. \ref{Q-rot}) is in general, for both $^{13}$C--MF isotopologues, slightly larger than the one for the approximated partition function (Eq. \ref{Qrot-approx}), as shown in Table~\ref{tabA1}.
For this reason, we used here the rotational partition function as a direct summation instead of using the approximated partition function (Eq. \ref{Qrot-approx}).
In Table~\ref{tabA1} it can be seen that the differences between $Q_{rot}$ and $Q_{rot}^{approx}$ can be around 1\% for $T=9.375$ K, decreasing for higher temperature to the error range estimated by \citet{Herzberg:1991}. 
 
%
\begin{deluxetable}{lllll}
\tablewidth{0.pt}
\tablenum{A1}
\tablecolumns{5}
\tablecaption{
Comparison between the rotational partition function\tablenotemark{a} computed as a direct sum $Q_{rot}$ (\ref{Q-rot}) and as the approximated expression $Q^{appr}_{rot}$ (\ref{Qrot-approx}) for $^{13}$C$_1$--MF and  $^{13}$C$_2$--MF.\label{tabA1}}
\tablehead{     
      & \multicolumn{2}{c}{$^{13}$C$_1$-MF} & \multicolumn{2}{c}{$^{13}$C$_2$-MF} \\
T(K) & $\rm Q^{appr}_{rot}$ & $\rm Q_{rot}$\tablenotemark{b} &  $\rm Q_{rot}^{appr}$ & $\rm Q_{rot}$\tablenotemark{b}}
\startdata
300.0  &32682.30 &32737.70  &33291.76 &33289.35 \\
225.0  &21227.78 &21308.82  &21623.63 &21672.46 \\
150.0  &11554.94 &11599.33  &11770.41 &11797.70 \\
 75.0  & 4085.29 & 4100.86  & 4161.47 & 4170.92 \\
 37.50 & 1444.37 & 1451.06  & 1471.30& 1475.82 \\
 18.75 &  510.66 &  514.10  &  520.18&  522.86 \\
  9.375&  180.55&  182.57   &  183.91 &  185.67\\
 \enddata
\tablenotetext{a}{The nuclear spin degeneracy was not considered in these calculations.} 
\tablenotetext{b}{The rotational partition function obtained as a direct sum of energy levels up to $J=79$ is the used in the final result.} 
     \end{deluxetable}

\clearpage
\subsection{Torsional partition function}
The torsional contribution $Q_{tor}$ to the partition function was obtained through the following formula:

\begin{equation}
\label{Q-tor}
Q_{tor}^{v_t^{max}} = \sum_{v_t=0}^{v_t^{max}} \left(e^{-{E^{(tor)}(v_t,A)
    \over k_B T}} + e^{-{E^{(tor)}(v_t,E) \over k_B T}} \right),
\end{equation}   
where $E^{(tor)}(v_t,A)$ and $E^{(tor)}(v_t,E)$ are the energies of the torsional states with quantum number $v_t$ for the $A$ ($A_1$
or $A_2$) and $E$ symmetries, respectively, referred to the  $v_t=0$ ground torsional state, i.e. $E^{(tor)}(v_t=0,A)=0$ cm$^{-1}$. Different approximations 
can be carried out depending on the maximum value $v_t^{max}$ considered in the equation. The torsional energies used in Eq.(\ref{Q-tor})
are the following:

\begin{itemize}
\item Torsional energies from $v_t=0$ to $v_t=2$ computed from the  Hamiltonian parameters of the RAM model. The torsional energies of  $^{13}$C$_1$-MF are computed with the parameters of \citet{Carvajal:2010} and of $^{13}$C$_2$-MF are computed with the parameters of \citet{Carvajal:2009}. These torsional energies are expected to be very reliable for the $^{13}$C$_1$-MF.
\item Torsional energies from $v_t=3$ to $v_t=4$ of main species of methyl formate given by \citet{Senent:2005} and considered as a good approximation for both $^{13}$C--MF isotopologues. 
\item Torsional energies from $v_t=5$ to $v_t=6$ were roughly estimated in the present work, where $E^{(tor)}(v_t=m,A)=m \times E^{(tor)}(v_t=1, A)$ and $E^{(tor)}(v_t=m,E)=m \times E^{(tor)}(v_t=1, E)$ and $m$ will take values as $5$ or $6$. This is only  an estimate to understand the contribution of these torsional levels  to the torsional partition function, whose contribution is of  $3$\% for $T=300$ K, $1.2$\% for $T=225$ K, $0.2$\% for $150$ K,
    etc \dots (see Tables~\ref{tabA2} and \ref{tabA3}).
\end{itemize}

As the torsional mode is very anharmonic, we cannot use the harmonic approximation for the torsional partition function. From our results (Tables~\ref{tabA2} and \ref{tabA3}), when computing the torsional partition function, the harmonic approximation could be assumed only for temperatures
T$<$100~K. Above  $T=100$ K,  the anharmonicity has the natural effect of increasing the estimated torsional partition function. This effect can be around $5$\% at $300$~K.

In Tables~\ref{tabA2} and \ref{tabA3} the torsional partition function at different approximations are shown for $^{13}$C$_1$--MF and $^{13}$C$_2$--MF isotopologues respectively. It can be noted that for $T=300$ K the convergence is reached to within 1$\%$ when the torsional states above $v_t=6$
are included. For temperatures T$<$200~K, the contribution of the torsional states above $v_t=4$ is insignificant.
In fact, at temperatures close to 100~K and below, the convergence is
reached (within 0.9$\%$ at T=100~K) when only $v_t=0,1$ and $2$ are considered.

%
\begin{deluxetable}{lllllll}
\tablewidth{0.pt}
\tablenum{A2}
\tablecolumns{7}
\tablecaption{Torsional partition function for $^{13}$C$_1$--MF at different approximations\tablenotemark{a}.\label{tabA2}}
\tablehead{     
T(K) & $\rm Q_{tor}^0$ & $\rm Q_{tor}^1$ & $\rm Q_{tor}^2$ & $\rm Q_{tor}^4$ & $\rm Q_{tor}^6$\tablenotemark{b}&$\rm Q_{tor}^{harm}$~\tablenotemark{c}}
\startdata
300.0  &1.99994 & 3.06540  & 3.71217  & 4.39180 & 4.52333 & 4.28020\\
225.0  &1.99991 & 2.86364  & 3.30761  & 3.68701 & 3.73003 & 3.52028\\
150.0  &1.99987 & 2.56748  & 2.77670  & 2.89625 & 2.90098 & 2.79253\\
 75.0  &1.99974 & 2.16083  & 2.18274  & 2.18678 & 2.18679 & 2.17520\\
 37.50 &1.99948 & 2.01246  & 2.01270  & 2.01270 & 2.01270 & 2.01306\\
 18.75 &1.99896 & 1.99905  & 1.99905  & 1.99905 & 1.99905 & 2.00008\\
  9.375&1.99793 & 1.99793  & 1.99793  & 1.99793 & 1.99793 & 2.00000\\
 \enddata
\tablenotetext{a}{The approximations are carried out by considering a number of torsional states up to a maximum quantum number $v_t^{max}$ in Eq.~(\ref{Q-tor}).}
\tablenotetext{b}{Computed torsional partition function used as a final result in the present work.} 
\tablenotetext{c}{Harmonic approximation for the torsional partition function: 
\begin{equation}
Q_{tor}^{harm}={1 \over 1 - e^{-{E^{(tor)}(v_t=1,A) \over k_B T}}} + {1 \over 1 - e^{-{E^{(tor)}(v_t=1,E) \over k_B T}}}
\end{equation}}
     \end{deluxetable}
     
%
\begin{deluxetable}{lllllll}
\tablewidth{0.pt}
\tablenum{A3}
\tablecolumns{7}
\tablecaption{Torsional partition function for $^{13}$C$_2$--MF at different approximations\tablenotemark{a}.\label{tabA3}}
\tablehead{  
T(K) & $\rm Q_{tor}^0$ & $\rm Q_{tor}^1$ & $\rm Q_{tor}^2$ & $\rm Q_{tor}^4$ & $\rm Q_{tor}^6$\tablenotemark{b}&$Q_{tor}^{harm}$~\tablenotemark{c}}
\startdata
300.0  &1.99994&3.07133   & 3.69952  & 4.37914  & 4.51465 & 4.30756\\
225.0  &1.99991&2.87006   & 3.29709  & 3.67650  & 3.72124 & 3.54028\\
150.0  &1.99987&2.57382   & 2.77116  & 2.89072  & 2.89573 & 2.80495\\
 75.0  &1.99974&2.16445   & 2.18394  & 2.18798  & 2.18798 & 2.17949\\
 37.50 &1.99948&2.01305   & 2.01324  & 2.01324  & 2.01324 & 2.01366\\
 18.75 &1.99896&1.99906   & 1.99906  & 1.99906  & 1.99906 & 2.00009\\
  9.375&1.99793&1.99793   & 1.99793  & 1.99793  & 1.99793 & 2.00000\\
 \enddata
\tablenotetext{a}{The approximations are carried out by considering a number of torsional states up to a maximum quantum number $v_t^{max}$ in Eq.~(\ref{Q-tor}).} 
\tablenotetext{b}{Computed torsional partition function used as a final result in the present work.} 
\tablenotetext{c}{Harmonic approximation for the torsional partition function: 
\begin{equation}
Q_{tor}^{harm}={1 \over 1 - e^{-{E^{(tor)}(v_t=1,A) \over k_B T}}} + {1 \over 1 - e^{-{E^{(tor)}(v_t=1,E) \over k_B T}}}
\end{equation}}
     \end{deluxetable}     

\subsection{Vibrational partition function}
In the calculation of the vibrational partition function $Q_{vib}$, it is expected that for ISM temperatures only the information of the vibrational frequencies at lower energies ($\approx 300$~cm$^{-1}$) is necessary. In order to check the convergence of the
vibrational partition function, the contribution of the remaining vibrational
modes has been taken into consideration. For this purpose, the harmonic
approximation of the vibrational partition function is considered in
general as:

\begin{equation}
\label{Qvib}
Q_{vib}=\Pi_{i=1}^{3N-7} {1 \over 1 - e^{E^{(vib)}_i/k_B T}}
\end{equation}

\noindent where $N$ is the number of atoms of the molecule, and $E^{(vib)}_i$
is the vibrational fundamental frequencies of each vibrational mode of the molecule. As the torsion is treated apart, the product in Eq.~(\ref{Qvib}) will only expand to the $3N-7$ small amplitude vibrational modes. It is important to note that no experimental vibrational frequencies exist for the $^{13}$C--MF species. Therefore, to take into account the vibrational contribution of the partition function, we assumed that the vibrational fundamental frequencies of $^{13}$C$_1$--MF and $^{13}$C$_2$--MF are approximately the same as for the main isotopologue, given the large experimental uncertainties \citep[mostly of 6--15~cm$^{-1}$,][]{Chao:1986}.  In this instance, the experimental vibrational energies for the main isotopologue taken from \citet{Chao:1986} are also valid for their other isotopologues.
The vibrational partition function computed with Eq.~(\ref{Qvib}) is given in Table~\ref{tabA4}.

In this work, for the temperature ranges considered, all the small amplitude vibrational fundamentals in Eq.~(\ref{Qvib}) are included. Nevertheless, when the temperatures are around $T=200$ K, all vibrational fundamentals could be omitted in the vibrational partition function except those of $\nu_{14}$ and $\nu_1$ modes (around $300$ cm$^{-1}$). Below $T=100$ K, inclusively $\nu_{14}$ and $\nu_1$ modes could be neglected.

\subsection{Rotational-torsional- vibrational partition function}
In addition, we have assessed that the partition function separated into functions of each rotational, torsional and vibrational contribution is a good enough approximation for temperatures at least under $300$ K. This assessment was set up after comparing our partition function calculation with that derived from its general expression, Eq.~3, for $v_t = 0$ and 1. 
Finally, Table ~\ref{tab3} summarizes  the rotational-torsional-vibrational partition function values that are used here for $^{13}$C$_1$--MF and $^{13}$C$_2$--MF.

%
\begin{deluxetable}{ll}
\tablewidth{0.pt}
\tablenum{A4}
\tablecolumns{2}
\tablecaption{Vibrational partition function for $^{13}$C$_1$--MF and $^{13}$C$_2$--MF. \label{tabA4}}
\tablehead{  
T(K) & $\rm Q_{vib}$}
\startdata
300.0  & 1.70330 \\
225.0  & 1.32486 \\
150.0  & 1.09599 \\
 75.0  & 1.00397 \\
 37.50 & 1.00001 \\
 18.75 & 1.00000 \\
  9.375& 1.00000 \\
 \enddata
     \end{deluxetable}

\clearpage

 \section{Transitions of $^{12}$C and $^{13}$C-methyl formate observed with the ALMA telescope toward Orion-KL}
 \label{AP:TAO}
Table \ref{tabB1} summarizes the line parameters for all detected, blended, or not detected transitions of $^{12}$C--MF, $^{13}$C$_1$--MF  and $^{13}$C$_2$--MF in all ALMA spectral windows.

\begin{deluxetable}{llrr}
\tablewidth{0pt}
\tabletypesize{\footnotesize}
\tablenum{B1}
\tablecolumns{4}
\tablecaption{Transitions of $^{12}$C and $^{13}$C-methyl formate observed with the ALMA telescope.\label{tabB1}}
\tablehead{
Frequency & Transition & E$_{up}$ &  S$\mu$$^{2}$ \\
(MHz) & & (K) & (D$^{2}$) \\}
\startdata
\multicolumn{4}{c}{HCOOCH$_3$\tablenotemark{a}} \\
\hline
214631.77	$^{*}$	&	$	17	_{	5	,	12	}$	--	$	16	_{	5	,	11	}$	(E, v$_{t}$=0)	&	108	&	41	\\
214652.63	$^{*}$	&	$	17	_{	5	,	12	}$	--	$	16	_{	5	,	11	}$	(A, v$_{t}$=0)	&	108	&	41	\\
214782.36	$^{*}$	&	$	18	_{	3	,	16	}$	--	$	17	_{	3	,	15	}$	(E, v$_{t}$=0)	&	106	&	46	\\
214792.55	$^{*}$	&	$	18	_{	3	,	16	}$	--	$	17	_{	3	,	15	}$	(A, v$_{t}$=0)	&	106	&	46	\\
214816.95		&	$	19	_{	2	,	18	}$	--	$	18	_{	2	,	17	}$	(A, v$_{t}$=1)	&	296	&	49	\\
214942.87		&	$	19	_{	1	,	18	}$	--	$	18	_{	1	,	17	}$	(A, v$_{t}$=1)	&	296	&	49	\\
215073.92		&	$	19	_{	2	,	18	}$	--	$	18	_{	2	,	17	}$	(E, v$_{t}$=1)	&	296	&	49	\\
\enddata
 \tablenotetext{a}{This Table lists the detected, blended, partially blended and optically thin lines of HCOOCH$_3$. Transitions that are emitting with an intensity less or equal to 3 times the noise level and that we used to constrain our model are also listed. Transitions with the $^{*}$ symbol, are transitions we estimate to be optically thick. }
     \end{deluxetable}
 \clearpage

 \section{Methyl formate emission towards the Orion-KL Compact Ridge as observed with ALMA}
 \label{AP:CR}
Figures~\ref{fga1}, \ref{fga2} and \ref{fga3} show the respective spectra of the detected HCOOCH$_{3}$, H$^{13}$COOCH$_{3}$ and HCOO$^{13}$CH$_{3}$ transitions observed with ALMA toward the Orion-KL Compact Ridge during the science verification program, along with our best models achieved using the XCLASS program.


\begin{figure}
\figurenum{C-1}
\includegraphics[angle=270,width=15cm]{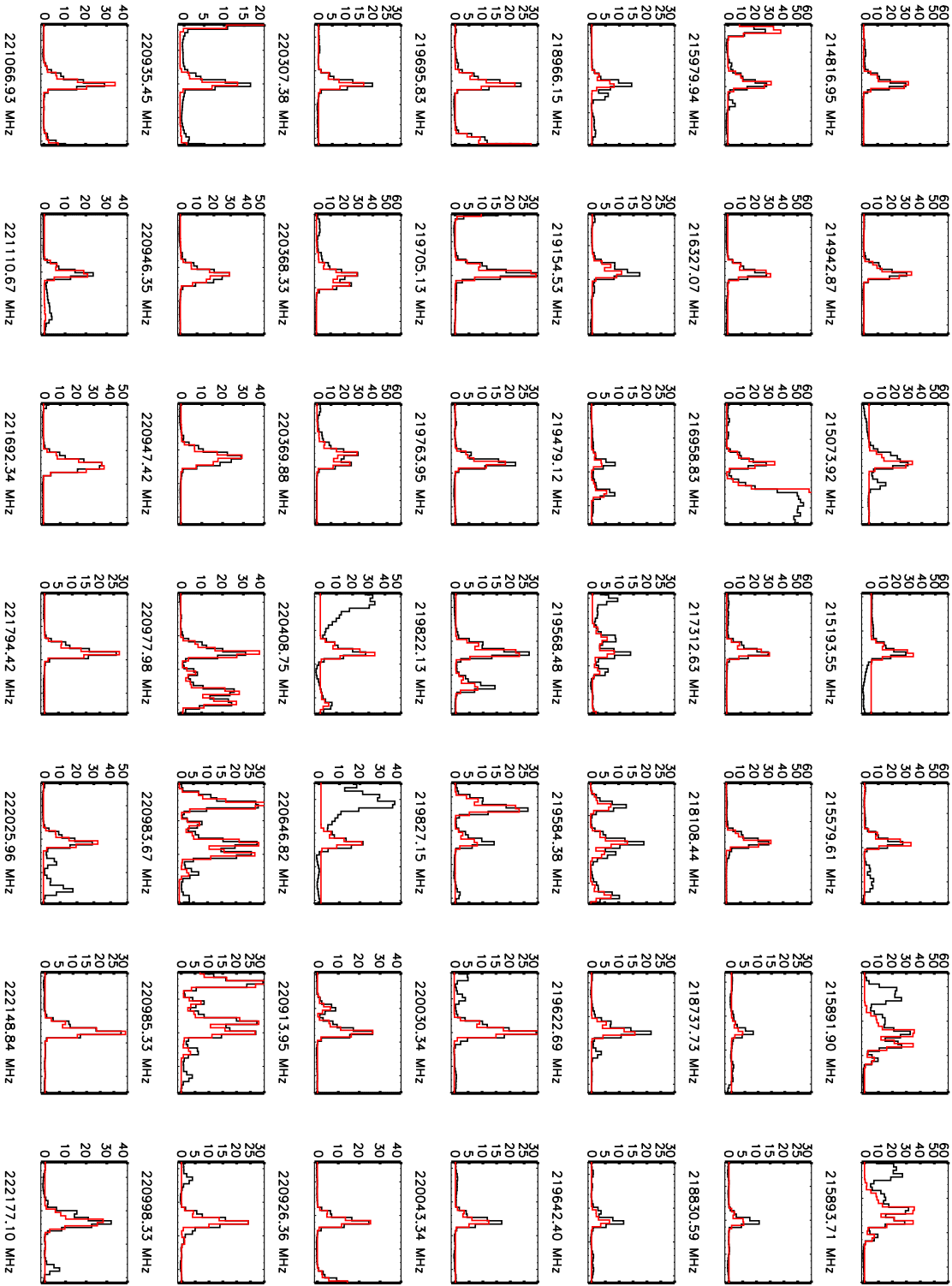}
\caption{HCOOCH$_{3}$ spectra (black) towards the Compact Ridge component associated with Orion-KL and model (red), as observed with ALMA. The intensity scale is in T$\rm_{MB}$ (K). The x-axis scale is about $\pm$9.5~MHz centered on the line rest frequency, which is indicated below each plot.\label{fga1}}
\end{figure} 

\begin{figure}
\figurenum{C-1}
\includegraphics[angle=270,width=15cm]{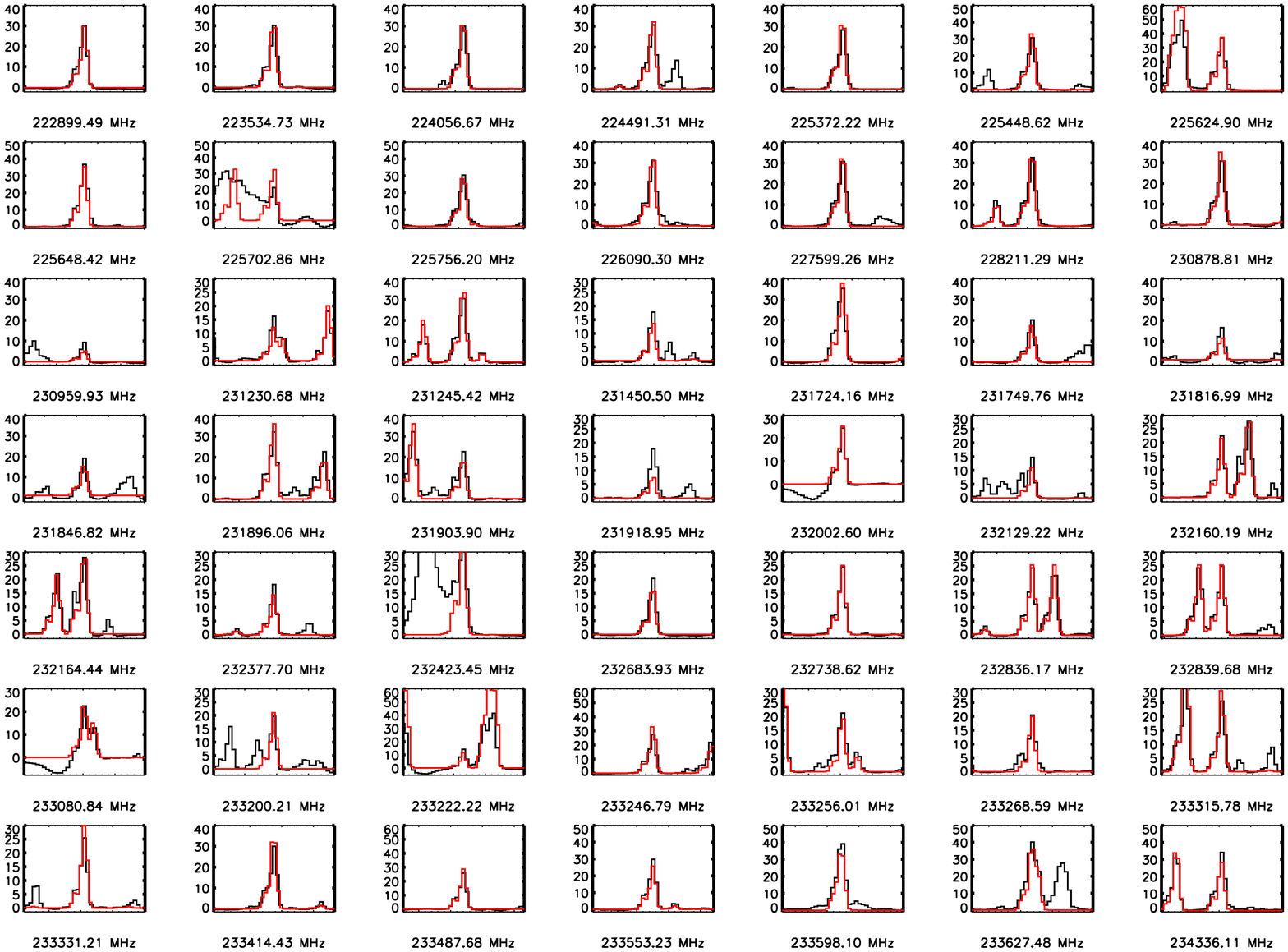}
\caption{Continue}
\end{figure} 

\begin{figure}
\figurenum{C-1}
\includegraphics[angle=270,width=15cm]{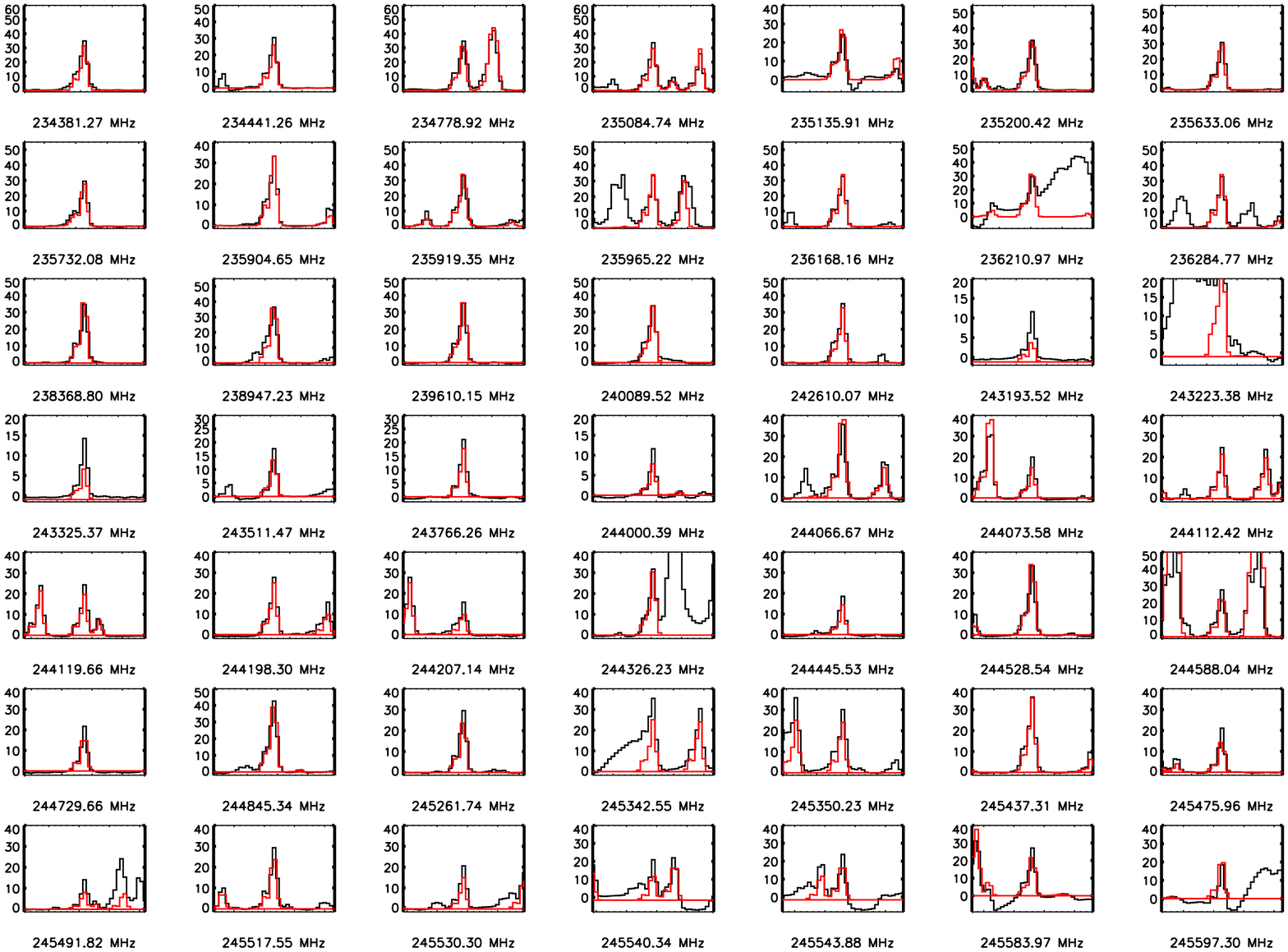}
\caption{Continue.}
\end{figure} 

\begin{figure}
\figurenum{C-1}
\includegraphics[angle=270,width=15cm]{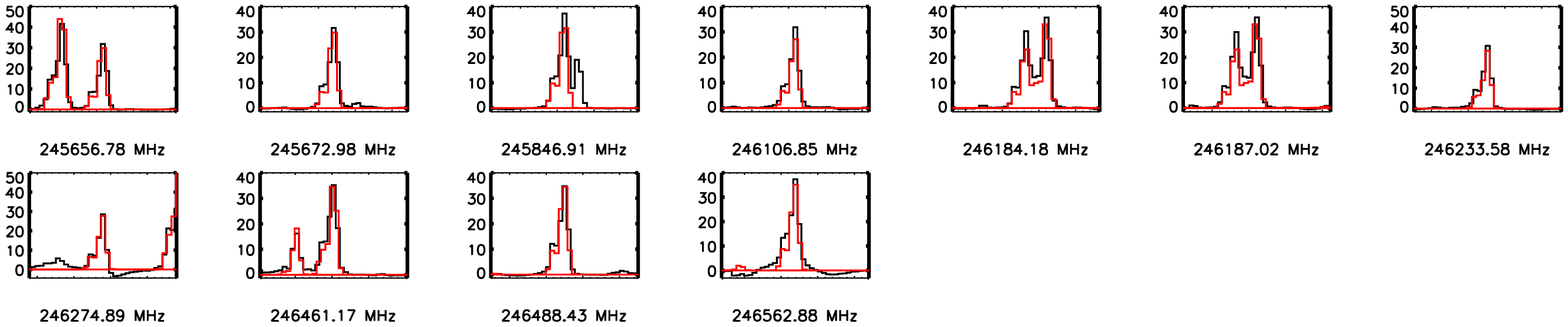}
\caption{Continue.}
\end{figure} 

\begin{figure}
\figurenum{C-2}
\includegraphics[angle=270,width=15cm]{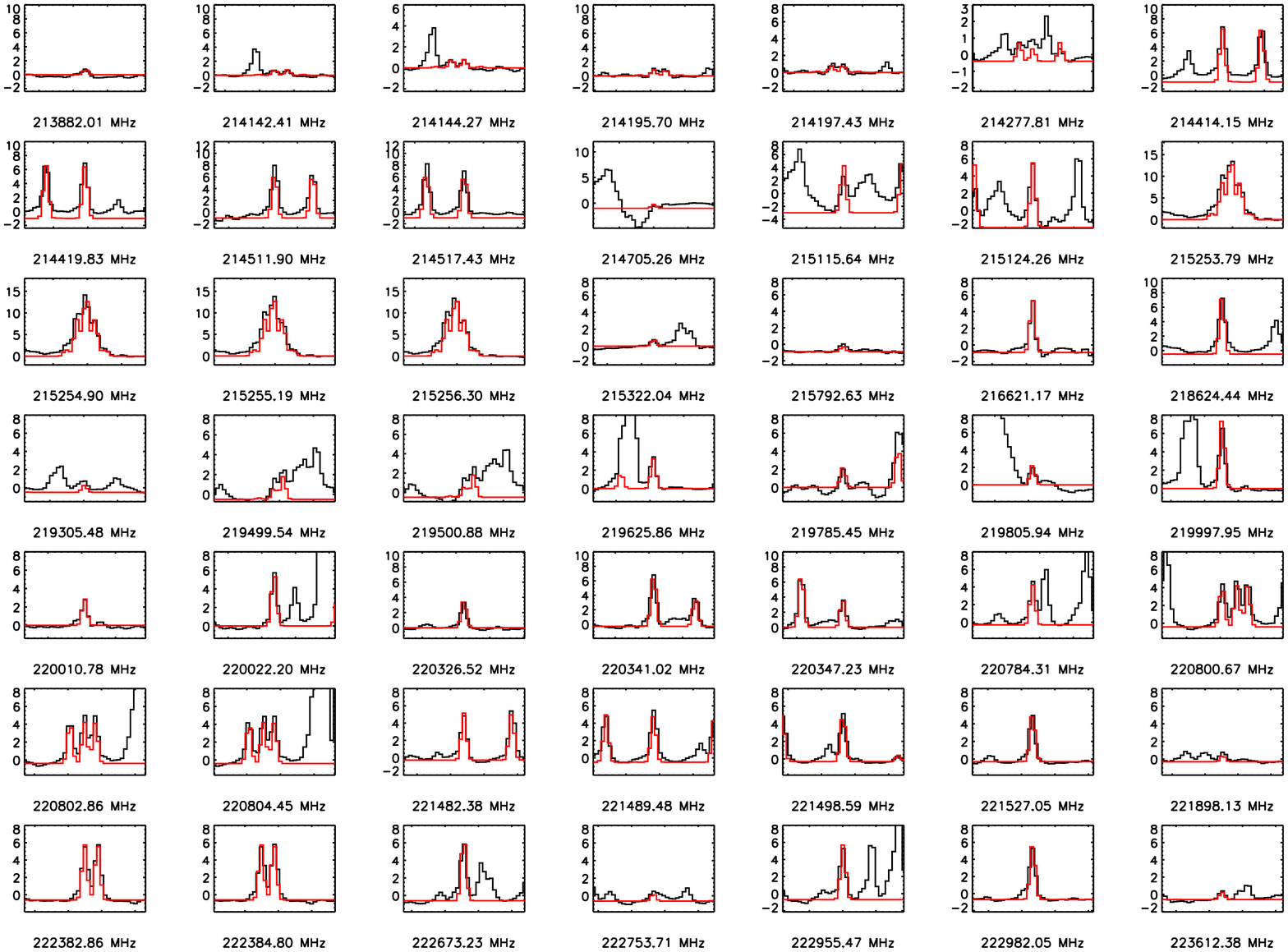}
\caption{H$^{13}$COOCH$_{3}$ spectra (black) towards the Compact Ridge component associated with Orion-KL and model (red), as observed with ALMA. The intensity scale is in T$\rm_{MB}$ (K). The x-axis scale is about $\pm$9.5~MHz centered on the line rest frequency, which is indicated below each plot.\label{fga2}}
\end{figure} 

\begin{figure}
\figurenum{C-2}
\includegraphics[angle=270,width=15cm]{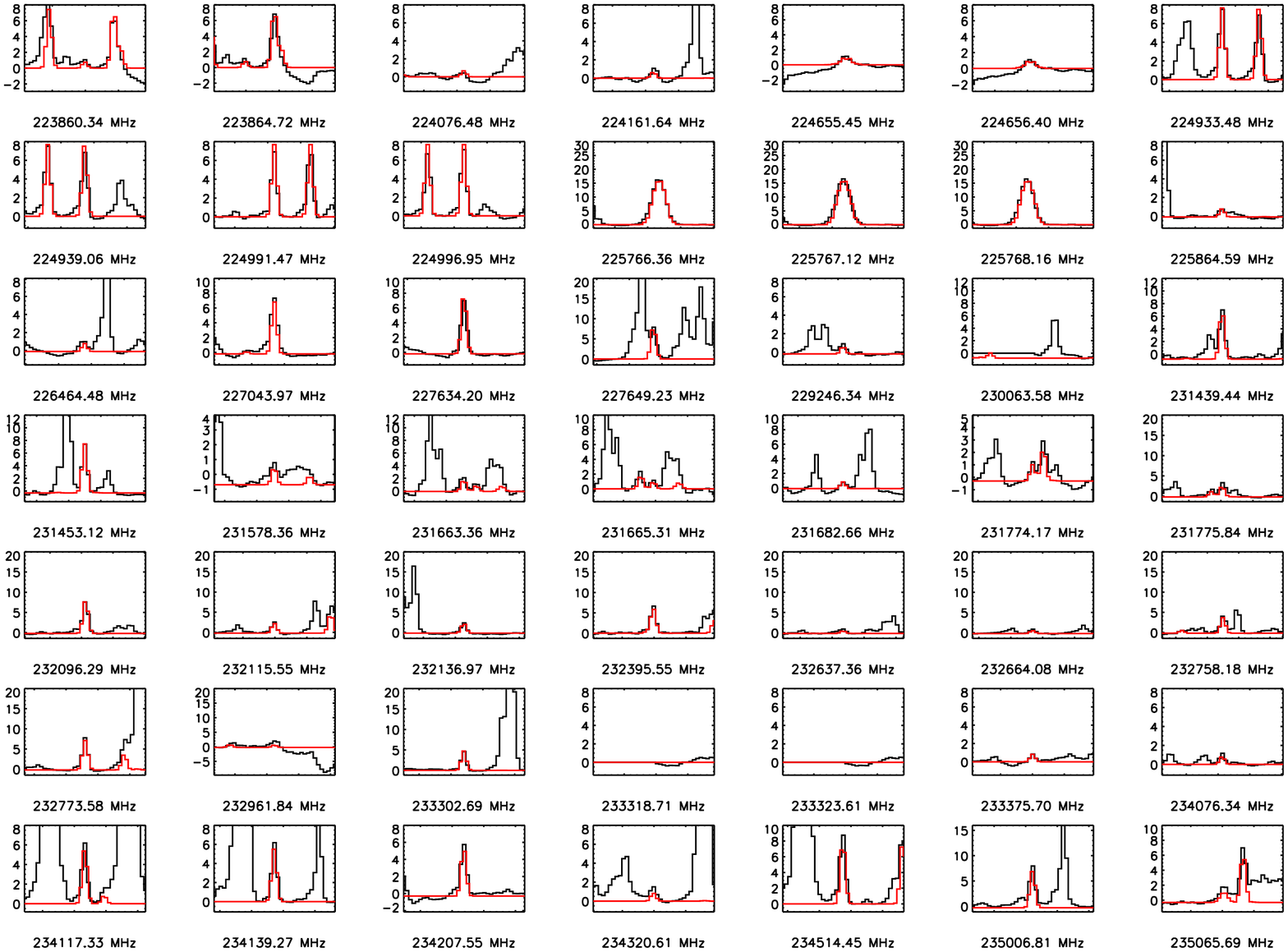}
\caption{Continue.}
\end{figure} 

\begin{figure}
\figurenum{C-2}
\includegraphics[angle=270,width=15cm]{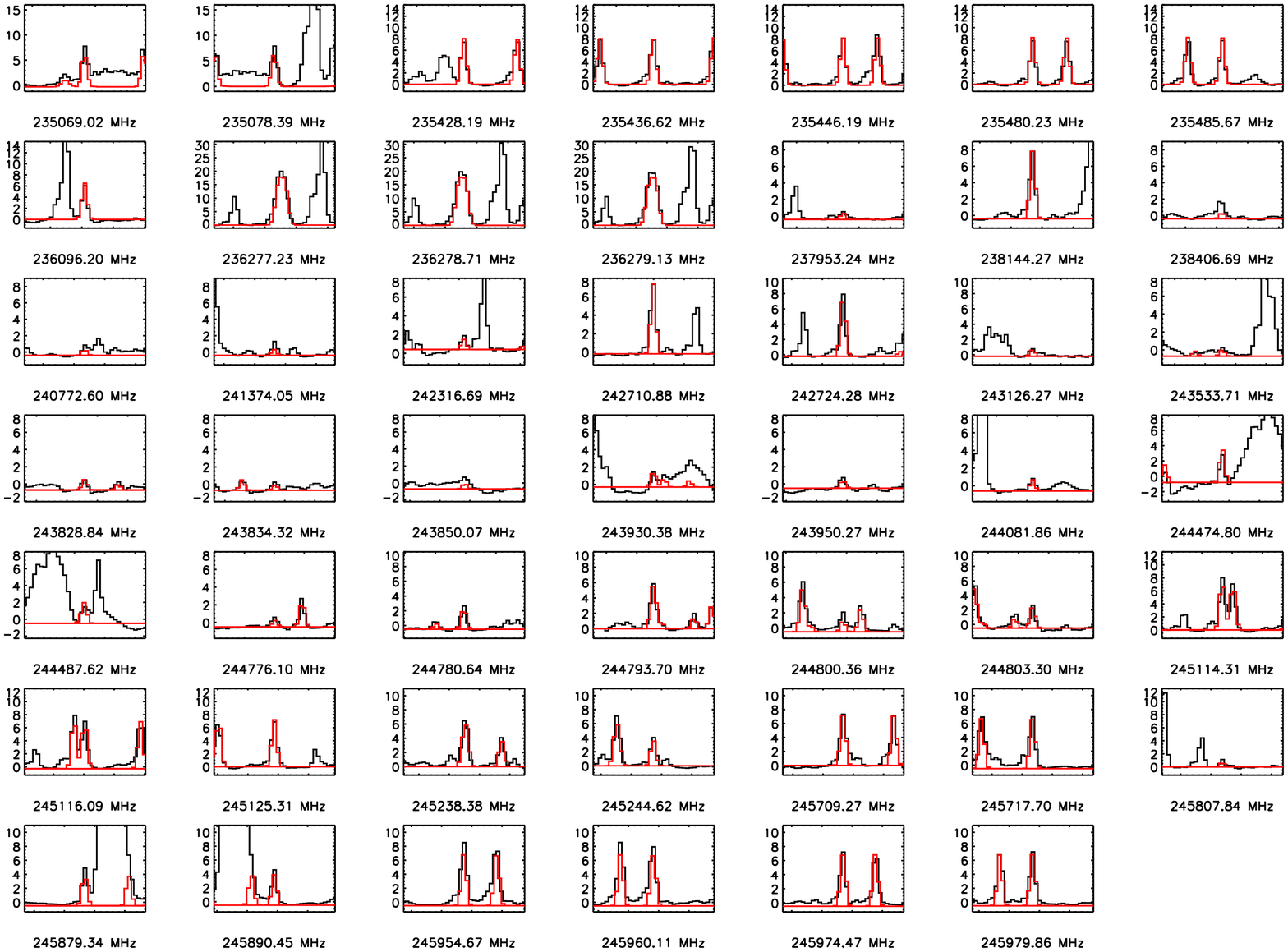}
\caption{Continue.}
\end{figure} 

\begin{figure}
\figurenum{C-3}
\includegraphics[angle=270,width=15cm]{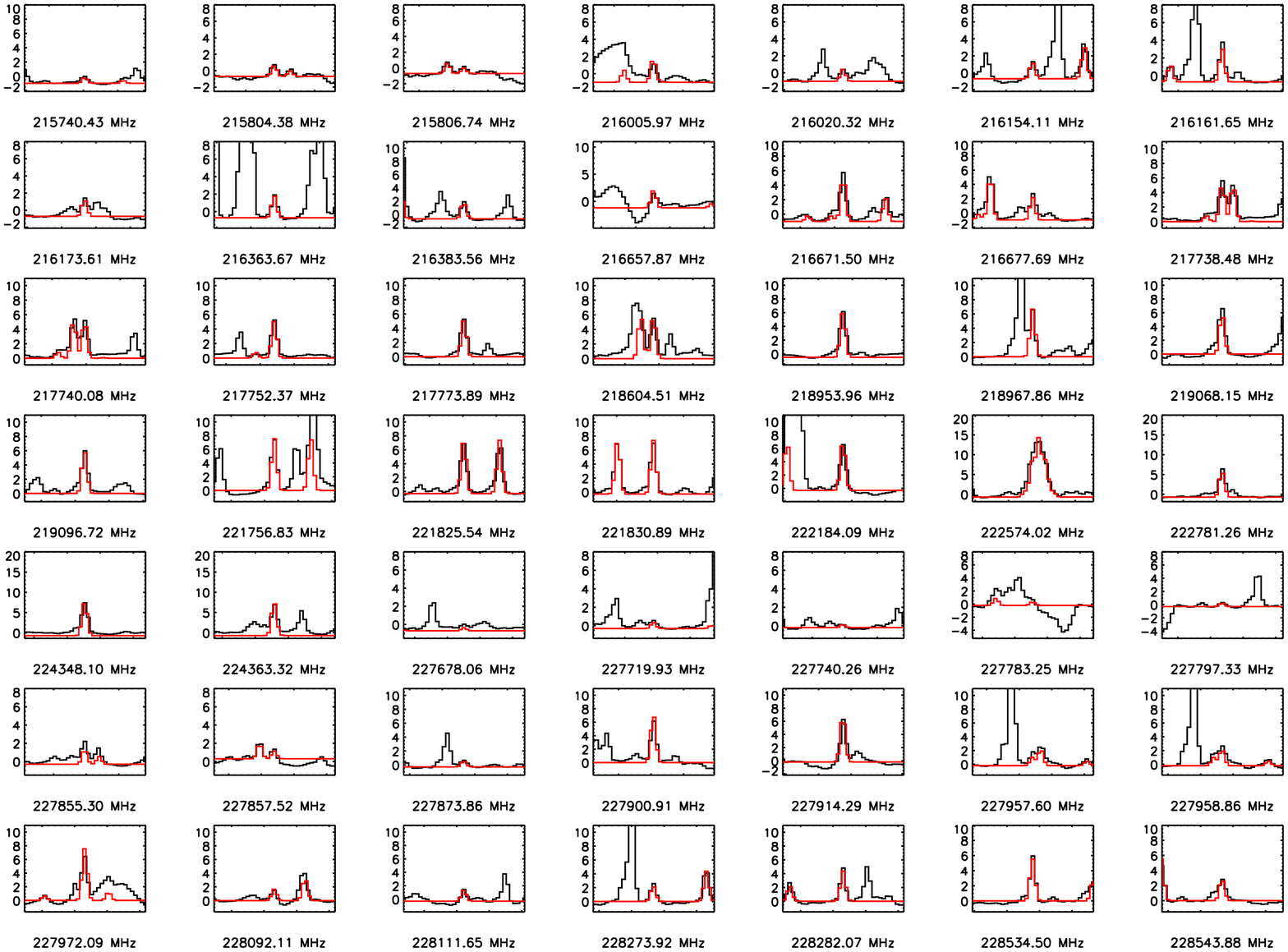}
\caption{HCOO$^{13}$CH$_{3}$ spectra (black) towards the Compact Ridge component associated with Orion-KL and model (red), as observed with ALMA. The intensity scale is in T$\rm_{MB}$ (K). The x-axis scale is about $\pm$9.5~MHz centered on the line rest frequency, which is indicated below each plot.\label{fga3}}
\end{figure} 

\begin{figure}
\figurenum{C-3}
\includegraphics[angle=270,width=15cm]{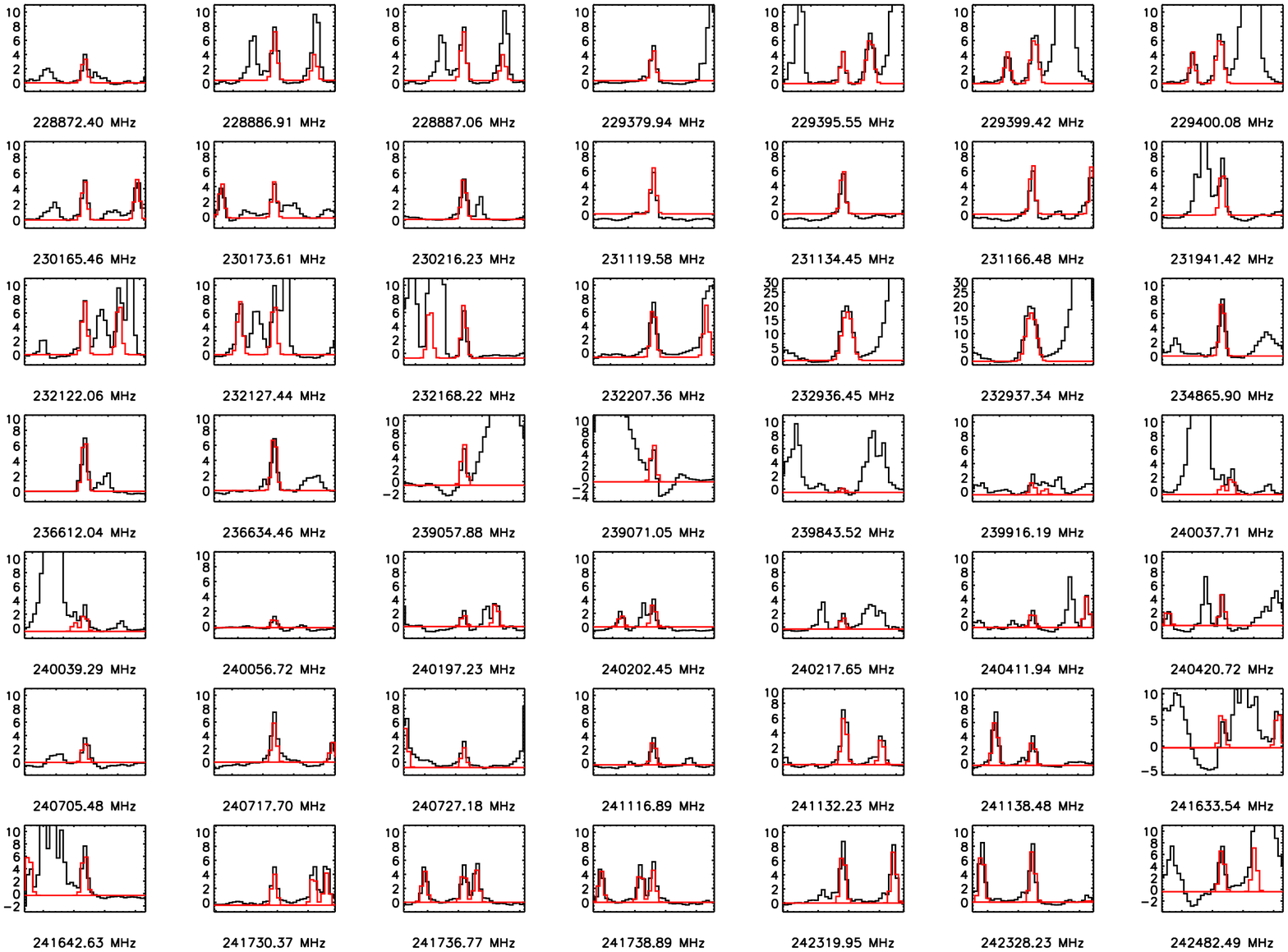}
\caption{Continue.}
\end{figure} 

\begin{figure}
\figurenum{C-3}
\includegraphics[angle=270,width=15cm]{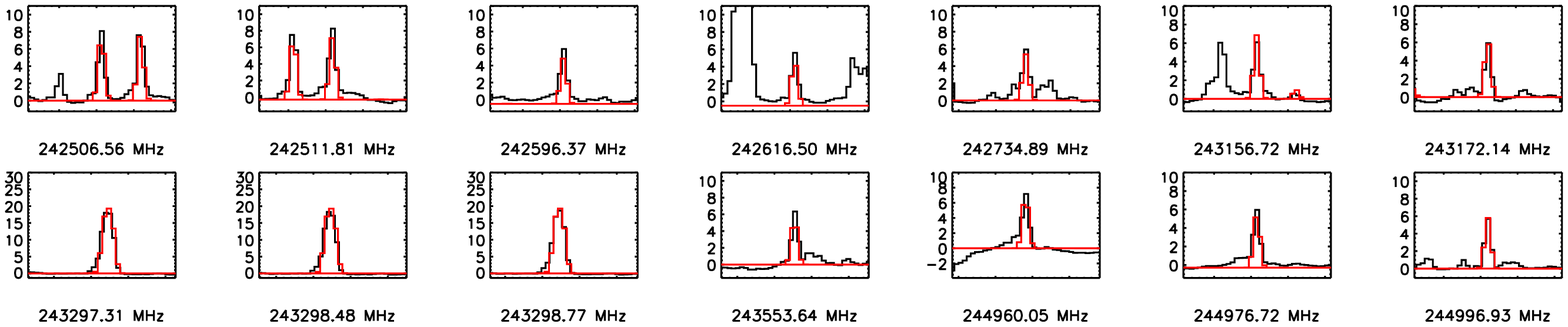}
\caption{Continue.}
\end{figure}

\newpage
 \section{Methyl formate emission towards the Orion-KL Hot Core-SW as observed with ALMA}
 \label{AP:HC}

Figures~\ref{fgb1}, \ref{fgb2} and \ref{fgb3} show the respective spectra of the detected HCOOCH$_{3}$, H$^{13}$COOCH$_{3}$ and HCOO$^{13}$CH$_{3}$ transitions observed with ALMA toward the Orion-KL Hot Core-SW during the science verification program, along with our best models achieved using the XCLASS program.

\begin{figure}
\figurenum{D-1}
\includegraphics[angle=270,width=15cm]{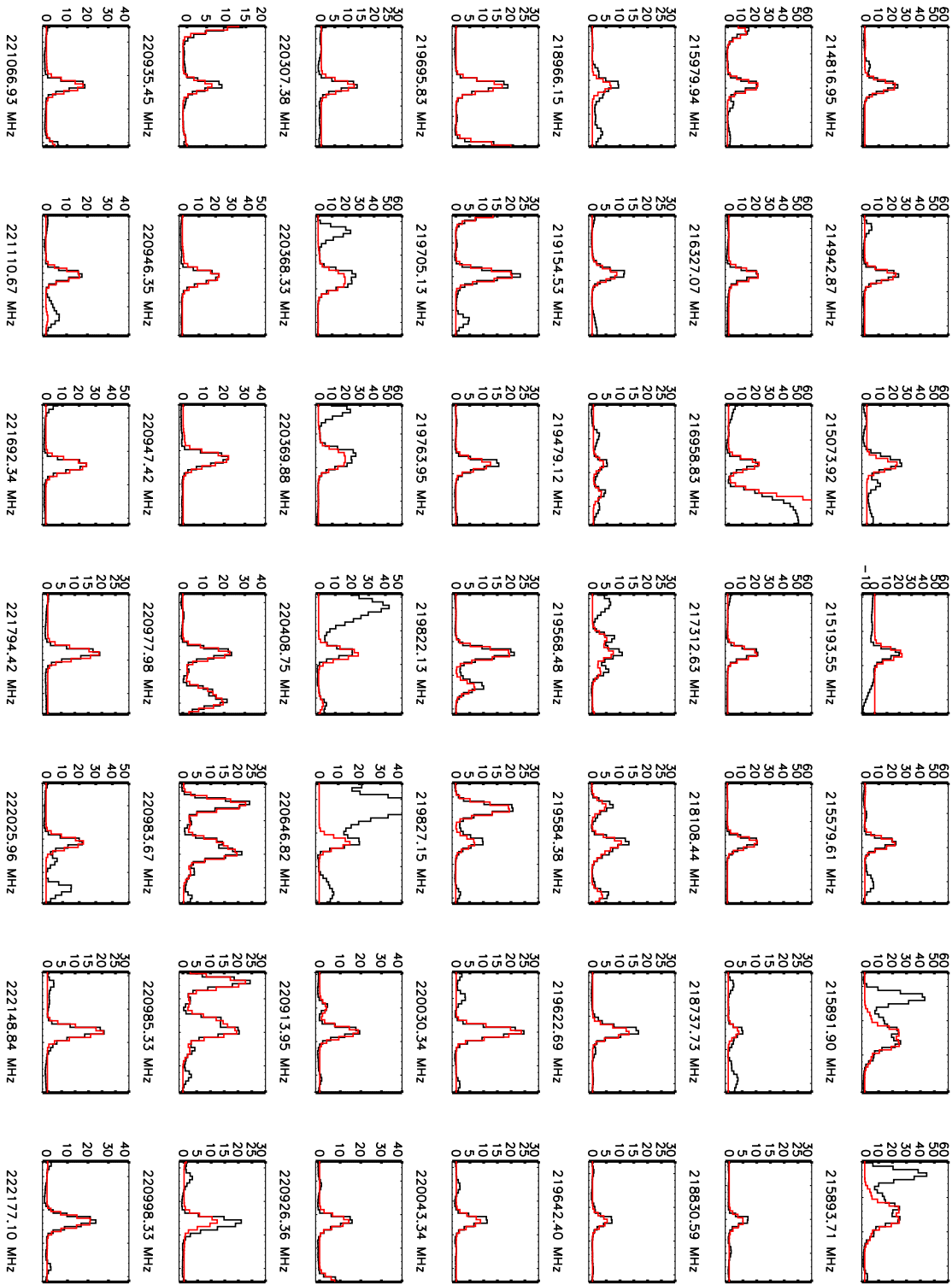}
\caption{HCOOCH$_{3}$ spectra (black) towards the Hot Core-SW component associated with Orion-KL and model (red), as observed with ALMA. The intensity scale is in T$\rm_{MB}$ (K). The x-axis scale is about $\pm$9.5~MHz centered on the line rest frequency, which is indicated below each plot.\label{fgb1}}
\end{figure} 

\begin{figure}
\figurenum{D-1}
\includegraphics[angle=270,width=15cm]{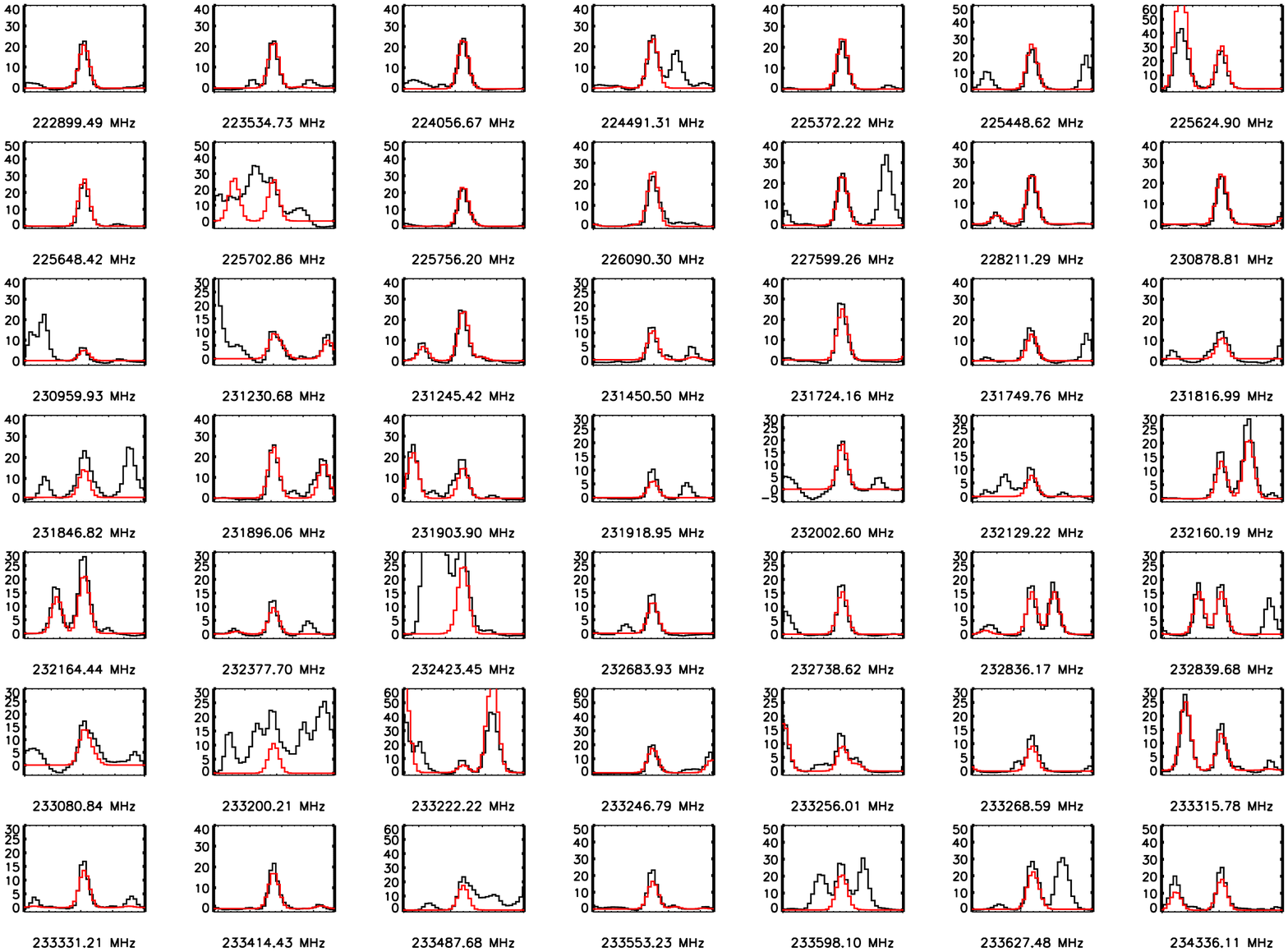}
\caption{Continue}
\end{figure} 

\begin{figure}
\figurenum{D-1}
\includegraphics[angle=270,width=15cm]{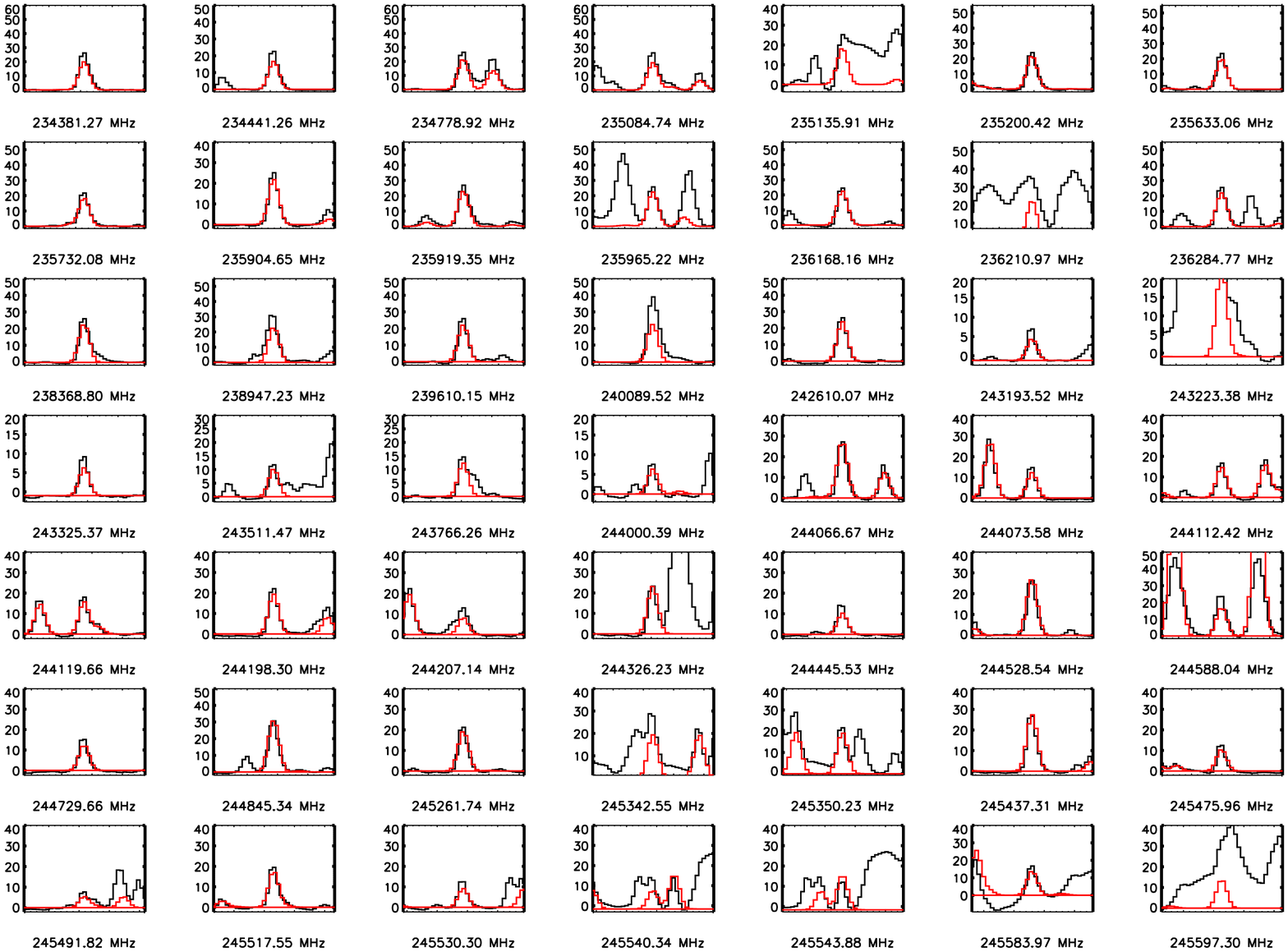}
\caption{Continue.}
\end{figure} 

\begin{figure}
\figurenum{D-1}
\includegraphics[angle=270,width=15cm]{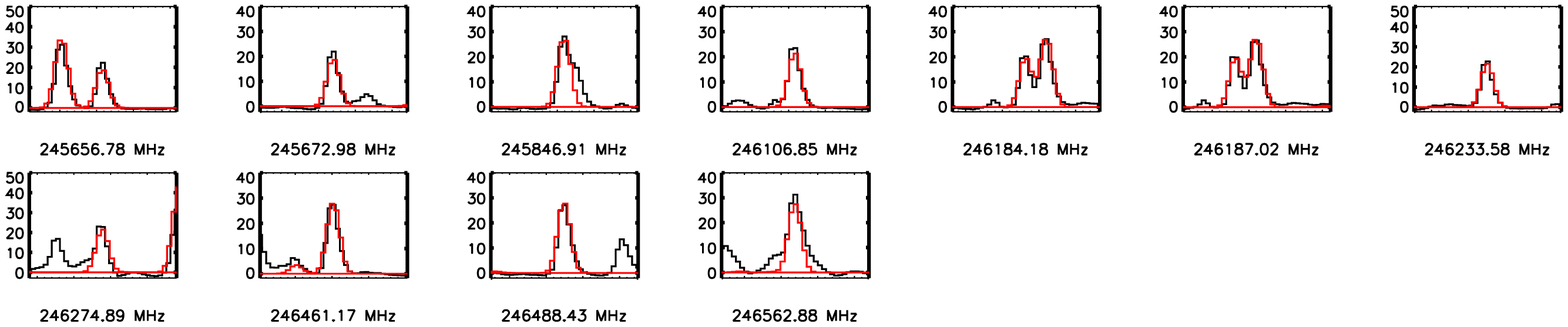}
\caption{Continue.}
\end{figure} 

\begin{figure}
\figurenum{D-2}
\includegraphics[angle=270,width=15cm]{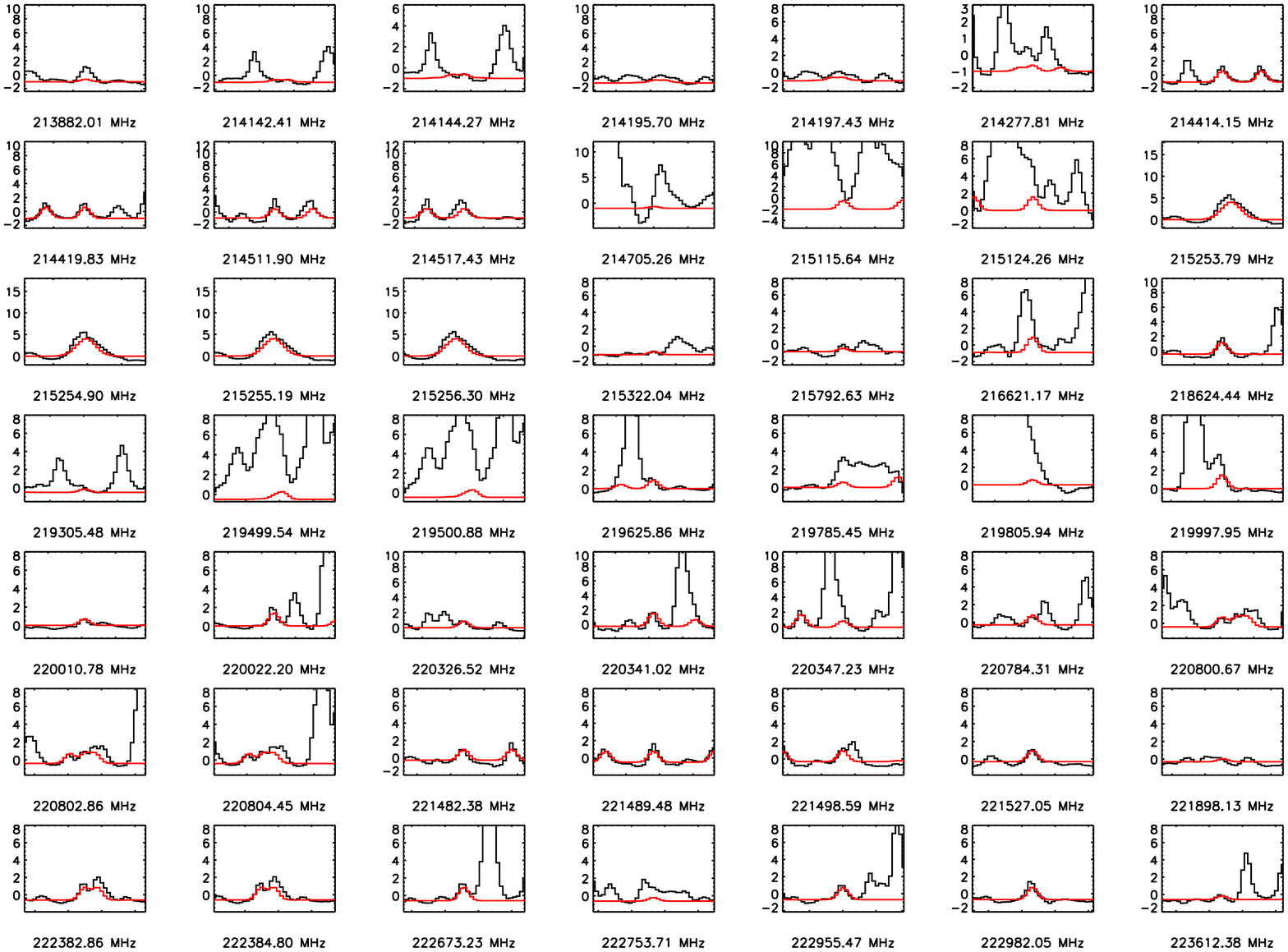}
\caption{H$^{13}$COOCH$_{3}$ spectra (black) towards the Hot Core-SW component associated with Orion-KL and model (red), as observed with ALMA. The intensity scale is in T$\rm_{MB}$ (K). The x-axis scale is about $\pm$9.5~MHz centered on the line rest frequency, which is indicated below each plot.\label{fgb2}}
\end{figure} 

\begin{figure}
\figurenum{D-2}
\includegraphics[angle=270,width=15cm]{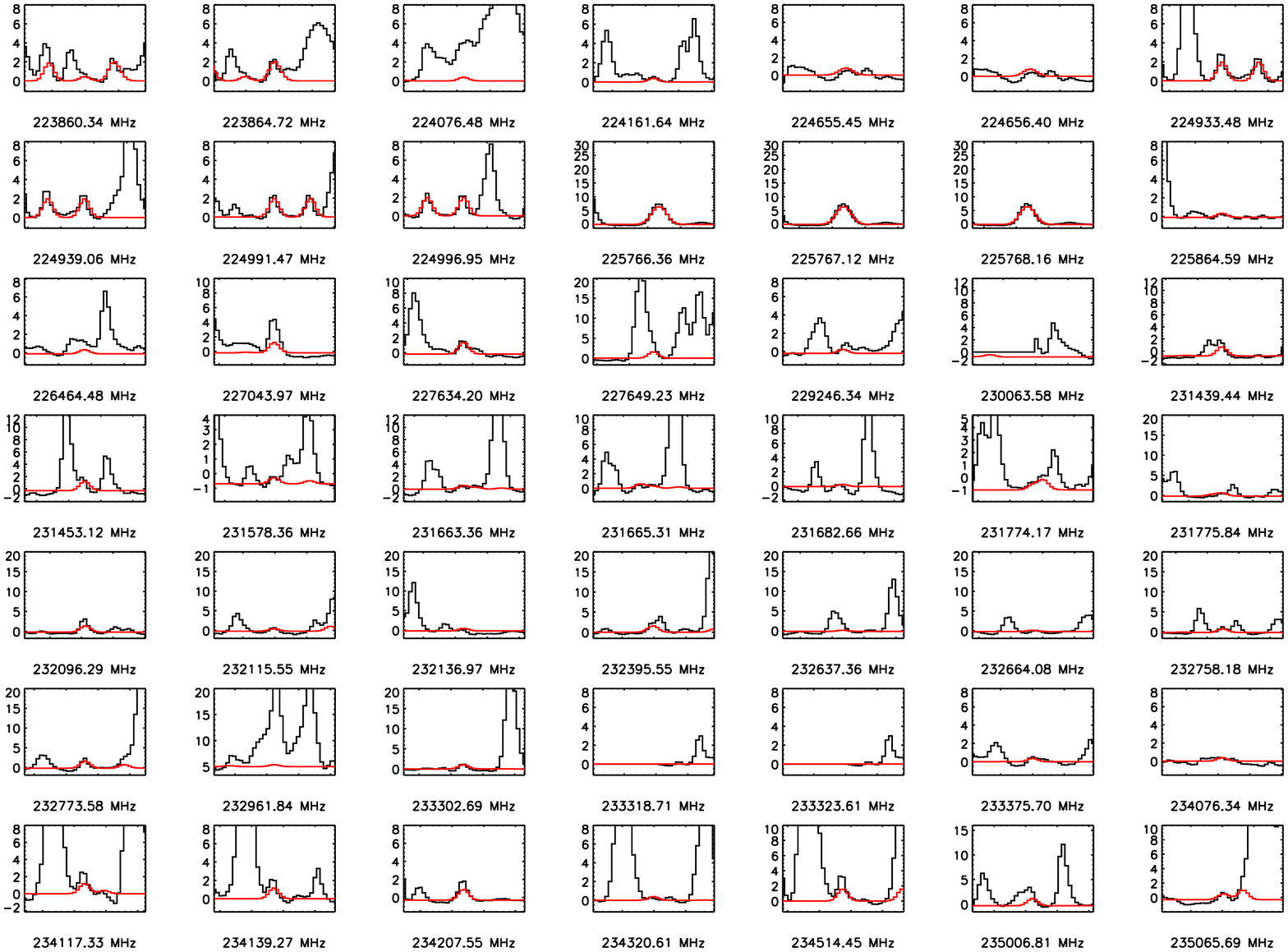}
\caption{Continue.}
\end{figure} 

\begin{figure}
\figurenum{D-2}
\includegraphics[angle=270,width=15cm]{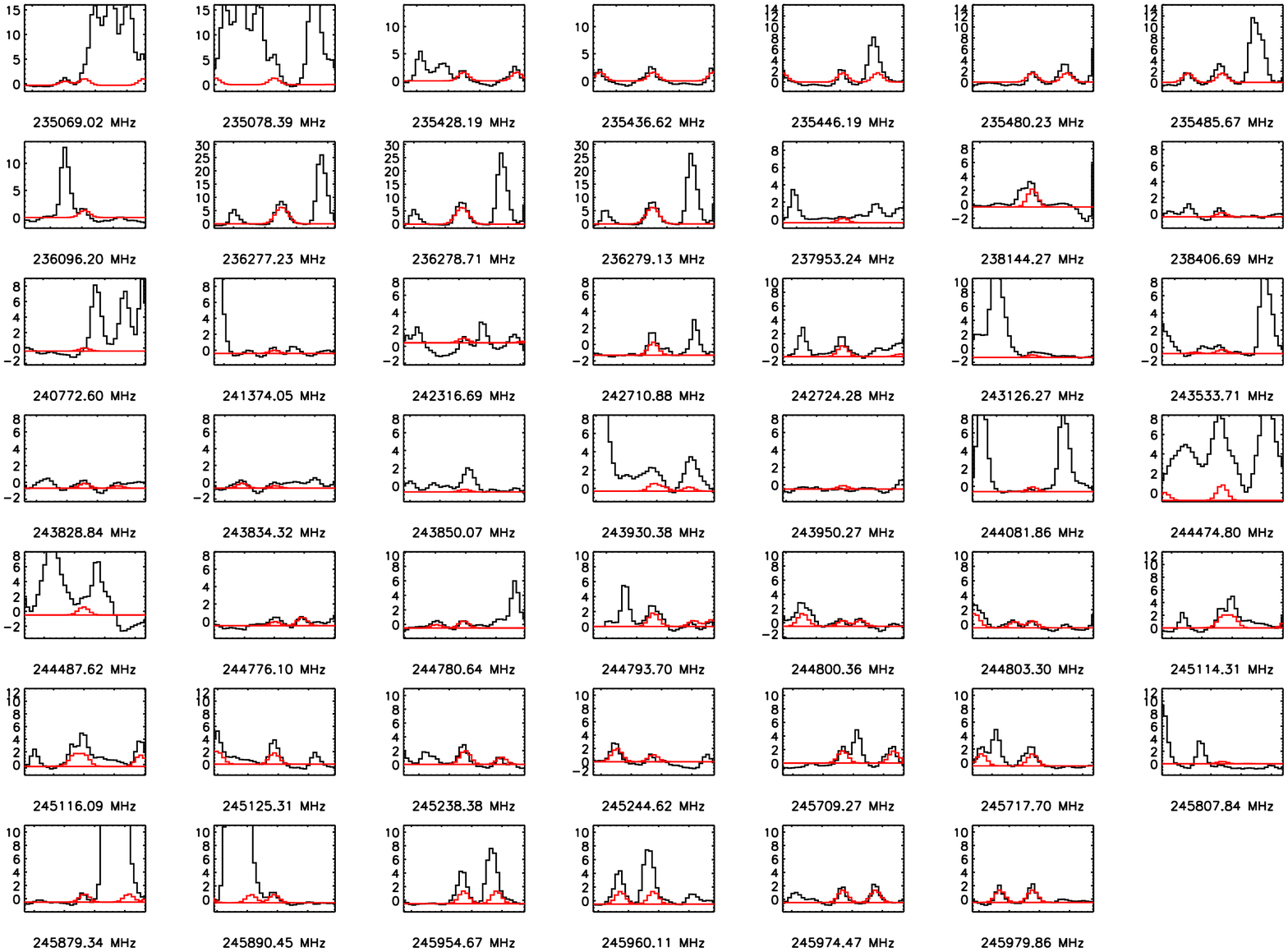}
\caption{Continue.}
\end{figure} 

\begin{figure}
\figurenum{D-3}
\includegraphics[angle=270,width=15cm]{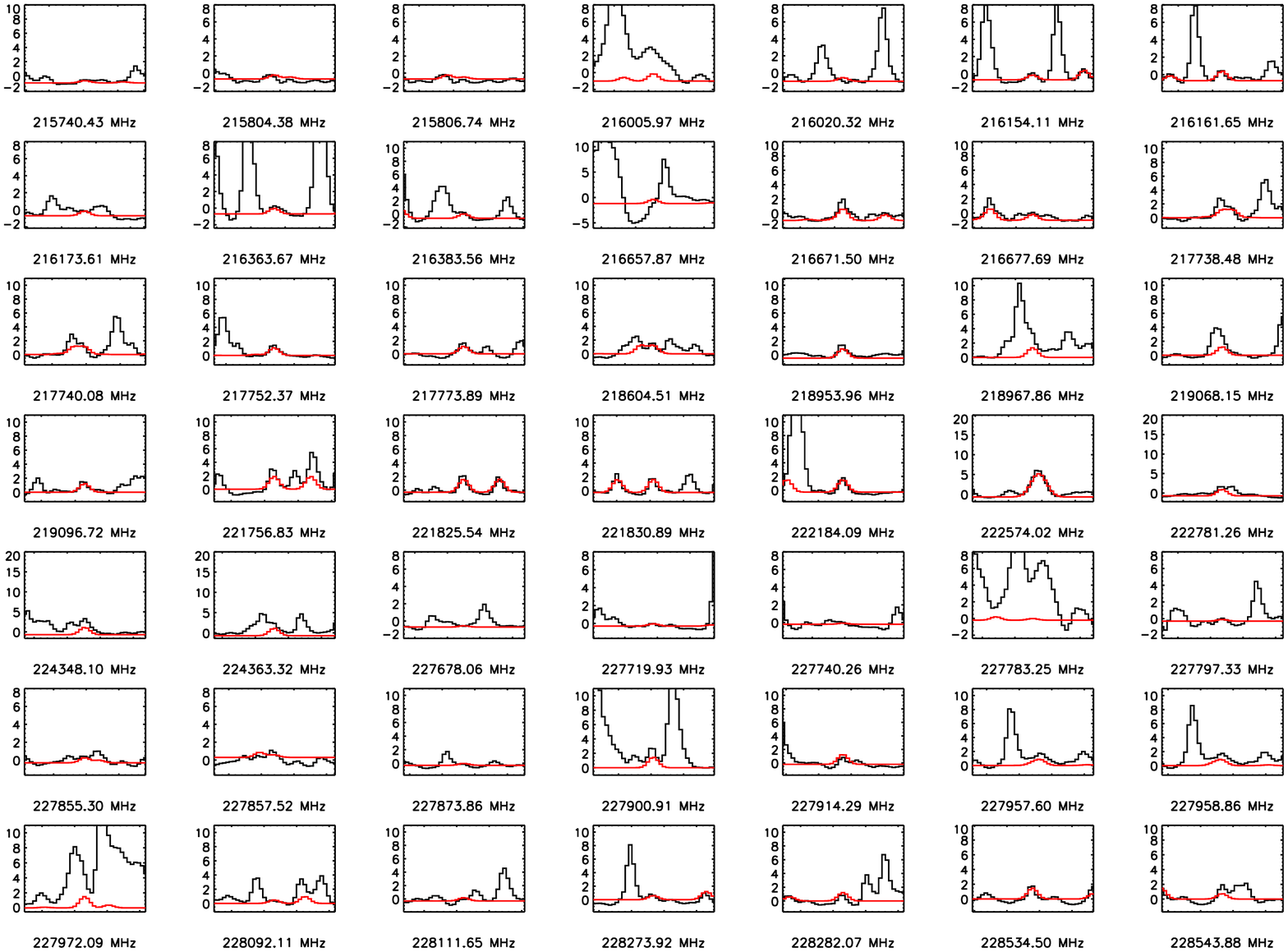}
\caption{HCOO$^{13}$CH$_{3}$ spectra (black) towards the Hot Core-SW component associated with Orion-KL and model (red), as observed with ALMA. The intensity scale is in T$\rm_{MB}$ (K). The x-axis scale is about $\pm$9.5~MHz centered on the line rest frequency, which is indicated below each plot.\label{fgb3}}
\end{figure} 

\begin{figure}
\figurenum{D-3}
\includegraphics[angle=270,width=15cm]{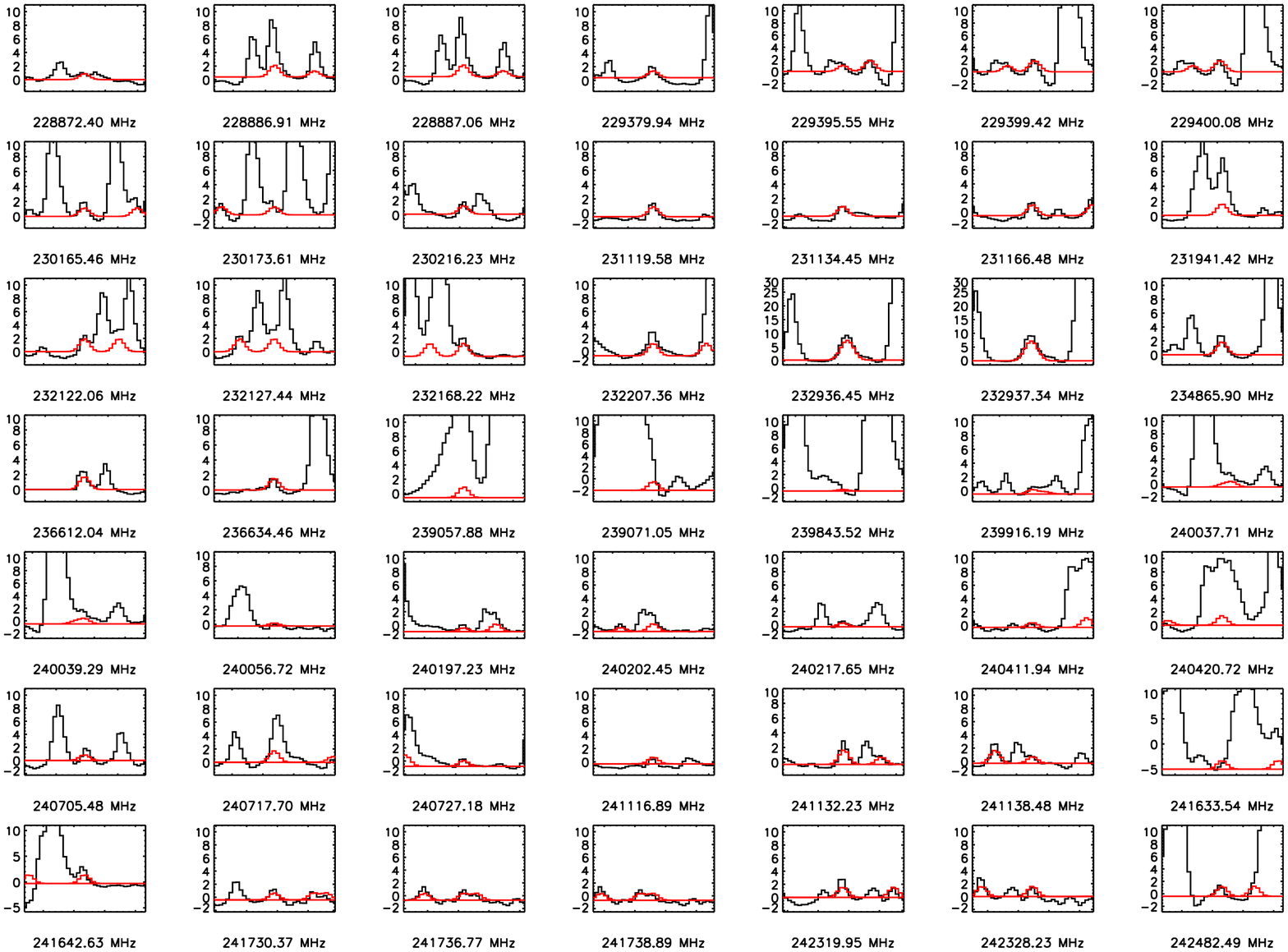}
\caption{Continue.}
\end{figure} 

\begin{figure}
\figurenum{D-3}
\includegraphics[angle=270,width=15cm]{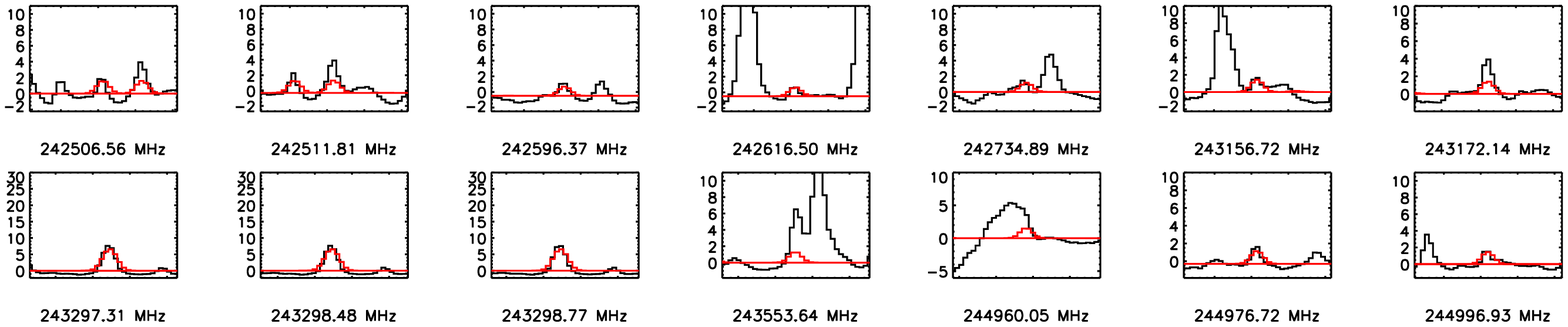}
\caption{Continue.}
\end{figure}

%
\bibliographystyle{apj}


\end{document}